%% file: area14VII2000.tex
\documentclass[12pt]{article}

\usepackage[mathscr]{eucal}
\usepackage{epsfig}
\usepackage{citesort,bibmods}
\usepackage{epic,eepic} 
\usepackage{amscd}
\usepackage{amssymb,latexsym}
\usepackage{theorem}      
\usepackage{a4}

\usepackage{eucal}                      
\def\greaterthansquiggle{\raise.3ex\hbox{$>$\kern-.75em\lower1ex\hbox{$\sim$}}}
\def\lessthansquiggle{\raise.3ex\hbox{$<$\kern-.75em\lower1ex\hbox{$\sim$}}}

\newcommand\GregS{\Sigma}
\newcommand\emb{}
\newcommand\Gregbeq{\begin{eqnarray}}
\newcommand\Gregeeq{\end{eqnarray}}
\newcommand\bbR{{\mathbb R}}
\newcommand\Gregd{{\partial}}
\newcommand\Gregn{{\nabla}}
\newcommand\Dgreg{{\nabla}}
\newcommand\nullhyp{\cH}
\newcommand\Lip{\mathrm{Lip}}

\newcommand{\reg}{\mbox{\scriptsize reg}}    
\newcommand{\la}{\langle}               
\newcommand{\ra}{\rangle}
 
\newcommand{\nnor}{{\mathbf{k}}}
\newcommand{\f}{\partial}
\renewcommand{\:}{\colon}

\newtheorem{thm}{Theorem}[section]
\newtheorem{lemma}[thm]{Lemma}
\newtheorem{prop}[thm]{Proposition}
\newtheorem{cor}[thm]{Corollary}

{\theorembodyfont{\rmfamily} \newtheorem{remark}[thm]{Remark} }

\newtheorem{Theorem} [thm]{Theorem} 
\newtheorem{Corollary} [thm] {Corollary}
\newtheorem{Lemma} [thm] {Lemma}
\newtheorem{Proposition} [thm] {Proposition}

\newtheorem{Definition}[thm]{Definition}

\newcommand{\Lpsi}{L_1}      
\newcommand{\Qpsi}{\alpha_1} 
\newcommand{\Lphif}{L_2}     
\newcommand{\Qphif}{\alpha_2} 

\newcommand{\BA}{B_\Al}

\newcommand{\Ad}{A_{\delta}}

\newcommand{\hs}{\cH_{\mbox{\scriptsize sing}}}

\newcommand{\sh}{\Sigma_{\cH}} 
\newcommand{\sdh}{\Sigma_{\dH}} 
 
\newcommand{\vn}{{\vec n}}

\newcommand{\myremark}{\begin{remark}\rm} 
\newcommand{\myendremark}{\end{remark}}

\newcommand{\dG}{\dot \Gamma}

\newcommand{\dH}{{\cH}_{\Al}} 
\newcommand{\Htwok}{\Hau_{\kaux}^{n-1}} 
\newcommand{\Hausone}{\Hau_h^1}
\newcommand{\Hsone}{\Hau_{\kaux}^1} 

\newcommand{\Hnk}{\Hau_{\kaux}^n}
\newcommand{\Hnmk}{\Hau_{\kaux}^{n-1}}

\newcommand{\Htwoh}{\Hau_{h}^{n-1}} 
\newcommand{\Htwohone}{\Hau_{h_1}^{n-1}} 
 
\newcommand{\Leb}{{\mathfrak L}}

\newcommand{\metricS}{h_S}
\newcommand{\LtwoS}{{\Leb}_{\metricS}^{n-1}} 
\newcommand{\Lnm}{{\Leb}_{h_{\cV}}^{n-1}} 
 
\newcommand{\Lthreenew}{{\Leb}_{h_0}^{n}} 
\newcommand{\Ar}{\mbox{\rm Area}} 
\newcommand{\Arm}{\mbox{\rm Area}_{S_2}} 
\newcommand{\pr}{\mbox{\rm pr}}

\newcommand{\hor}{\cH} 
\newcommand{\Hau}{{\mathfrak H}}  

\newcommand{\hmtwo}{\Hau_{h_2}}
\newcommand{\hmone}{\Hau_{h_1}}

\title{Regularity of Horizons and The Area Theorem}

\author{Piotr T.\ Chru\'sciel\thanks{ Supported in part by KBN grant
    \# 2 P03B 130 16. \emph{E--mail}: Chrusciel@Univ-Tours.fr}\qquad
  Erwann Delay\thanks{Current adress: Department of Mathematics,
Royal Institute of Technology, S--10044 Stockholm. \emph{E--mail}:
    Delay@gargan.math.Univ-Tours.fr } \\ D\'epartement de
  Math\'ematiques\\ Facult\'e des Sciences\\ Parc de Grandmont\\ 
  F37200 Tours, France\\ \\ Gregory J. Galloway\thanks{Supported in
    part by NSF grant \# DMS-9803566.  \emph{E--mail}:
    galloway@math.miami.edu}\\ Department of Mathematics and Computer
  Science\\ University of Miami\\ Coral Gables FL 33124, USA\\ \\ 
  Ralph Howard\thanks {Supported in part by DoD Grant \#
    N00014-97-1-0806 \emph{E--mail}: howard@math.sc.edu} \\ Department
  of Mathematics\\ University of South Carolina \\ Columbia S.C.
  29208, USA}

\newcommand{\A}{{\mathscr  A}}

\newcommand{\cH}{{\mathscr H}}

\newcommand{\Al}{{\mathcal Al}}

\newcommand{\cB}{{\mathscr  B}}
\newcommand{\cW}{{\mathscr  W}}
\newcommand{\cU}{{\mathscr  U}}
\newcommand{\cV}{{\mathscr  V}}

\newcommand{\cN}{{{\mathscr  N}}{}}

\newcommand{\cD}{{\mathscr  D}}

\newcommand{\cO}{{\mathscr  O}}
\newcommand{\loc}{\mbox{\scriptsize \rm loc}}

\newcommand{\text}[1]{\mbox{\rm #1}}
\newcommand{\qed}{\hfill $\Box$ \medskip} 

\newcommand{\gnor}{\mathbf{m}}
\newcommand{\mygnor}{\mathbf{m}}
\newcommand{\bn}{\mathbf{n}}

\newcommand{\kaux}{\sigma}   
\newcommand{\sdist}{\mbox{\rm {$\kaux$-dist}}}
\newcommand{\Ric}{\mbox{\rm Ric}}

\renewcommand{\emptyset}{\varnothing} 

\newcommand{\Cloc}{C_{{\text{\scriptsize\rm loc}}}}
\newcommand{\nor}{{\mathbf{n}}}

\newcommand{\e}{\varepsilon}    

\newcommand{\nothing}{\varnothing}

\newcommand{\rems}{\myremark}

\newcommand{\beq}{\begin{equation}}

\newcommand{\eeq}{\end{equation}}
\newcommand{\ee}{\end{equation}}
\newcommand{\beqa}{\begin{eqnarray}}
\newcommand{\eeqa}{\end{eqnarray}}
\newcommand{\beqan}{\begin{eqnarray*}}
\newcommand{\eeqan}{\end{eqnarray*}}
\newcommand{\ba}{\begin{array}}
\newcommand{\ea}{\end{array}}

\newcommand{\ol}{\overline}

\def\nz{\ifmmode {I\hskip -3pt N} \else {\hbox {$I\hskip -3pt N$}}\fi}
\def\zz{\ifmmode {Z\hskip -4.8pt Z} \else
       {\hbox {$Z\hskip -4.8pt Z$}}\fi}
\def\qz{\ifmmode {Q\hskip -5.0pt\vrule height6.0pt depth 0pt
       \hskip 6pt} \else {\hbox
       {$Q\hskip -5.0pt\vrule height6.0pt depth 0pt\hskip 6pt$}}\fi}
\def\rz{\ifmmode {I\hskip -3pt R} \else {\hbox {$I\hskip -3pt R$}}\fi}
\def\cz{\ifmmode {C\hskip -4.8pt\vrule height5.8pt\hskip 6.3pt} \else
       {\hbox {$C\hskip -4.8pt\vrule height5.8pt\hskip 6.3pt$}}\fi}
\def\au{{\setbox0=\hbox{\lower1.36775ex\hbox{''}\kern-.05em}\dp0=.36775ex\hs
kip0pt\box0}}
\def\ao{{}\kern-.10em\hbox{``}}

{\catcode `\@=11 \global\let\AddToReset=\@addtoreset}
\AddToReset{equation}{section}

\newcounter{mnotecount}[section]

\newcommand{\mnote}[1]{}

\DeclareFontFamily{OT1}{rsfs}{} 
\DeclareFontShape{OT1}{rsfs}{m}{n}{ <-7> rsfs5 <7-10> rsfs7 <10->
rsfs10}{} 
\DeclareMathAlphabet{\mycal}{OT1}{rsfs}{m}{n} 
\def\scri{{\mycal I}}%
\def\scrip{\scri^{+}}%
\def\Scri{\scri}
\def\Scrip{\scrip}

\newcommand{\tA}{\theta_\Al}

\newcommand{\eq}[1]{(\ref{#1})}

\newcommand{\commentout}[1]{}

\newcommand{\bea}{\begin{eqnarray}}
\newcommand{\eea}{\end{eqnarray}}
\newcommand{\beaa}{\begin{eqnarray*}}
\newcommand{\eeaa}{\end{eqnarray*}}

\newcommand{\N}{{\mathbb N}}
\newcommand{\R}{{\mathbb R}}

\newcommand{\Gint}{\Gamma_{\mbox{\rm\scriptsize{int}}}}

\newcommand{\calS}{{\mathscr S}}
\newcommand{\tAcS}{\theta^{\calS\cap\cH}_{\Al}}
\newcommand{\BAcS}{B^{\calS\cap\cH}_{\Al}}

\newcommand{\myi}{i}%
\newcommand{\myj}{j}%

\newcommand{\transversally}{properly transversally}

\begin{document}

\maketitle

\begin{abstract} We prove that the area of sections of
  future event horizons in space--times satisfying the null energy
  condition is non--decreasing towards the future under the following
  circumstances: 1) the horizon is future geodesically complete; 2)
  the horizon is a black hole event horizon in a globally hyperbolic
  space--time and there exists a conformal completion with a
  ``$\cH$--regular'' $\scri^+$; 3) the horizon is a black hole event
  horizon in a space--time which has a globally hyperbolic conformal
  completion. (Some related results under less restrictive hypotheses
  are also established.) This extends a theorem of Hawking, in which
  piecewise smoothness of the event horizon seems to have been
  assumed. We prove smoothness or analyticity of the relevant part of
  the event horizon when equality in the area inequality is attained
  --- this has applications to the theory of stationary black holes,
  as well as to the structure of compact Cauchy horizons. In the
  course of the proof we establish several new results concerning the
  differentiability properties of horizons.
\end{abstract}
\tableofcontents
\section{Introduction}
\label{introduction}
The thermodynamics of black holes rests upon
\emph{Hawking's area theorem} \cite{HE} which asserts that in
appropriate space--times the area of cross--sections of black hole
horizons is non--decreasing towards the future. In the published
proofs of this result \cite{HE,Waldbook} there is considerable
vagueness as to the hypotheses of differentiability of the event
horizon (see, however, \cite{GiuliniArea}).  Indeed, it is known that
black hole horizons can be pretty rough \cite{ChGalloway}, and it is
not immediately clear that the area of their cross--sections can even
be defined. The reading of the proofs given in \cite{HE,Waldbook}
suggests that those authors have assumed the horizons under
consideration to be piecewise $C^2$. Such a hypothesis is certainly
incompatible with the examples constructed in \cite{ChGalloway} which
are \emph{nowhere} $C^2$.  The aim of this paper is to show that the
monotonicity theorem holds without any further differentiability
hypotheses on the horizon, in appropriate space--times, for a large
class of cross--sections of the horizon.  More precisely we show the
following:
  \begin{Theorem}[The area theorem]
    \label{Tarea}
    Let $\cH$ be a black hole event horizon in a smooth space--time
    $(M,g)$.  Suppose that either
     \begin{enumerate}\renewcommand{\theenumi}{\alph{enumi}}
\renewcommand{\labelenumi}{\theenumi)}
 \item \label{firsthyp} $(M,g)$ is globally hyperbolic, and there exists
   a conformal completion $(\bar M,\bar g)=(M\cup \scri^+,\Omega^2g)$
   of $(M,g)$ with a $\cH$--regular\footnote{See Section~\ref{Sarea2}
     for definitions.}  $\scri^+$.  Further the null
   energy condition holds on the past $I^-(\scri^+;\bar M)\cap M$ of
   $\scri^+$ in $M$, or
     \item\label{secondpoint} the generators of $\cH$ are future
       complete and the null energy condition holds on $\cH$, or 
     \item \label{thirdhyp}there exists a globally hyperbolic
       conformal completion $(\bar M,\bar g)=(M\cup
       \scri^+,\Omega^2g)$ of $(M,g)$, with the null energy condition
       holding on $I^-(\scri^+;\bar M)\cap M$. 
     \end{enumerate}
     Let $\Sigma_a$, $a=1,2$ be two achronal spacelike embedded
     hypersurfaces of $C^2$ differentiability class, set $S_a=
     \Sigma_a\cap\cH$. Then:
     \begin{enumerate}
 \item The area $\Ar(S_a)$ of $S_a$ is well defined.
   \item If $$S_1\subset J^-(S_2)\ , 
     $$ then the area of $S_2$ is larger than or equal to that of
     $S_1$.  (Moreover, this is true even if the area of $S_1$ is
     counted with multiplicity of generators provided that $S_1\cap
     S_2 = \nothing$, see Theorem \ref{thm:local-area}.)
\end{enumerate}
\end{Theorem}

Point~1.\ of Theorem~\ref{Tarea} is Proposition \ref{PHaus} below, see
also Proposition~\ref{Phow}. Point~2.\ above follows immediately from
Proposition~\ref{Pghwb}, Corollary~\ref{Cglobal} and
Proposition~\ref{Ptheta2}, as a special case of the first part of
Theorem~\ref{thm:local-area} below.\footnote{The condition $S_1\cap
  S_2 = \nothing$ of Theorem~\ref{thm:local-area} is needed for
  monotonicity of ``area with multiplicity'', but is not needed if one
  only wants to compare standard areas, see Remark~\ref{R5.2}.} The
question of how to define the area of sections of the horizon is
discussed in detail in Section~\ref{Ssections}.  It is suggested there
that a notion of area, appropriate for the identification of area with
the entropy, should include the multiplicity of generators of the
horizon.

We stress that we are \emph{not} assuming that $\scrip$ is null --- in
fact it could be even changing causal type from point to point --- in
particular Theorem~\ref{Tarea} also applies when the cosmological
constant does not vanish. In point \ref{thirdhyp}) global
hyperbolicity of $(\bar M,\bar g)$ should be understood as that of a
manifold with boundary, \emph{cf.} Section \ref{Sarea2}.  Actually in
points \ref{firsthyp}) and \ref{thirdhyp}) of Theorem~\ref{Tarea} we
have assumed global hyperbolicity of $(M,g)$ or that of $(\bar M,\bar
g)$ for simplicity only: as far as Lorentzian causality hypotheses are
concerned, the assumptions of Proposition~\ref{Ptheta1} are sufficient
to obtain the conclusions of Theorem~\ref{Tarea}.  We show in Section
\ref{Sarea2} that those hypotheses will hold under the conditions of
Theorem~\ref{Tarea}.  Alternative sets of causality conditions, which
do not require global hyperbolicity of $(M,g)$ or of its conformal
completion, are given in Propositions~\ref{Pglobal1} and
\ref{Pglobal2}.

It seems useful to compare our results to other related ones existing
in the literature \cite{HE,Waldbook,Krolak:bh}.  First, the hypotheses
of point \ref{firsthyp}) above are fulfilled under the conditions of
the area theorem of \cite{Waldbook} (after replacing the space--time
$M$ considered in \cite{Waldbook} by an appropriate subset thereof),
and (disregarding questions of differentiability of $\cH$) are
considerably weaker than the hypotheses of \cite{Waldbook}.  Consider,
next, the original area theorem of \cite{HE}, which we describe in
some detail in Appendix~\ref{HEarea} for the convenience of the
reader.  We note that we have been unable to obtain a proof of the
area theorem without some condition of causal regularity of $\Scri^+$
(\emph{e.g.} the one we propose in point a) of Theorem \ref{Tarea}),
and we do not know\footnote{The proof of Proposition 9.2.1 in
  \cite{HE} would imply that the causal regularity condition assumed
  here holds under the conditions of the area theorem of \cite{HE}.
  However, there are problems with that proof (this has already been
  noted by Newman \cite{Newman:coscen}).} whether or not the area
theorem holds under the original conditions of \cite{HE} without the
modifications indicated above, or in Appendix~\ref{HEarea}; see
Appendix~\ref{HEarea} for some comments concerning this point.
 
Let us make a few comments about the strategy of the proof of
Theorem~\ref{Tarea}.  It is well known that event horizons are
Lipschitz hypersurfaces, and the examples constructed in
\cite{ChGalloway,BKK} show that much more cannot be expected. We start
by showing that horizons are {\em semi--convex}\footnote{Actually this
  depends upon the time orientation: \emph{future} horizons are {\em
    semi--convex}, while \emph{past} horizons are {\em
    semi--concave}.}. This, together with Alexandrov's theorem
concerning the regularity of convex functions shows that they are
twice differentiable in an appropriate sense (\emph{cf.\/}\ 
Proposition~\ref{Pdiff} below) almost everywhere. This allows one to
define notions such as the divergence $\theta_\Al$ of the generators
of the horizon $\cH$, as well as the divergence
$\theta_\Al^{\cH\cap\calS}$ of sections ${\cH\cap\calS}$ of $\cH$. The
existence of the second order expansions at Alexandrov points leads
further to the proof that, under appropriate conditions, $\theta_\Al$
or $\theta_\Al^{\cH\cap\calS}$ have the right sign.  Next, an
approximation result of
  Whitney type, Proposition~\ref{C11-extend}, 
  allows one to embed certain subsets of the horizon into $C^{1,1}$
  manifolds. The area theorem then follows
  from the change--of--variables theorem for Lipschitz maps proved in
  \cite{Federer:measures}. We note that some further effort is
  required to convert the information that $\theta_\Al$ has the
  correct sign into an inequality concerning the Jacobian that appears
  in the change of variables formula.
  
  Various authors have considered the problem of defining black holes
  in settings more general than standard asymptotically flat
  spacetimes or spacetimes admitting a conformal infinity; see
  especially \cite{GibbonsHawkingCEH,Krolak:scc,Krolak:coscen} and
  references cited therein.  It is likely that proofs of the area
  theorem given in  more general settings, for horizons assumed
  to be piecewise $C^2$, which are based on establishing the
  positivity of the (classically defined) expansion $\theta$ of the
  null generators can be adapted, using the methods of Section
  \ref{Sarea1} ({\em cf.\/} especially Proposition \ref{Ptheta1} and
  Lemma \ref{Ltheta2}), to obtain proofs which do not require the
  added smoothness.  \mnote{remark}(We show in Appendix \ref{krolak}
  that this is indeed the case for the area theorems of Kr\'olak
  \cite{Krolak:bh,Krolak:scc,Krolak:coscen}.) In all situations which
  lead to the positivity of $\theta$ in the weak Alexandrov sense
  considered here, the area theorem follows from Theorem
  \ref{thm:local-area} below.
 
  It is of interest to consider the equality case: as discussed in
  more detail in Section~\ref{Sconclusions}, this question is relevant
  to the classification of stationary black holes, as well as to the
  understanding of compact Cauchy horizons.  Here we prove the
  following:

  \begin{Theorem}
    \label{Trigidity}
Under the hypotheses of Theorem~\ref{Tarea}, suppose that 
the area of $S_1$ equals that of $S_2$. Then
$$(J^+(S_1)\setminus S_1)\cap (J^-(S_2)\setminus S_2)$$ is a
\emph{smooth} (analytic if the metric is analytic) null hypersurface
with vanishing null second fundamental form.  Moreover, if $\gamma$ is
a null generator of $\hor$ with $\gamma(0)\in S_1$ and $\gamma(1)\in
S_2$, then the curvature tensor of $(M,g)$ satisfies
$R(X,\gamma'(t))\gamma'(t)=0$ for all $t\in [0,1]$ and $X\in
T_{\gamma(t)}\cH$.
  \end{Theorem}

  Theorem~\ref{Trigidity} follows immediately from
  Corollary~\ref{Cglobal} and
  Proposition~\ref{Ptheta2}, as a special case of the second part of
  Theorem~\ref{thm:local-area} below.  The key step of the proof here
  is Theorem~\ref{Tsmoothness}, which has some interest in its own. An
  application of those results to stationary black holes is given in
  Theorem \ref{Tbh}, Section \ref{Sconclusions}.

  As already pointed out, one of the steps of the proof of
  Theorem~\ref{Tarea} is to establish that a notion of divergence
  $\tA$ of the generators of the horizon, or of sections of the
  horizon, can be
  defined almost everywhere, and that $\tA$ so defined is positive. We
  note that $\tA$ coincides with the usual divergence $\theta$ for
  horizons which are twice differentiable. Let us show, by means of an
  example, that the positivity of $\theta$ might fail to hold in
  space--times $(M,g)$ which do not satisfy the hypotheses of
  Theorem~\ref{Tarea}: Let $t$ be a standard time coordinate on the
  three dimensional Minkowski space--time $\R^{1,2}$, and let
  $K\subset \{t=0\}$ be an open conditionally compact set with smooth
  boundary $\partial K$.  Choose $K$ so that the mean curvature $H$ of
  $\partial K$ has changing
sign. Let 
$M=I^{-}(K)$, with the metric conformal to the Minkowski metric by a
conformal factor which is one in a neighborhood of $\partial
\cD^{-}(K;\R^{1,2})\setminus K$, and which makes $\partial
J^-{(K;\R^{1,2})}$ into $\Scri^+$ in the completion $\bar M\equiv M
\cup \partial J^-{(K;\R^{1,2})}$.
We have $M \setminus J^{-}(\Scri^+;\bar M)= 
\cD^{-}(K;\R^{1,2})\ne \nothing$, thus $M$ contains a black hole
region, with the event horizon being the Cauchy horizon $\partial
\cD^{-}(K;\R^{1,2})\setminus K$. The generators of the event horizon
coincide with the generators of $\partial \cD^{-}(K;\R^{1,2})$, which
are null geodesics normal to $\partial K$.  Further for $t$ negative
and close to zero the divergence $\theta$ of those generators is well
defined in a classical sense (since the horizon is smooth there) and
approaches, when $t$ tends to zero along the generators, the mean
curvature $H$ of $\partial K$.  Since the conformal factor equals one
in a neighborhood of the horizon the null energy condition holds
there, and $\theta$ is negative near those points of $\partial K$
where $H$ is negative. This implies also the failure of the area
theorem for some (local) sections of the horizon. We note that
condition b) of Theorem~\ref{Tarea} is not satisfied because the
generators of $\cH$ are not future geodesically complete. On the other
hand condition a) does not hold because $g$ will not satisfy the null
energy condition throughout $I^{-}(\Scri^+;\bar M)$ whatever the
choice of the conformal factor\footnote{This follows from
  Theorem~\ref{Tarea}.}.

This paper is organized as follows: In Section \ref{Shorizons} we show
that future horizons, as defined there, are always \emph{semi--convex}
(Theorem \ref{T1}). This allows us to define such notions as the
\emph{Alexandrov divergence} $\tA$ of the generators of the horizon,
and their \emph{Alexandrov} null second fundamental form.  In Section
\ref{Ssections} we consider sections of horizons and their geometry,
in particular we show that sections of horizons have a well defined
area. We also discuss the ambiguities which arise when defining the
area of those sections when the horizon is not \emph{globally smooth}.
In fact, those ambiguities have nothing to do with ``very low''
differentiability of horizons and arise already for piecewise smooth
horizons. In Section \ref{Sarea1} we prove positivity of the
Alexandrov divergence of generators of horizons --- in Section
\ref{Sarea2} this is done under the hypothesis of existence of a
conformal completion satisfying a regularity condition, together with
some global causality assumptions on the space--time; in Section
\ref{sScomplete} positivity of $\tA$ is established under the
hypothesis that the generators of the horizon are future complete. In
Section \ref{Sfr} we show that Alexandrov points \emph{``propagate to
  the past''} along the generators of the horizon. This allows one to
show that the optical equation holds on ``almost all'' generators of
the horizon. We also present there a theorem (Theorem \ref{TCfr})
which shows that ``almost all generators are Alexandrov''; while this
theorem belongs naturally to Section \ref{Sfr}, its proof uses methods
which are developed in Section \ref{Smonotonicity} only, therefore it
is deferred to Appendix \ref{ACfr}. In Section \ref{Smonotonicity} we
prove our main result -- the monotonicity theorem, Theorem
\ref{thm:local-area}.  This is done under the assumption that $\tA$ is
non--negative.  One of the key elements of the proof is a new (to us)
extension result of Whitney type (Proposition \ref{C11-extend}), the
hypotheses of which are rather different from the usual ones; in
particular it seems to be much easier to work with in some situations.
Section \ref{Sconclusions} discusses the relevance of the rigidity
part of Theorem \ref{thm:local-area} to the theory of black holes and
to the differentiability of compact Cauchy horizons.  Appendix
\ref{apR} reviews the geometry of $C^2$ horizons, we also prove there
a new result concerning the relationship between the (classical)
differentiability of a horizon \emph{vs} the (classical)
differentiability of sections thereof, Proposition \ref{C2-null}.  In
Appendix \ref{HEarea} some comments on the area theorem of \cite{HE}
are made.

\section{Horizons}
\label{Shorizons}
 Let $(M,g)$ be a smooth spacetime, that is, a smooth
paracompact Hausdorff time-oriented Lorentzian manifold, of dimension
$n+1\ge 3$, with a smooth Lorentzian metric $g$. Throughout this paper
hypersurfaces are assumed to be embedded. A hypersurface $\cH\subset
M$ will be said to be {\em future null geodesically ruled} if every
point $p\in \cH $ belongs to a future inextensible null geodesic
$\Gamma \subset \cH $; those geodesics will be called {\em the
  generators} of $\cH $. We emphasize that the generators are allowed
to have past endpoints on $\cH $, but no future endpoints.  {\em Past
  null geodesically ruled} hypersurfaces are defined by changing the
time orientation.  We shall say that $\cH$ is \emph{a future (past)
  horizon} if $\cH $ is an \emph{achronal, closed, future (past) null
  geodesically ruled topological hypersurface}. A hypersurface $\cH$
will be called a \emph{horizon} if $\cH$ is a future or a past
horizon.  Our terminology has been tailored to the black hole setting,
so that a future black hole event horizon $\partial J^-(\scri^+)$ is a
future horizon in the sense just described \cite[p.~312]{HE}. The
terminology is somewhat awkward in a Cauchy horizon setting, in which
a \emph{past Cauchy horizon} ${\cD }^-(\Sigma)$ of an achronal
edgeless set $\Sigma$ is a \emph{future} horizon in our terminology
\cite[Theorem~5.12]{PenroseDiffTopo}.

Let $\mathrm{dim} M=n+1$ and suppose that $\cO$ is a domain in $\R^n$.
Recall that a continuous function $f\: \cO\to \R$ is called
semi--convex iff each point $p$ has a convex neighborhood
$\mathcal{U}$ in $\cO$ so that there exists a $C^2$ function
$\phi\:\mathcal{U} \to \R$ such that $f+\phi$ is convex in
$\mathcal{U}$.  We shall say that the graph of $f$ is a semi--convex
hypersurface if $f$ is semi--convex. A hypersurface $\cH$ in a
manifold $M$ will be said to be semi--convex if $\cH $ can be covered
by coordinate patches ${\cU}_\alpha$ such that $\cH \cap {\cU}_\alpha$
is a semi--convex graph for each $\alpha$. The interest of this notion
stems from the specific differentiability properties of such
hypersurfaces:

\begin{Proposition}\textbf{\em  (Alexandrov \cite[Appendix
    E]{FlemingSoner})} 
  \label{Pdiff}
Let $B$ be an open subset of $\R^p$ and let $f\:B \to \R$ be 
semi--convex. Then there exists a set $\BA\subset B$ such that: 
\begin{enumerate}
\item the $p$~dimensional Lebesgue measure ${\Leb}^p(B\setminus \BA)$
  of $ B\setminus \BA$ vanishes.
\item $f$ is differentiable at all points $x\in \BA$, \emph{i.e.},
  \bea&\forall\ x\in \BA\ \exists x^*\in (\R^p)^* \ 
  \mbox{\rm such that}\phantom{x(y-x)+r1(x,y)}&\nonumber\\ &\forall \ y\in B \quad f(y)-f(x)=
  x^*(y-x)+r_1(x,y)\ , &\label{alex1}\eea with
  $r_1(x,y)=o(|x-y|)$. The linear map 
  $x^*$ above will be denoted by $df(x)$. 
\item $f$ is twice--differentiable at all points  $x\in \BA$ in the
  sense that
  \bea&\forall\ x\in \BA\ \exists 
  \ Q\in (\R^p)^*\otimes (\R^p)^*\ \mbox{\rm such that}\ \forall \ 
  y\in B \phantom{df(x)(y-x)=}&\nonumber\\ &\quad f(y)-f(x) -
  df(x)(y-x)=Q(x-y,x-y)+ r_2(x,y)\ ,& \label{alex2}\eea with
  $r_2(x,y)=o(|x-y|^2)$.  The symmetric quadratic form $Q$ above will
  be denoted by $\frac{1}{2}D^2f(x)$, and will be called the \emph{second
  Alexandrov derivative} of $f$ at $x$.\footnote{Caffarelli and Cabr\`e
    \cite{CaffarelliCabre} use the term ``second punctual
    differentiability of $f$ at $x$'' for \eq{alex2}; Fleming and
    Soner \cite[Definition 5.3, p.~234]{FlemingSoner} use the name
    ``point of twice--differentiability'' for points at which
    \eq{alex2} holds.}
\end{enumerate}
\end{Proposition}

The points $q$ at which Equations~\eq{alex1}--\eq{alex2} hold will
be called \emph{Alexandrov points} of $f$, while the points $(q,f(q))$
will be called \emph{Alexandrov points of the graph of} $f$ (it will
be shown in Proposition~\ref{Pcov} below that if $q$ is an Alexandrov
point of $f$, then $(q,f(q))$ will project to an Alexandrov point of
any graphing function of the graph of $f$, so this terminology is
meaningful).

We shall say that $\cH$ is \emph{locally achronal} if for every point
$p\in \cH$ there exists a neighborhood $\cO$ of $p$ such that $\cH\cap
\cO$ is achronal in $\cO$. We have the following:

\begin{Theorem}
\label{T1}
Let $\cH \ne \nothing$ be a locally achronal future null geodesically
ruled hypersurface.  Then $\cH$ is semi--convex.
\end{Theorem}
\begin{remark}
  An alternative proof of Theorem~\ref{T1} can be given using a
variational characterization of horizons (compare
\cite{AP,GMP,Perlick}).
\end{remark}
\begin{remark} 
\label{rem2T1}  Recall 
for a real valued function $f\:B\to\R$ on a set the epigraph of $f$ is
 $\{(x,y)\in B\times \R: y\ge f(x)\}$.  We note that while the notion
 of convexity of a function and its epigraph is coordinate dependent,
 that of semi--convexity is not; indeed, the proof given below is
 based on the equivalence of semi--convexity and of existence of lower
 support hypersurfaces with locally uniform one side Hessian bounds.
 The latter is clearly independent of the coordinate systems used to
 represent $f$, or its graph, as long as the relevant orientation is
 preserved (changing $y^{n+1}$ to $-y^{n+1}$ transforms a lower
 support hypersurface into an upper one). Further, if the future of
 $\cH$ is represented as an epigraph in two different ways,
\begin{eqnarray}
  J^+(\cH)& =& \{x^{n+1}\ge f(x^1,\ldots,x^n)\}
\nonumber \\ & = & \{y^{n+1}\ge
  g(y^1,\ldots,y^n)\} 
\ ,
\label{twoways}
\end{eqnarray}
then semi--convexity of $f$ is equivalent to that of $g$. This follows
immediately from the considerations below. Thus, the notion of a
semi--convex hypersurface is not tied to a particular choice of
coordinate systems and of graphing functions used to represent it.
\end{remark}


\begin{proof}
Let $\cO$ be as in the definition of local achronality; passing
to a subset of $\cO$ we can without loss of generality assume that
$\cO$ is globally hyperbolic. Replacing the space--time $(M,g)$ by
$\cO$ with the induced metric we can without loss of generality assume
that $(M,g)$ is globally hyperbolic.  Let $t$ be a time function on
$\cO$ which induces a diffeomorphism of $\cO$ with $\R \times \Sigma$
in the standard way \cite{GerochDoD,Seifert}, with the level sets
$\Sigma_\tau\equiv \{p\ |\ t(p)=\tau\}$ of $t$ being Cauchy surfaces. As
usual we identify $\Sigma_0$ with $\Sigma$, and in the identification
above the curves $\R\times \{q\}$, $q\in\Sigma$, are integral curves
of $\nabla t$. Define
\begin{equation}  \label{sh}
  \sh=\{q\in \Sigma\ |\ \R\times \{q\}\ \mbox{\rm intersects}\ \cH \}\ .
\end{equation}
For $q\in\sh$ the set $(\R\times \{q\} )\cap \cH$ is a point by
achronality of $\cH$, we shall denote this point by $(f(q),q)$. Thus
an achronal hypersurface $\cH$ in a globally hyperbolic space--time is
a graph over $\sh$ of a function $f$. The map which to a point
$p\in\cH$ assigns $q\in\Sigma$ such that $p=(f(q),q)$ is injective, so
that the hypothesis that $\cH$ is a topological hypersurface together
with the invariance of the domain theorem (\emph{cf., e.g.,}
\cite[Prop. 7.4, p. 79]{Dold}) imply that $\sh$ is open. We wish to
use \cite[Lemma 3.2]{AGH}\footnote{The result in that Lemma actually
  follows from \cite[Lemma 2.15]{AGHCPAM} together with
  \cite[Prop.~5.4, p.~24]{EkelandTemam}.} to obtain semi--convexity of
the (local) graphing function $f$, this requires a construction of
lower support hypersurfaces at $p$.  Let $\Gamma$ be a generator of
$\cH$ passing through $p$ and let $p_+\in\Gamma\cap J^+(p)$. By
achronality of $\Gamma$ there are no points on $J^-(p_+)\cap J^+(p)$
which are conjugate to $p_+$, \emph{cf.\/}\ \cite[Prop.~4.5.12,
p.~115]{HE} or \cite[Theorem~10.72, p.~391]{Beem-Ehrlich:Lorentz2}. It
follows that the intersection of a sufficiently small neighborhood of
$p$ with the past light cone $\partial J^-(p_+)$ of $p_+$ is a smooth
hypersurface contained in $J^{-}(\cH)$. This provides the appropriate
lower support hypersurface at $p$.  In particular, for suitably chosen
points $p_+$, these support hypersurfaces have null second fundamental
forms (see Appendix A) which are locally (in the point $p$) uniformly
bounded from below.  This in turn implies that the Hessians of the
associated graphing functions are locally bounded from below, as is
needed to apply Lemma 3.2 in \cite{AGH}.  \qed\end{proof}

Theorem~\ref{T1} allows us to apply Proposition~\ref{Pdiff} to $f$ to
infer twice Alexandrov differentiability of $\cH$ almost everywhere,
in the sense made precise in Proposition~\ref{Pdiff}.  Let us denote
by $\Lthreenew$ the $n$~dimensional Riemannian measure\footnote{In
  local coordinates, $d\Lthreenew=\sqrt{\det h_0} d^nx$.} on
$\Sigma\equiv \Sigma_0$, where $h_0$ is the metric induced on
$\Sigma_0$ by $g$, then there exists a set $\sdh\subset\sh$ on which
$f$ is twice Alexandrov differentiable, and such that
$\Lthreenew(\sh\setminus \sdh)=0 $.  Set
\begin{equation}
  \label{Haldef}
\dH\equiv \mbox{graph of $f$ over $\sdh$}\ .
\end{equation}
Point 1 of Proposition~\ref{Pdiff} shows that $\dH$ has a tangent
space at every point $p\in\dH$. For those points define
\begin{equation}
  \label{kdef}
   k(p)= k_\mu(p)dx^\mu= -dt+df(q)\ , \qquad p=(f(q),q),\qquad q\in
\sdh\ .
\end{equation}
 A theorem of Beem and Kr\'olak shows that
$\cH$ is differentiable precisely at those points $p$ which belong to
exactly one generator $\Gamma$ \cite{BK2,Chendpoints}, with
$T_p\cH\subset T_pM$ being the null hyper-plane containing $\dG$.  It
follows that
\begin{equation}
  \label{Kdef}
 K\equiv g^{\mu\nu}k_\mu\partial_\nu
\end{equation} is null, future pointing, and tangent to the generators
of $\cH$, wherever defined. 

We define the generators of $\dH$ as the intersections of the
generators of $\cH$ with $\dH$.  Point 2 of Proposition~\ref{Pdiff}
allows us to define the \emph{ divergence} $\tA$ of those generators,
as follows: Let $e_i$, $i=1,\ldots,n$ be a basis of $T_p\cH$ such that
\begin{equation}\label{or}g(e_1,e_1)=\cdots=g(e_{n-1},e_{n-1})=1\ ,
  \qquad g(e_i,e_j)=0\  
, i\ne j\ , \qquad e_n=K\ . 
\end{equation} We further choose the $e_a$'s, $a=1,\ldots,n-1$ to be
orthogonal to $\nabla t$. It follows that the vectors $e_a=e_a^\mu
\partial_\mu $'s have no $\partial/\partial t$ components in the
coordinate system used in the proof of Theorem~\ref{T1}, thus
$e_a^0=0$. Using this coordinate system for $p\in \dH$ we
set
\begin{eqnarray}
&  i,j=1,\ldots,n\qquad \nabla_i k_j = D^2_{ij}f - \Gamma^{\mu}_{ij}
k_\mu \ ,& \label{nablak}\\ 
& \tA  =(e_1^i e_1^j + \cdots +e_{n-1}^i e_{n-1}^j) \nabla_i k_j \
.&\label{theta} 
\end{eqnarray}
It is sometimes convenient to set $\tA=0$
on $\cH\setminus \dH$. The function $\tA $ so defined on $\cH$ will be
called the divergence (towards the future) of both the generators of
$\dH$ and of $\cH$.
It coincides with the usual divergence of the generators of $\cH$ at
every set $\cU\subset\cH$ on which the horizon is of $C^2$
differentiability class: Indeed, in a space--time neighborhood of
$\cU$ we can locally extend the $e_i$'s to $C^1$ vector fields, still
denoted $e_i$, satisfying \eq{or}. It is then easily checked
that the set of Alexandrov points of $\cU$ is $\cU$, and that the
divergence $\theta$ of the generators as defined in \cite{HE,Waldbook}
or in Appendix~\ref{apR} coincides on $\cU$ with $\tA $ as defined by
\eq{theta}.

The {\em null second fundamental form} $\BA$ of $\cH$, or of $\dH$, is
defined as follows: in the basis above, if $ X = X^a e_a$, $Y = Y^b
e_b$ (the sums are from 1 to $n-1$), then at Alexandrov points we
set
 \begin{equation}
\label{nullform}
\BA(X,Y) = X^aY^b e^i_a e^j_b \nabla_i k_j.  \end{equation} with
$\nabla_i k_j$ defined by \eq{nablak}. This coincides\footnote{More
  precisely, when $\cH$ is $C^2$ equation \eq{nullform} defines, in
  local coordinates, a tensor field which reproduces $b$ defined in
  Appendix~\ref{apR} when passing to the quotient $\nullhyp
  /K$.} with the usual definition of $B$ as
discussed \emph{e.g.\/} in Appendix~\ref{apR} on any subset of $\cH$
which is $C^2$.  In this definition of the null second fundamental
form $B$ with respect to the null direction $K$, $B$ measures
expansion as positive and contraction as negative.

The definitions of $\tA$ and $\BA$ given in Equations
\eq{theta}--\eq{nullform} have, so far, been only given for horizons
which can be globally covered by an appropriate coordinate system. As
a first step towards a globalization of those notions one needs to
find out how those objects change when another representation is
chosen. We have the following:

\begin{Proposition}
  \label{Pcov} 
  \begin{enumerate}
  \item Let $f$ and $g$ be two locally Lipschitz graphing functions
    representing $\cH$ in two coordinate systems
    $\{x^i\}_{i=1,\ldots,n+1}$ and $\{y^i\}_{i=1,\ldots,n+1}$, related
    to each other by a $C^2$ diffeomorphism $\phi$. If $(x_0^1,\ldots
    x^n_0)$ is an Alexandrov point of $f$ and $$(y_0^1,\ldots, y^n_0,
    g(y_0^1,\ldots, y^n_0))=\phi(x_0^1,\ldots, x^n_0, f( x_0^1,\ldots,
    x^n_0))\ ,$$ then $(y_0^1,\ldots, y^n_0)$ is an Alexandrov point of
    $g$.
  \item The null second fundamental form $\BA$ of Equation
    \eq{nullform} is invariantly defined modulo a point dependent
    multiplicative factor.  In particular the sign of $\tA$ defined in
    Equation~\eq{theta} does not depend upon the choice of the
    graphing function used in \eq{theta}.
  \end{enumerate}
\end{Proposition}

\rems  Recall that the standard divergence $\theta$ of generators is
defined up to a multiplicative function (constant on the generators)
only, so that (essentially) the only geometric invariant associated to
$\theta$ is its sign.
\end{remark}
\rems We emphasize that we do not assume that $\partial/\partial
x^{n+1}$ and/or $\partial/\partial y^{n+1}$ are timelike.
\end{remark}

\begin{proof}
1. Let $\vec y = (y^1,\ldots, y^n)$, $\vec x_0 = (x_0^1,\ldots,
x_0^n)$, \emph{etc.}, set $f_0=f(\vec x_0 )$, $g_0=g(\vec y_0 )$. 
Without loss of generality we may assume $(\vec x_0,f(\vec x_0))=(\vec
y_0,f(\vec y_0))=0$. To establish \eq{alex2} for $g$, write $\phi$ as
$(\vec \phi,\phi^{n+1})$, and let
$$
\vec\psi(\vec x):=\vec\phi(\vec x,f(\vec x))\ .
$$ As $(id,f)$ and $(id,g)$, where $id$ is the identity map, are
bijections between neighborhoods of zero and open subsets of the
graph, invariance of domain shows that $\vec\psi$ is a homeomorphism
from a neighborhood of zero to its image.  Further, $\vec \psi$ admits
a Taylor development of order two at the origin:
\begin{equation}
\label{tayl1}
\vec\psi(\vec x)=\Lpsi \vec x + \Qpsi(\vec x, \vec x) + o(|\vec x|^2)\
,
\end{equation}
where
$$
\Lpsi=D_{\vec x}\vec \phi (0)+D_{x^{n+1}}\vec \phi (0) \, Df(0)\ .
$$ Let us show that $\Lpsi$ is invertible: suppose, for contradiction,
that this is not the case, let $\vec x\neq 0$ be an element of the
kernel $\ker \Lpsi$, and for $|s|<<1$ let ${\vec
  y}_s:=\vec \psi (s\vec x)=O(s^2)$.  As ${\vec y}_s=\vec\phi(s\vec
x,f(s\vec x))$ we have
$$
g({\vec y}_s)=\phi^{n+1}(s\vec x,f(s\vec x))
=s [D_{\vec x} \phi^{n+1} (0)+D_{x^{n+1}} \phi^{n+1} (0) Df(0)]\vec x
+o(s)\ .
$$ The coefficient of the term linear in $s$ does not vanish,
otherwise $(\vec x, Df(0)\vec x)$ would be a non--zero vector of 
$\ker\; D\phi(0)$.  We then obtain $|g({\vec y}_s)|/|{\vec y}_s|\ge
C/|s|\to_{s\to 0}\infty$, which contradicts the hypothesis that $g$ is
locally Lipschitz.

To finish the proof, for $\vec y$ close to the origin let $\vec
x:=\psi^{-1}(\vec y)$, thus $\vec y=\psi (\vec x)$, and from
(\ref{tayl1}) we infer 
\begin{equation}\label{invtayl1}
\vec x=\Lpsi^{-1} \vec y - \Lpsi^{-1} \Qpsi( \Lpsi^{-1} \vec
y,\Lpsi^{-1} \vec y) + o(|\vec y|^2)\ .  
\end{equation} 
Equation~\eq{alex2} and
twice differentiability of $\phi$ show that $\phi^{n+1}(\cdot,f(\cdot))$ has a
second order Alexandrov expansion at the origin: \begin{equation}
\label{tayl2}
\phi^{n+1}(\vec x,f(\vec x))=\Lphif \vec x + \Qphif(\vec x, \vec x) +
o(|\vec x|^2)\ .
\end{equation}
Finally from $g(\vec y)= \phi^{n+1}(\vec x,f(\vec x))$ and Equations 
(\ref{invtayl1}) and (\ref{tayl2}) we get
$$ g(\vec y)=\Lphif\Lpsi^{-1}\vec y
+(\Qphif-\Lphif\Lpsi^{-1}\Qpsi)(\Lpsi^{-1} \vec y, \Lpsi^{-1} \vec
y)+o(|\vec y|^2)\ .
$$ 

2. The proof of point 1.\  shows that under changes of graphing
functions the Alexandrov derivatives $D_{ij}f$ transform into the
$D_{ij}g$'s exactly as they would if $f$ and $g$ were twice
differentiable. The proof of point 2.\  reduces therefore to that of the
analogous statement for $C^2$ functions, which is well known. 
\qed\end{proof}

Proposition~\ref{Pcov} shows that Equation \eq{nullform} defines an
equivalence class of tensors $\BA$ at every Alexandrov point of $\cH$,
where two tensors are identified when they are proportional to each
other with a positive proportionality factor.  Whenever $\cH$ can not
be covered by a global coordinate system required in
Equation~\eq{nullform}, one can use Equation~\eq{nullform} in a local
coordinate patch to define a representative of the appropriate
equivalence class, and the classes so obtained will coincide on the
overlaps by Proposition~\ref{Pcov}. From this point of view $\tA(p)$
can be thought as the assignment, to an Alexandrov point of the
horizon $p$, of the number $0,\pm1$, according to the sign of
$\tA(p)$.

\mnote{a paragraph which I like but Greg doesn't, and so the referee,
has been removed}

The generators of a horizon $\cH$ play an important part in our
results.  The following shows that most points of a horizon are on
just one generator.

\begin{prop}\label{one-gen}
  The set of points of a horizon $\cH$ that are on more than one
  generator has vanishing $n$~dimensional Hausdorff measure.
\end{prop}

\begin{proof}
Let $\mathcal{T}$ be the set of points of $\cH$ that are on two or
more generators of $\cH$.  By a theorem  of Beem and
Kr\'olak~\cite{BK2,Chendpoints} a point of $\cH$ is differentiable if
and only if it belongs to exactly one generator of $\cH$.  Therefore no
point of $\mathcal{T}$ is differentiable and thus no point of
$\mathcal{T}$ is an Alexandrov point of $\cH$.  Whence
Theorem~\ref{T1} and  Proposition~\ref{Pdiff} imply
$\cH^{n}(\mathcal{T})=0$.\qed
\end{proof}

\begin{remark}
For a future horizon $\cH$ in a spacetime $(M,g)$ let $\mathcal{E}$ be
the set of past endpoints of generators of $\cH$.  Then the set of
points $\mathcal{T}$ of $\cH$ that are on two or more generators is a
subset of $\mathcal{E}$.  From the last Proposition we know that
$\Hau^n(\mathcal{T})=0$ and it is tempting to conjecture
$\Hau^n(\mathcal{E})=0$.  However this seems to be an open question.
We note that there are examples~\cite{ChGalloway} of horizons $\cH$ so
that $\mathcal{E}$ is dense in $\cH$.
\end{remark}

\section{Sections of horizons}
\label{Ssections}

Recall that in the standard approach to the area theorem \cite{HE} one
considers sections $\cH_\tau$ of a black hole event horizon $\cH$
obtained by intersecting $\cH$ with the level sets $\Sigma_\tau$ of a
time function $t$, \begin{equation}\label{Hsection} \cH_\tau=\cH\cap
  \Sigma_\tau \ . \end{equation} We note that Theorem~\ref{T1} and
Proposition~\ref{Pdiff} do not directly yield any information about
the regularity of all the sections $\cH_\tau$, since Proposition
\ref{Pdiff} gives regularity away from a set of zero measure, but
those sections are expected to be of zero measure anyway.  Now,
because the notion of semi--convexity is independent of the choice of
the graphing function, and it is preserved by taking restrictions to
lower dimensional subsets, we have:

\begin{Proposition}\emb
  \label{Pinter} Let $\Sigma$ be an embedded $C^2$ spacelike hypersurface and
  let $\cH$ be a horizon in $(M,g)$, set
  $$S=\Sigma\cap \cH\ . $$ There exists a semi--convex topological
  submanifold $S_{\reg}\subset S$ of co--dimension two in $M$ which
  contains all points $p\in S$ at which $\cH$ is differentiable.
\end{Proposition}
\begin{remark} ``Most of the time'' $S_{\reg}$ has full measure in $S$.
There exist, however, sections of horizons for which this is not the
case; \emph{cf.}\ the discussion at the end of this section, and the
example discussed after Equation~\eq{defs} below.
\end{remark}

\begin{proof}
Let $p\in S$ be such that $\cH$ is differentiable at $p$, let
$\cO_p$ be a coordinate neighborhood of $p$ such that
$\Sigma\cap\cO_p$ is given by the equation $t=0$, and such that
$\cH\cap\cO_p$ is the graph $t=f$ of a semi--convex function $f$. If
we write $p=(f(q),q)$, then $f$ is differentiable at $q$. We wish,
first, to show that $S$ can be locally represented as a Lipschitz
graph, using the Clarke implicit function theorem \cite[Chapter
7]{Clarke:optimization}.  Consider the Clarke differential $\partial
f$ of $f$ at $q$ (\emph{cf.}~\cite[p.  27]{Clarke:optimization}); by
\cite[Prop. 2.2.7]{Clarke:optimization} and by standard properties of
convex (hence of semi--convex) functions we have $\partial
f(q)=\{df(q)\}$.  By a rigid rotation of the coordinate axes we may
assume that $df(q)(\partial/\partial x^n)\ne 0$. If we write
$q=(0,q^1,\dots,q^n)$, then Clarke's implicit function theorem
\cite[Corollary, p.~256]{Clarke:optimization} shows that there exists
a neighborhood $\cW_q\subset\{(t,x^1,\ldots,x^{n-1})\in \R^n\}$ of
$(0,q^1,\dots,q^{n-1})$ and a Lipschitz function $g$ such that,
replacing $\cO_p$ by a subset thereof if necessary, $\cH\cap \cO_p$
can be written as a graph of $g$ over $\cW_q$. Remark~\ref{rem2T1}
establishes semi--convexity of $g$.  Now, semi--convexity is a
property which is preserved when restricting a function to a smooth
lower dimensional submanifold of its domain of definition, which shows
that $h\equiv g\big|_{t=0}$ is semi--convex.  It follows that the graph
$\cV_q\subset \cH$ of $h$ over $\cW_q\cap\{t=0\}$ is a semi--convex
topological submanifold of $\cH$.  The manifold
$$S_{\reg}\equiv \cup_q \cV_q\ , $$ where the $q$'s run over the set
of differentiability points of $\cH$ in $\Sigma$, has all the desired
properties. 
\qed\end{proof} 

Let $S$, $S_{\reg}$, $\cH$ and $\Sigma$ be as in Proposition
\ref{Pinter}. Now $S_{\reg}$ and $ \cH$ are both semi--convex, and
have therefore their respective Alexandrov points; as $S_{\reg}\subset
\cH$, it is natural to inquire about the relationship among those. It
should be clear that if $p\in S$ is an Alexandrov point of $ \cH$,
then it also is an Alexandrov point of $S$. The following example
shows that the converse is not true: Let $\cH\subset \R^{1,2}$ be
defined as the set $\cH=\{(t,x,y)\ | \ t=|x|\}$, and let $\Sigma_\tau$
be the level sets of the Minkowskian time in $ \R^{1,2}$,
$\Sigma_\tau=\{(t,x,y)\ | \ t=\tau\}$. Each section
$S_\tau=\cH\cap\Sigma_\tau$ is a smooth one dimensional submanifold of
$\R^{1,2}$ (not connected if $\tau >0$, empty if $\tau < 0$).  In
particular all points of $S_0=\{t=x=0\}$ are Alexandrov points
thereof, while none of them is an Alexandrov point of $\cH$. In this
example the horizon $\cH$ is not differentiable on any of the points
of $S_0$, which is a necessary condition for being an Alexandrov point
of $\cH$. It would be of some interest to find out whether or not
Alexandrov points of sections of $\cH$ which are also points of
differentiability of $\cH$ are necessarily Alexandrov points of $\cH$.

Following \cite{Greg:nullsplit}, we shall say that a differentiable
embedded hypersurface $\calS$ meets $\cH$ \emph{\transversally} if for
each point $p\in \calS\cap\cH$ for which $T_p\cH$ exists the tangent
hyperplane $T_p\calS$ is transverse to $T_p\cH$. If $\calS$ is
spacelike and intersects $\cH$ proper transversality will always hold;
on the other hand if $\calS$ is timelike this might, but does not have
to be the case.  If $\cH$ and $\calS$ are $C^1$ and $\calS$ is either
spacelike or timelike intersecting $\cH$ transversely then $\cH\cap
\calS$ is a $C^1$ spacelike submanifold.  Therefore when $\calS$ is
timelike there is a spacelike $\calS_1$ so that $\cH\cap \calS=\cH\cap
\calS_1$.  It would be interesting to know if this was true (even
locally) for the transverse intersection of general horizons $\cH$
with timelike $C^2$ hypersurfaces $\calS$.

Let $\calS$ be any spacelike or timelike $C^2$ hypersurface in $M$
meeting $\cH$ \transversally.\footnote{Our definitions below makes use
  of a unit normal to $\calS$, whence the restriction to spacelike or
  timelike, properly transverse $\calS$'s.  Clearly one should be able
  to give a definition of $\tAcS$, {\em etc.}, for any hypersurface
  $\calS$ intersecting $\cH$ properly transversally. Now for a smooth
  $\cH$ and for smooth properly transverse $\calS$'s the intersection
  $\calS\cap\cH$ will be a smooth spacelike submanifold of $M$, and it
  is easy to construct a spacelike $\calS'$ so that
  $\calS'\cap\cH=\calS\cap\cH$.  Thus, in the smooth case, no loss of
  generality is involved by restricting the $\calS$'s to be spacelike.
  For this reason, and because the current setup is sufficient for our
  purposes anyway, we do not address the complications which arise
  when $\calS$ is allowed to be null, or to change type.} Suppose,
first, that $\calS$ is covered by a single coordinate patch such that
$\calS\cap\cH$ is a graph $x^{n}=g(x^1,\ldots,x^{n-1})$ of a
semi--convex function $g$.  Let $\bn$ be the field of unit normals to
$\calS$; at each point $(x^1,\ldots,x^{n-1})$ at which $g$ is
differentiable and for which $T_p\cH$ exists, where
$$p\equiv(x^1,\ldots,x^{n-1},g(x^1,\ldots,x^{n-1}))\ ,$$ there is a
unique number $a(p)\in \R$ and a unique future pointing null vector
$K\in T_p\cH$ such that 
\begin{equation}
\label{Kequation}
\langle K-a\bn,\cdot\rangle =-dx^n+dg\;.
\end{equation}
\mnote{sign changed; comment} Here we have assumed that the coordinate
$x^n$ is chosen so that $J^-(\cH)\cap \calS$ lies under the graph of
$g$. We set
$$
k_\mu dx^\mu = \langle K, \cdot\rangle \in T^*_pM\ .
$$
Assume moreover that
$(x^1,\ldots,x^{n-1})$ is an Alexandrov point of $g$, we thus have the
Alexandrov second derivatives $D^2g$ at our disposal. Consider $e_i$
--- a basis of $T_p\cH$ as in Equation~\eq{or}, satisfying further
$e_i\in T_p\calS\cap T_p\cH$, $i=1,\ldots,n-1$. In a manner completely
analogous to \eq{nablak}--\eq{theta} we set
\begin{eqnarray}
&  i,j=1,\ldots,n-1\qquad \nabla_i k_j = D^2_{ij}g - \Gamma^{\mu}_{ij}
k_\mu 
\ ,& \label{nablakS}\\ 
& \tAcS  =(e_1^i e_1^j + \cdots +e_{n-1}^i e_{n-1}^j) \nabla_i k_j \
.&\label{thetaS} 
\end{eqnarray}
(Here, as in \eq{nablak}, the $\Gamma^{\mu}_{ij}$'s are the
Christoffel symbols of the space--time metric $g$.) 
For $X,Y\in T_p\calS\cap T_p\cH$ we set, analogously to \eq{nullform},
 \begin{equation}
\label{nullformS}
\BAcS(X,Y) = X^aY^b e^i_a e^j_b \nabla_i k_j.  \end{equation}
Similarly to the definitions of $\tA$ and $\BA$, the vector $K$ with
respect to which $\tAcS$ and $\BAcS$ have been defined has been tied
to the particular choice of coordinates used to represent
$\calS\cap\cH$ as a graph. 
In order to globalize this definition it might be convenient to regard
$\BAcS(p)$ as an equivalence class of tensors defined up to a positive
multiplicative factor. Then $\tAcS(p)$ can be thought as the
assignment to a point $p$ of the number $0,\pm1$, according to the
sign of $\tAcS(p)$. This, together with Proposition~\ref{Pcov}, can
then be used to define $\BAcS$ and $\tAcS$ for $\calS$ which are not
globally covered by a single coordinate patch.

\mnote{paragraph removed}

As already pointed out, if $p\in\calS\cap\cH$ is an Alexandrov point
of $\cH$ then $p$ will also be an Alexandrov point of $\calS\cap\cH$.
In such a case the equivalence class of $\BA$, defined at $p$ by
Equation~\eq{nullform}, will coincide with that defined by
\eq{nullformS}, when \eq{nullform} is restricted to vectors $X,Y\in
T_p\calS\cap T_p\cH$. Similarly the sign of $\tA(p)$ will coincide
with that of $\tAcS(p)$, and $\tA(p)$ will vanish if and only if
$\tAcS(p)$ does.

Let us turn our attention to the question, how to define the area of
sections of horizons.  The monotonicity theorem we prove in Section
\ref{Smonotonicity} uses the Hausdorff measure, so let us start by
pointing out the following:
\begin{Proposition}
  \label{PHaus} Let $\cH$ be a horizon and let $\calS$ be an
  embedded hypersurface in $M$.  Then $\calS\cap\cH$ is a Borel set,
  in particular it is $\nu$--Hausdorff measurable for any
  $\nu\in\R^+$.
\end{Proposition}
\begin{proof}
  Let $\kaux$ be any complete Riemannian metric on $M$, we can cover
  $\calS$ by a countable collection of sets ${\cal O}_i\subset\calS$
  of the form ${\cal O}_i=B_\kaux(p_i,r_i)\cap\calS$, where the
  $B_\kaux(p_i,r_i)$ are open balls of $\kaux$--radius $r_i$ centered
  at $p_i$ with compact closure. We have ${\cal O}_i= \left(
    \overline{B_\kaux(p_i,r_i) \cap\calS}\right) \setminus \partial
  B_\kaux(p_i,r_i)$, which shows that the ${\cal O}_i$'s are Borel
  sets.  Since $\calS\cap\cH =\cup_i\left({\cal O}_i\cap\cH\right)$,
  the Borel character of $\calS\cap\cH$ ensues. The Hausdorff
  measurability of $\calS\cap\cH$ follows now from the fact that Borel
  sets are Hausdorff measurable\footnote{\label{fBorel}Let $(X,d)$ be
    a metric space.  Then an outer measure $\mu$ defined on the class
    of all subsets of $X$ is a \emph{metric outer measure} iff
    $\mu(A\cup B)=\mu(A)+\mu(B)$ whenever the distance (\emph{i.e.}
    $\inf_{a\in A, b\in B}d(a,b)$) is positive.  For a metric outer
    measure, $\mu$, the Borel sets are all $\mu$-measurable
    (\emph{cf.} \cite[p.~48 Prob.~8]{Halmos:measure} or \cite[p.~188
    Exercise~1.48]{Hewitt-Stromberg}).  The definition of the
    Hausdorff outer measures implies they are metric outer measures
    (\emph{cf.} \cite[p.~188 Exercise~1.49]{Hewitt-Stromberg}). See
    also \cite[p.~7]{SimonGMT} or \cite[p.~147]{Edgar:fractals}.}.
\end{proof}\qed

Proposition \ref{PHaus} is sufficient to guarantee that a notion of
area of sections of horizons --- namely their $(n-1)$--dimensional
Hausdorff area --- is well defined. Since the Hausdorff area is not
something very handy to work with in practice, it is convenient to
obtain more information about regularity of those sections. Before we
do this let us shortly discuss how area can be defined, depending upon
the regularity of the set under consideration.  When $S$ is a
piecewise $C^1$, $(n-1)$~dimensional, paracompact, orientable
submanifold of $M$ this can be done by first defining
$$
d^{n-1}\mu=e_1\wedge\cdots\wedge e_{n-1}\ ,
$$ where the $e_a$'s form an oriented orthonormal basis of the
cotangent space $T^*S$; obviously $d^{n-1}\mu$ does not depend upon
the choice of the $e_a$'s.  Then one sets
\begin{equation}\label{standardarea}\Ar(S)= \int_{S} d^{n-1}\mu\ .
\end{equation} Suppose, next, that $S$ is the image by a Lipschitz map
$\psi$ of a $C^1$ manifold $N$. Let $\metricS$ denote any complete
Riemannian metric on $N$; by \cite[Theorem 5.3]{SimonGMT} for every
$\epsilon > 0$ there exists a $C^1$ map $\psi_\epsilon$ from $N$ to
$M$ such that
 \begin{equation}
\label{a2}
\LtwoS(\{\psi\ne \psi_\epsilon\}) \le \epsilon\ .  \end{equation} Here
and throughout $\LtwoS$ denotes the $(n-1)$~dimensional Riemannian
measure associated with a metric $\metricS$.  One then sets
$S_\epsilon=\psi_\epsilon(N)$ and \begin{equation}
\label{a3}
\Ar(S)=\lim_{\epsilon\to 0} \Ar(S_\epsilon)\ .  \end{equation} (It is
straightforward to check that $\Ar(S)$ so defined is independent of
the choice of the sequence $\psi_\epsilon$. In particular if $\psi$ is
$C^1$ on $N$ one recovers the definition \eq{standardarea} using
$\psi_\epsilon=\psi$ for all $\epsilon$.)

It turns out that for general sections of horizons some more work is
needed.  Throughout this paper $\kaux$ will be an auxiliary Riemannian
metric such that $(M,\kaux)$ is a complete Riemannian manifold. Let
$S$ be a Lipschitz $(n-1)$~dimensional submanifold of $M$ such that
$S\subset\cH$; recall that, by Rademacher's theorem, $S$ is
differentiable $\Htwok$ almost everywhere.  Let us denote by
$\Hau^s_{\kaux}$ the $s$~dimensional Hausdorff measure
\cite{FedererMeasureTheory} defined using the distance function of
$\kaux$.  Recall that $S$ is called $n$~\emph{rectifiable}
\cite{FedererMeasureTheory} iff $S$ is the image of a bounded subset
of $\R^n$ under a Lipschitz map.  A set is \emph{countably}
$n$~\emph{rectifiable} iff it is a countable union of $n$~rectifiable
sets.  (\emph{Cf}.\/~\cite[p.~251]{FedererMeasureTheory}.)
(Instructive examples of countably rectifiable sets can be found in
\cite{Morgan}.) We have the following result \cite{Howard} (The proof
of the first part of Proposition \ref{Phow} is given in
Remark~\ref{R:S1-rect} below):

\begin{Proposition} 
  \label{Phow} Let $S$ be as in Proposition~\ref{Pinter}, then $S$ is
  countably $(n-1)$~rectifiable.  
  If $S$ is compact then it is $(n-1)$~rectifiable.\qed    
\end{Proposition}

\begin{figure}[ht]
\centering
\setlength{\unitlength}{0.00083333in}
\begingroup\makeatletter\ifx\SetFigFont\undefined
\def\x#1#2#3#4#5#6#7\relax{\def\x{#1#2#3#4#5#6}}%
\expandafter\x\fmtname xxxxxx\relax \def\y{splain}%
\ifx\x\y   
\gdef\SetFigFont#1#2#3{%
  \ifnum #1<17\tiny\else \ifnum #1<20\small\else
  \ifnum #1<24\normalsize\else \ifnum #1<29\large\else
  \ifnum #1<34\Large\else \ifnum #1<41\LARGE\else
     \huge\fi\fi\fi\fi\fi\fi
  \csname #3\endcsname}%
\else
\gdef\SetFigFont#1#2#3{\begingroup
  \count@#1\relax \ifnum 25<\count@\count@25\fi
  \def\x{\endgroup\@setsize\SetFigFont{#2pt}}%
  \expandafter\x
    \csname \romannumeral\the\count@ pt\expandafter\endcsname
    \csname @\romannumeral\the\count@ pt\endcsname
  \csname #3\endcsname}%
\fi
\fi\endgroup
{\renewcommand{\dashlinestretch}{30}
\begin{picture}(4816,1241)(0,-10)
\path(1808,613)(3008,613)
\path(1808,613) (1745.429,614.894)
        (1687.227,617.033)
        (1633.172,619.443)
        (1583.047,622.152)
        (1536.630,625.188)
        (1493.703,628.579)
        (1454.045,632.350)
        (1417.438,636.531)
        (1352.492,646.230)
        (1297.109,657.895)
        (1249.531,671.745)
        (1208.000,688.000)

\path(1208,688) (1141.895,734.565)
        (1079.085,806.109)
        (1048.005,848.195)
        (1016.653,892.863)
        (984.667,938.893)
        (951.680,985.064)
        (917.328,1030.154)
        (881.247,1072.944)
        (843.070,1112.212)
        (802.435,1146.737)
        (758.975,1175.299)
        (712.325,1196.678)
        (662.122,1209.652)
        (608.000,1213.000)

\path(608,1213) (548.689,1208.137)
        (490.625,1197.704)
        (434.165,1181.963)
        (379.663,1161.175)
        (327.477,1135.599)
        (277.962,1105.498)
        (231.473,1071.133)
        (188.368,1032.764)
        (149.000,990.652)
        (113.727,945.059)
        (82.904,896.245)
        (56.888,844.471)
        (36.033,789.999)
        (20.696,733.089)
        (11.233,674.002)
        (8.000,613.000)

\path(8,613)    (11.233,551.998)
        (20.696,492.911)
        (36.033,436.001)
        (56.888,381.529)
        (82.904,329.755)
        (113.727,280.941)
        (149.000,235.348)
        (188.368,193.236)
        (231.473,154.867)
        (277.962,120.502)
        (327.477,90.401)
        (379.663,64.825)
        (434.165,44.037)
        (490.625,28.296)
        (548.689,17.863)
        (608.000,13.000)

\path(608,13)   (662.120,16.348)
        (712.322,29.322)
        (758.971,50.701)
        (802.430,79.263)
        (843.066,113.788)
        (881.242,153.056)
        (917.324,195.846)
        (951.676,240.936)
        (984.664,287.107)
        (1016.651,333.137)
        (1048.003,377.805)
        (1079.084,419.891)
        (1141.894,491.435)
        (1208.000,538.000)

\path(1208,538) (1249.531,554.255)
        (1297.109,568.105)
        (1352.492,579.770)
        (1417.438,589.469)
        (1454.045,593.650)
        (1493.703,597.421)
        (1536.630,600.812)
        (1583.047,603.848)
        (1633.172,606.557)
        (1687.227,608.967)
        (1745.429,611.106)
        (1808.000,613.000)

\path(3008,613) (3070.571,614.894)
        (3128.773,617.033)
        (3182.828,619.443)
        (3232.953,622.152)
        (3279.370,625.188)
        (3322.297,628.579)
        (3361.955,632.350)
        (3398.562,636.531)
        (3463.508,646.230)
        (3518.891,657.895)
        (3566.469,671.745)
        (3608.000,688.000)

\path(3608,688) (3674.105,734.565)
        (3736.915,806.109)
        (3767.995,848.195)
        (3799.347,892.863)
        (3831.333,938.893)
        (3864.320,985.064)
        (3898.672,1030.154)
        (3934.753,1072.944)
        (3972.930,1112.212)
        (4013.565,1146.737)
        (4057.025,1175.299)
        (4103.675,1196.678)
        (4153.878,1209.652)
        (4208.000,1213.000)

\path(4208,1213)        (4267.311,1208.137)
        (4325.375,1197.704)
        (4381.835,1181.963)
        (4436.337,1161.175)
        (4488.523,1135.599)
        (4538.038,1105.498)
        (4584.527,1071.133)
        (4627.632,1032.764)
        (4667.000,990.652)
        (4702.273,945.059)
        (4733.096,896.245)
        (4759.112,844.471)
        (4779.967,789.999)
        (4795.304,733.089)
        (4804.767,674.002)
        (4808.000,613.000)

\path(4808,613) (4804.767,551.998)
        (4795.304,492.911)
        (4779.967,436.001)
        (4759.112,381.529)
        (4733.096,329.755)
        (4702.273,280.941)
        (4667.000,235.348)
        (4627.632,193.236)
        (4584.527,154.867)
        (4538.038,120.502)
        (4488.523,90.401)
        (4436.337,64.825)
        (4381.835,44.037)
        (4325.375,28.296)
        (4267.311,17.863)
        (4208.000,13.000)

\path(4208,13)  (4153.880,16.348)
        (4103.678,29.322)
        (4057.029,50.701)
        (4013.570,79.263)
        (3972.934,113.788)
        (3934.758,153.056)
        (3898.676,195.846)
        (3864.324,240.936)
        (3831.336,287.107)
        (3799.349,333.137)
        (3767.997,377.805)
        (3736.916,419.891)
        (3674.106,491.435)
        (3608.000,538.000)

\path(3608,538) (3566.469,554.255)
        (3518.891,568.105)
        (3463.508,579.770)
        (3398.562,589.469)
        (3361.955,593.650)
        (3322.297,597.421)
        (3279.370,600.812)
        (3232.953,603.848)
        (3182.828,606.557)
        (3128.773,608.967)
        (3070.571,611.106)
        (3008.000,613.000)

\end{picture} } \caption{A section of the event horizon in a $2+1$
dimensional space--time with ``two black holes merging".}
\label{curve.fig} \end{figure}
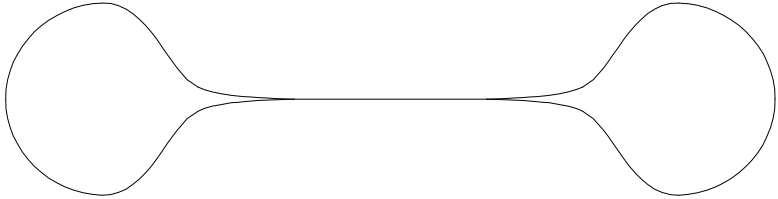

Consider, then, a family of sets $\cV_{q_i}$ which are Lipschitz
images and form a partition of $S$ up to a set of $\Htwoh$ Hausdorff
measure zero, where $h$ is the metric on $\Sigma$ induced by $g$.  One
sets
\begin{equation}
\label{a.4} 
   \Ar(S)=\sum_i  \Ar(\cV_{q_i})\ .
\end{equation} 
We note that $\Ar(S)$ so defined again does not depend on the choices
made, and reduces to the previous definitions whenever applicable.
Further, $\Ar(S)$ so defined is\footnote{ The equality \eq{areah} can
  be established using the area formula. In the case of $C^1$
  submanifolds of Euclidean space this is done explicitly in
  \cite[p.~48]{SimonGMT}.  This can be extended to countable
  $n$~rectifiable sets by use of more general version of the area
  theorem \cite[p.~69]{SimonGMT} (for subsets of Euclidean space) or
  \cite[Theorem~3.1]{Federer:measures} (for subsets of Riemannian
  manifolds).} precisely the $(n-1)$~dimensional Hausdorff measure
$\Htwoh$ of $S$:
\begin{equation}
  \label{areah}
  \Ar(S)=\int_{S} \, d\Htwoh(p)= \Htwoh(S)\ .
\end{equation}
As we will see, there is still another quantity which appears
naturally in the area theorem, Theorem~\ref{thm:local-area} below: the
area \emph{counting multiplicities}. Recall that the
\emph{multiplicity} $N(p)$ of a point $p$ belonging to a horizon $\cH$
is defined as the number (possibly infinite) of generators passing
through $p$. Similarly, given a subset $S$ of $\cH$, for $p\in \cH$ we
set
\begin{eqnarray}
  & N(p,S)=\mbox{the number of generators of $\cH$ passing through
    \phantom{thof gener } } & \nonumber \\ & \phantom{the numbener }
  \mbox{ $p$ which meet $S$ when followed to the (causal) future}\ . &
  \label{nmult}\end{eqnarray} 
Note that $N(p,S)=N(p)$ for $p\in S$.  Whenever $N(p,S_2)$ is
$\Htwohone$ measurable on the intersection $S_1$ of $\cH$ with a
spacelike hypersurface $\Sigma_1$ we set
\begin{equation}
  \label{aream}
  \Arm(S_1)=\int_{S_1}N(p,S_2) \, d\Htwohone(p)\ ,
\end{equation}
where $h_1$ is the metric induced by $g$ on $\Sigma_1$. Note that
$N(p,S_2)=0$ at points of $S_1$ which have the property that the
generators through them do not meet $S_2$; thus the area $\Arm(S_1)$
only takes into account those generators that are seen from $S_2$.  If
$S_1\subset J^-{(S_2)}$, as will be the case \emph{e.g.} if $S_2$ is
obtained by intersecting $\cH$ with a Cauchy surface $\Sigma_2$ lying
to the future of $S_1$, then $N(p,S_2)\ge 1$ for all $p\in S_1$
(actually in that case we will have $N(p,S_2)= N(p)$), so that
$$\Arm(S_1)\ge \Ar(S_1)\ .$$ Let us show, by means of an example, that
the inequality can be strict in some cases. Consider a black hole in a
three dimensional space--time, suppose that its section by a spacelike
hypersurface $t=0$ looks as shown in Figure~\ref{curve.fig}.

As we are in three dimensions area should be
replaced by length. (A four dimensional analogue of Figure
\ref{curve.fig} can be obtained by rotating the curve from Figure
\ref{curve.fig} around a vertical axis.)  When a slicing by spacelike
hypersurfaces is appropriately chosen, the behavior depicted can occur
when two black holes merge together\footnote{\label{amb}The four
  dimensional analogue of Figure~\ref{curve.fig} obtained by rotating
  the curve from Figure~\ref{curve.fig} around a \emph{vertical} axis
  can occur in a time slicing of a black--hole space--time in which a
  ``temporarily toroidal black hole'' changes topology from toroidal
  to spherical.  Ambiguities related to the definition of area and/or
  area discontinuities do occur in this example. On the other hand,
  the four dimensional analogue of Figure~\ref{curve.fig} obtained by
  rotating the curve from Figure~\ref{curve.fig} around a
  \emph{horizontal} axis can occur in a time slicing of a black--hole
  space--time in which two black holes merge together. There are
  neither obvious ambiguities related to the definition of the area
  nor area discontinuities in this case.}.  When measuring the length
of the curve in Figure~\ref{curve.fig} one faces various options: a)
discard the middle piece altogether, as it has no interior; b) count
it once; c) count it twice --- once from each side.  The purely
differential geometric approach to area, as given by
Equation~\eq{standardarea} does not say which choice should be made.
The Hausdorff area approach, Equation~\eq{areah}, counts the middle
piece once. The prescription \eq{aream} counts it twice.  We wish to
argue that the most reasonable prescription, from an entropic point of
view, is to use the prescription \eq{aream}.  In order to do that, let
$(M,g)$ be the three dimensional Minkowski space--time and consider a
thin long straw $R$ lying on the $y$ axis in the hypersurface $t=0$:
\begin{equation}
  \label{defs}
  R=\{t=x=0, y\in [-10,10]\}\ . 
\end{equation}
Set $K=J^+(R)\cap \{t=1 \}$, then $K$ is a convex compact set which
consists of a strip of width $2$ lying parallel to the straw with two
half--disks of radius $1$ added at the ends.  Let $\hat M= M\setminus
K$, equipped with the Minkowski metric, still denoted by $g$. Then
$\hat M$ has a black hole region $\cB$ which is the past domain of
dependence of $K$ in $M$ (with $K$ removed).  The sections $\cH_\tau$ of
the event horizon $\cH$ defined as $\cH\cap\{t=\tau\}$ are empty for
$\tau<0$ and $\tau\ge 1$. Next, $\cH_0$ consists precisely of the straw.
Finally, for $t\in (0,1)$ $\cH_t$ is the boundary of the
 union of a strip of width $2t$
lying parallel to the straw with two half--disks of radius $t$ added at
the ends of the strip.  Thus
\begin{equation}
  \label{comp}
  \Ar(\cH_t)=\int_{\cH_t} d\Hausone=\cases{0\ , & $t< 0$ , \cr 10\ , &
    $t=0$ , \cr 20+2\pi t\ , & $0<t<1$\ , }
\end{equation}
while
\begin{equation}
  \label{comp1}
  \int_{\cH_t}N(p)\, d\Hausone=\cases{0\ , & $t< 0$ , \cr
    20\ , & $t=0$ , \cr
    20+2\pi t\ , & $0<t<1$\ . }
\end{equation}
(The end points of the straw, at which the multiplicities are
infinite, do not contribute to the above integrals with $t=0$, having
vanishing measure.)  We see that $\Ar(\cH_t)$ jumps once when reaching
zero, and a second time immediately thereafter, while the ``area
counting multiplicities'' \eq{comp1} jumps only once, at the time of
formation of the black hole. From an entropy point of view the
existence of the first jump can be explained by the formation of a
black hole in a very unlikely configuration: as discussed below events
like that can happen only for a negligible set of times. However, the
second jump of $\Ar(\cH_t)$ does not make any sense, and we conclude
that \eq{comp1} behaves in a more reasonable way. We note that the
behavior seen in Figure~\ref{curve.fig} is obtained by intersecting
$\cH$ by a spacelike surface which coincides with $t=0$ in a
neighborhood of the center of the straw $R$ and smoothly goes up in
time away from this neighborhood.

As already mentioned it footnote \ref{amb}, this example easily
generalizes to $3+1$~dimensions: To obtain a $3+1$ dimensional model
with a similar discontinuity in the cross sectional area function, for
each fixed $t$, rotate the curve $\mathcal{H}_t$ about a
\emph{vertical} axis in $x$--$y$--$z$ space.  This corresponds to
looking at the equi--distant sets to a disk.  The resulting model is a
flat space model for the Hughes \emph{et al.\/} temporarily toroidal
black hole \cite{HKSTWW}, \emph{cf.\/}\ also
\cite{STW,GSWW,HusaWinicour}.

We now show that the behavior exhibited in Figure~\ref{curve.fig},
where there is a set of points of positive measure with multiplicity
$N(p)\ge 2$, can happen for at most a \emph{negligible set of times}.

\begin{prop}\label{nice-times}
Let $M$ be a spacetime with a global $C^1$ time function $\tau \:M\to
\R$ and for $t\in \R$ let $\Sigma_t=\{p\in M: \tau(p)=t\}$ be a level
set of $\tau$. Then for any past or future horizon $\cH\subset M$ 
$$
\Hau^{n-1}_{h_t}(\{p\in \Sigma_t: N(p)\ge 2\})=0
$$
for almost all $t\in \R$. (Where $h_t$ is the induced Riemannian
metric on $\Sigma_t$.)  For these $t$ values
$$\int_{\cH\cap\Sigma_{t}}N(p)\,d\Hau^{n-1}(p) =
\Hau^{n-1}_{h_t}(\cH\cap\Sigma_{t})\;,$$ so that for almost all $t$
the area of $\cH\cap \Sigma_t$ counted with multiplicity is the same
as the $n-1$~dimensional Hausdorff measure of $\cH\cap \Sigma_t$.
\end{prop}

\begin{proof}
Let $\hs$ be the set of points of $\cH$ that are on more than one
generator.  By Proposition~\ref{one-gen} $\Hau_\sigma^{n}(\hs)=0$
(where $\kaux$ is an auxiliary complete Riemannian metric $\kaux$ on
$M$).  The Hausdorff measure version of Fubini's theorem, known as the
``co--area formula''
\cite[Eq.~10.3, p.~55]{SimonGMT}, gives
\begin{equation}\label{slicing}
  \int_{\R}\Hau^{n-1}_{\kaux}(\hs \cap \Sigma_t)\,
  dt=\int_{\R}\Hau^{n-1}_{\kaux}\big(\tau\big|_{\cH}^{-1}[t]\big)\,dt=
  \int_{\hs}J(\tau\big|_{\cH})\,d\Hau^n_{\kaux}=0\ .
\end{equation}
Here $J(\tau\big|_{\cH})$ is the Jacobian of the function $\tau$ restricted
to $\cH$ and the integral over $\hs$ vanishes as $\Hau^n_{\kaux}(\hs
)=0$. 

Equation~\eq{slicing} implies that for almost all $t\in \R$ we have
$\Hau_\kaux^{n-1}(\cH\cap \Sigma_t)=0$.  But for any such $t$ we also
have $\Hau_{h_t}^{n-1}(\cH\cap \Sigma_t)=0$.  (This can be seen by
noting that the identity map between $\Sigma_t$ with the metric of
$\kaux$ restricted to $\Sigma_t$ and $(\Sigma_t,h_t)$ and is locally
Lipschitz. And locally Lipschitz maps send sets of $n-1$~dimensional
Hausdorff measure zero to sets with $n-1$~dimensional Hausdorff
measure zero.)  This completes the proof.
\qed
\end{proof}

 We note that the equation $\Hau^{n-1}_{\kaux}(\hs\cap \Sigma_t)=0$
 for almost all $t$'s shows that the set $(S_t)_{\reg} \subset
 S_t\equiv \Sigma_t\cap \cH$ given by Proposition~\ref{Pinter} has
 full measure in $S_t$ for almost all $t$'s. It is not too difficult
 to show (using the Besicovitch covering theorem) that $(S_t)_{\reg} $
 is countably rectifiable which gives a proof, alternative to
 Proposition~\ref{Phow}, of countable rectifiablity (up to a
 negligible set) of $S_t$ for almost all $t$'s. We emphasize, however,
 that Proposition~\ref{Phow} applies to all sections of $\cH$.

\section{Non-negativity of $\tA$}
\label{Sarea1}

The proof of the area theorem consists of two rather distinct steps:
the first is to show the non-negativity of the divergence of the
generators of event horizons under appropriate conditions, the other
is to use this result to conclude that the area of sections is
nondecreasing towards the future.  In this section we shall establish
non-negativity  of $\tA$.

\subsection{Causally regular conformal completions}

\label{Sarea2}
A pair $(\bar M,\bar g)$ will be called a \emph{conformal completion}
of $(M,g)$ if $\bar M$ is a manifold with boundary such that $M$ is
the interior of $\bar M$. The boundary of $\bar M$ will be called Scri
and denoted $\Scri$.  We shall further suppose that there exists a
function $\Omega$, positive on $M$ and differentiable on $\bar M$,
which vanishes on $\Scri$, with $d\Omega$ \emph{nowhere vanishing} on
$\Scri$.  We emphasize that no assumptions about the causal nature of
Scri are made.  We shall also require that the metric $\Omega^{-2}g$
extends by continuity to $\Scri$, in such a way that the resulting
metric $\bar g$ on $\bar M$ is differentiable. We set
$$\Scri^+=\{p\in\Scri\ |\ I^{-}(p;\bar M)\cap M \ne \nothing\}\ .$$ 
In the results presented below  $\scrip$ is not required to be connected.

Recall that a space--time is said to satisfy the \emph{null energy
  condition}, or the \emph{null convergence condition} if
\begin{equation}
  \label{nec}
  \Ric(X,X)\ge 0
\end{equation}
for all null vectors $X$. 
The following Proposition spells out
some conditions which guarantee non-negativity of the Alexandrov
divergence of the generators of $\cH$; as already pointed out, we do
not assume that $\scrip$ is null:
\begin{Proposition}
  \label{Ptheta1} Let  $(M,g)$ be a smooth
  space--time with a conformal completion $(\bar M,\bar g)=(M\cup
  \scri^+,\Omega^2g)$ of $(M,g)$ and suppose that the null energy
  condition holds on the past $I^-(\scri^+;\bar M)\cap M$ of $\scri^+$
  in $M$. Set
$$\cH =\partial J^-(\scri^+;\bar M)\cap M\ .$$
Suppose that there exists a neighborhood $\cal O$ of $\cH$ with the
following property: for every compact set $C\subset \cal O$ that meets
$I^-(\scrip; \bar M)$ there exists a future inextendible (in $M$) null
geodesic $\eta \subset \partial J^+(C; M)$ starting on $C$ and having
future end point on $\scrip$. Then
$$\tA\ge 0\ . $$
Further, if $\calS$ is any $C^2$ spacelike or timelike hypersurface
which meets $\cH$ \transversally, then
$$\tAcS\ge 0\quad  \mbox{\rm on $\calS\cap\cH$}\ .$$
\end{Proposition}


\medskip

\begin{proof}
  Suppose that the result does not hold.  Let, first, $p_0$ be an
  Alexandrov point of $\cH$ with $\tA<0$ at $p_0$ and consider a
  neighborhood $\mathcal{N}$ of $p_0$ of the form $\Sigma\times\R$,
  constructed like the set $\cO$ (not to be confused with the set
  $\cO$ in the statement of the present proposition) in the proof of
  Theorem~\ref{T1}, so that $\cH\cap \mathcal{N}$ is a graph over
  $\Sigma$ of a function $f\:\Sigma\to\R$, define $x_0$ by
  $p_0=(x_0,f(x_0))$.  By point 3.\ of Proposition~\ref{Pdiff} in a
  coordinate neighborhood $\cU\subset \Sigma$ the function $f$ can be
  written in the form
\begin{equation}
  \label{alex3}
  f(x)=f(x_0) + df(x_0)(x-x_0)+\frac{1}{2}D^2f(x_0)(x-x_0,x-x_0)+
  o(|x-x_0|^2)\ .
\end{equation}
After a translation, a rotation and a rescaling we will have
\begin{equation}
  \label{fde0}
x_0=0\ , \quad f(x_0)= 0\ , \quad df(x_0) =dx^n\ . 
\end{equation}
Let $B^{n-1}(\delta)\subset \R^{n-1}$ be the $(n-1)$~dimensional open
ball of radius $\delta$ centered at the origin, for $q\in
B^{n-1}(\delta)$, set $x=(q,0)$ and 
\begin{equation}
  \label{fde}
  f_{\epsilon,\eta}(q)=f(x_0) +
  df(x_0)(x-x_0)+\frac{1}{2}D^2f(x_0)(x-x_0,x-x_0)+\epsilon|x-x_0|^2+\eta
  \ .
\end{equation}
Define
\begin{equation}
  \label{Sgraphdef}
  S_{\epsilon,\eta,\delta}=\{\mbox{ graph over
$(B^{n-1}(\delta)\times \{0\})\cap \cU$ of 
$f_{\epsilon,\eta}$}\}\ .
\end{equation}
If its parameters are chosen small enough, $S_{\epsilon,\eta,\delta}$
will be a smooth spacelike submanifold of $M$ of co--dimension two.
Let $\theta_{\epsilon,\eta,\delta}$ be the mean curvature of
$S_{\epsilon,\eta,\delta}$ with respect to the null vector field $K$,
normal to $S_{\epsilon,\eta,\delta}$, defined as in \eq{Kequation}
with $g$ there replaced by $f_{\epsilon,\eta}$ (compare
\eq{eq:2a})\mnote{rewordings}.  From \eq{fde} when all the parameters
are smaller in absolute value than some thresholds one will
have\mnote{equation corrected}
$$ |\theta_{\epsilon,0,\delta}(p_0) - \tA(p_0) |\le C \epsilon $$
for some constant $C$, so that if $\epsilon$ is chosen small enough we
will have 
\begin{equation}
  \label{sign}
  \theta_{\epsilon,\eta,\delta}< 0
\end{equation}
at $p_0$ and $\eta=0$; as $ \theta_{\epsilon,\eta,\delta}$ is
continuous in all its relevant arguments,\mnote{some rewordings
  associated to the previous correction} Equation~\eq{sign} will hold
throughout $S_{\epsilon,\eta,\delta}$ if $\delta$ is chosen small
enough, for all sufficiently small $\epsilon$'s and $\eta$'s.  If
$\delta $ is small enough and $\eta=0$ we have
$$ \overline{S_{\epsilon,\eta=0,\delta}} \subset J^{+}(\cH)\ , \qquad
\overline{S_{\epsilon,\eta=0,\delta}}\cap \cH = \{p_0\}\ .$$ Here
$\overline{S_{\epsilon,\eta,\delta}}$ denotes the closure of
$S_{\epsilon,\eta,\delta}$. It follows that for all sufficiently small
strictly negative $\eta$'s we will have\begin{eqnarray}
  \label{out1}&
  S_{\epsilon,\eta,\delta}\cap J^-(\Scri^+)\ne \nothing\ , & \\ 
 \label{out2} &  \overline{S_{\epsilon,\eta,\delta}} \setminus
  S_{\epsilon,\eta,\delta} \subset J^{+}(\cH)\ , &
\end{eqnarray}
Making $\delta$ smaller if necessary so that
$S_{\epsilon,\eta,\delta}\subset \cal O$, our condition on $\cal O$
implies that 
there exists a null geodesic $\Gamma\:[0,1]\to\bar M$ such that
$\Gamma(0)\in \overline{S_{\epsilon,\eta,\delta}}$,
$\Gamma|_{[0,1)}\subset \partial J^+(S_{\epsilon,\eta,\delta}; M)$,
and $\Gamma(1)\in \Scri^+$. The behavior of null geodesics under
conformal rescalings of the metric (\emph{cf., e.g.,}
\cite[p.~222]{HE}) guarantees that $\Gamma|_{[0,1)}$ is a complete
geodesic in $M$. Suppose that $\Gamma(0)\in
\overline{S_{\epsilon,\eta,\delta}} \setminus
S_{\epsilon,\eta,\delta}$, then, by Equation~\eq{out2}, $\Gamma$ would
be a causal curve from $J^{+}(\cH)$ to $\Scri^+$, contradicting the
hypothesis that $\cH=\partial J^{-}(\scrip)$. It follows that
$\Gamma(0)\in S_{\epsilon,\eta,\delta}$.  As the null energy condition
holds along $\Gamma$ Equation~\eq{sign} implies that
$S_{\e,\eta,\delta}$ will have a focal point, $\Gamma(t_0)$, to the
future of $\Gamma(0)$. But then (\emph{cf}.~\cite[Theorem 51, p.
298]{ONeill}) the points of $\Gamma$ to the future of $\Gamma(t_0)$
are in $I^+(S_{\e,\eta,\delta})$, and thus $\Gamma$ does not lie in
$\partial J^+(S_{\epsilon,\eta,\delta}; M)$.  This is a contradiction
and the non-negativity of $\tA$ follows.


The argument to establish non-negativity of $\tAcS$ is an essentially
identical (and somewhat simpler) version of the above. Suppose, thus,
that the claim about the sign of $\tAcS$ is wrong, then there exists a
point $q_0\in\calS$ which is an Alexandrov point of $\calS\cap\cH$ and
at which $\tAcS$ is negative. Consider a coordinate patch around
$q_0=0$ such that $\calS\cap\cH$ is a graph
$x^{n}=g(x^1,\ldots,x^{n-1})$ of a semi--convex function $g$.
Regardless of the spacelike/timelike character of $\calS$ we can
choose the coordinates, consistently with semi--convexity of $g$, so
that $J^-(\cH)\cap\calS$ lies {under} the graph of $g$ . The
definition~\eq{fde} is replaced by
\begin{equation}
  \label{fdeS}
  f_{\epsilon,\eta}(q)=g(q_0) +
  dg(q_0)(q-q_0)+\frac{1}{2}D^2g(q_0)(q-q_0,q-q_0)+\epsilon|q-q_0|^2+\eta\ 
  ,
\end{equation}
while \eq{Sgraphdef} is replaced by
\begin{equation}
  \label{SgraphdefS}
  S_{\epsilon,\eta,\delta}=\{\mbox{ graph over
$B^{n-1}(\delta)$ of 
$f_{\epsilon,\eta}$}\}\ .
\end{equation}
The other arguments used to prove that $\tA\ge 0$ go through without
modifications.
\qed\end{proof}

Proposition \ref{Ptheta1} assumes the existence of a neighborhood of
the horizon with some precise global properties, and it is natural to
look for global conditions which will ensure that such a neighborhood
exists.  The simplest way to guarantee that is to assume that the
conformal completion $(\bar M,\bar g)$ is globally hyperbolic, perhaps
as a manifold with boundary. To be precise, we shall say that a
manifold $(\bar M,\bar g)$, with or without boundary, is 
globally hyperbolic if there exists a smooth time function $t$ on
$\bar M$, and if $J^+(p)\cap J^-(q)$ is compact for all $p,q\in\bar M$
(compare \cite{GSWW}). For example, Minkowski space--time with the
standard $\scrip$ attached is a globally hyperbolic manifold with
boundary. (However, if both $\scrip$ and $\scri^{-}$ are attached,
then it is not. Note that if $\scrip$ and $\scri^{-}$ and $i^0$ are
attached to Minkowski space--time, then it is not a manifold with
boundary any more. Likewise, the conformal completions considered in
\cite{Waldbook} are not manifolds with boundary.)  Similarly
Schwarzschild
 space--time with the standard $\scrip$ attached to it is
a globally hyperbolic manifold with boundary. Further, the standard
conformal completions of de--Sitter, or anti--de Sitter space--time
\cite{GibbonsHawkingCEH} as well as those of the Kottler space--times
\cite{Kottler} (sometimes called Schwarzschild -- de Sitter and
Schwarzschild -- anti--de Sitter space--times) and their
generalizations considered in \cite{BLP} are globally hyperbolic
manifolds with boundary. 

It is well known that in globally hyperbolic manifolds the causal
hypotheses of Proposition \ref{Ptheta1} are satisfied, therefore we
have proved:

\begin{Proposition}
  \label{Pghwb} Under the condition \ref{thirdhyp}) of Theorem
  \ref{Tarea}, the conclusions of Proposition \ref{Ptheta1} hold.\qed
\end{Proposition}

The hypothesis of global hyperbolicity of $(\bar M,\bar g)$ is
esthetically unsatisfactory, as it mixes conditions concerning the
physical space--time $(M,g)$ together with conditions concerning an
artificial boundary one attaches to it. On the other hand it seems
sensible to treat on a different footing the conditions concerning
$(M,g)$ and those concerning $\scrip$. We wish to indicate here a
possible way to do this. In order to proceed further some terminology
will be needed:

\begin{Definition}\label{Dreg}
\begin{enumerate}
\item A point $q$ in a set $A\subset B$ is said to be a {\em past}
  point\footnote{A very similar notion has already been used in
    \cite[p.~102]{CBY}.} of $A$ with respect to $B$ if 
  ${J^-(q;B)}\cap {A}=\{q\}$.
\item We shall say that $\scrip$ is $\cH$--\emph{regular} if there
  exists a neighborhood $\cal O$ of $\cH$ such that for every compact
  set $C\subset\cal O$ satisfying $I^+(C;\bar M) \cap \Scri^+ \ne
  \emptyset$ there exists a past point \emph{with respect to
    $\Scri^+$} in $\partial I^+(C; \bar M)\cap \Scri^+$.
\end{enumerate}
 \end{Definition}

 We note, as is easily verified\footnote{The inclusion ``$\subset$"
   makes use of the fact that if $q\in \scrip$, then for any $p\in M$
   near $q$ there exists $p'\in \scrip\cap I^+(p; \bar M)$ near $q$.
   If $\scrip$ is null (and hence possibly type changing) at $q$ one
   sees this as follows.  From the definition of $\scrip$, there
   exists at $q$ a future directed outward pointing timelike vector
   $X_0$.  This can be extended in a neighborhood $U$ of $q$ to a
   future directed timelike vector field $X$ everywhere transverse to
   $\scrip$.  By flowing along the integral curves of $X$, we see that
   any point $p\in M$ sufficiently close to $q$ is in the timelike
   past of a point $p' \in \scrip$ close to $q$.}, that $\partial
 I^+(C;\bar M)\cap\Scri^+ = \partial (I^+(C;\bar M)\cap \scrip)$,
 where the boundary on the right hand side is meant as a subset of
 $\scrip$.
\begin{remark}
  In the null and timelike cases the purpose of the condition is to
  exclude pathological situations in which the closure of the domain
  of influence of a compact set $C$ in $M$ contains points which are
  arbitrarily far in the past on $\Scri^+$ in an uncontrollable way.
  A somewhat similar condition has been first introduced in
  \cite{galloway:woolgar} for null $\Scrip$'s in the context of
  topological censorship, and has been termed ``the $i^0$ avoidance
  condition'' there. We shall use the term ``$\cH$--regular'' instead,
  to avoid the misleading impression that we assume existence of an
  $i^0$.
\end{remark}
\begin{remark}\label{r4.5}
  When $\Scri^+$ is \emph{null} throughout the condition of
  $\cH$--\emph{regularity} is equivalent to the following requirement:
  there exists a neighborhood $\cal O$ of $\cH$ such that for every
  compact set $C\subset\cal O$ satisfying $I^+(C;\bar M) \cap \Scri^+
  \ne \emptyset$ there exists at least one generator of $\Scrip$ which
  intersects $\overline{I^+(C; \bar M)}$ and leaves it when followed
  to the past\footnote{\label{leavefoot}Those points at which the
    generators of $\scrip$ exit $\overline{I^+(C; \bar M)}$ are past
    points of $\partial{I^+(C; \bar M)}\cap \scrip$.}.  This condition
  is satisfied by the standard conformal completions of Minkowski
  space--time, or of the Kerr--Newman space--times. In fact, in those
  examples, when $\cal O$ is suitably chosen, for every compact set
  $C\subset\cal O$ satisfying $I^+(C;\bar M) \cap \Scri^+ \ne
  \emptyset$ it holds that \emph{every} generator of $\Scrip$
  intersects $\overline{I^+(C; \bar M)}$ and leaves it when followed
  to the past.
\end{remark}

\begin{remark}
  When $\Scri^+$ is \emph{timelike} throughout the condition of
  $\cH$--\emph{regularity} is equivalent to the following requirement:
  there exists a neighborhood $\cal O$ of $\cH$ such that for every
  compact set $C\subset\cal O$ satisfying $I^+(C;\bar M) \cap \Scri^+
  \ne \emptyset$, there is a point $x$ in $\overline{I^+(C; \bar M)}
  \cap \scrip$ such that every past inextendible causal curve in
  $\scrip$ from $x$ leaves $\overline{I^+(C; \bar M)}$.  (The standard
  conformal completions of anti--de Sitter space--time
  \cite{GibbonsHawkingCEH} as well as those of the negative $\Lambda$
  Kottler space--times \cite{Kottler} and their generalizations
  considered in \cite{BLP} all satisfy this condition.)  Under this
  condition the existence of a past point in $\partial I^+(C;\bar
  M)\cap \scrip$ can be established as follows: Let $p\in
  \partial{I^+(C; \bar M)}\cap J^-(x;\scrip)$, let $\gamma$ be a
  causal curve entirely contained in $\partial{I^+(C; \bar M)}\cap
  \scrip$ through $p$; if no such curves exist then $p$ is a past
  point of $\partial{I^+(C; \bar M)}\cap \scrip$ and we are done; if
  such curves exist, by a standard construction that involves Zorn's
  lemma we can without loss of generality assume that $\gamma$ is past
  inextendible \emph{in} $\partial{I^+(C; \bar M)}\cap \scrip$. Let
  $\Gamma$ be a past inextendible causal curve \emph{in} $\scrip$
  which contains $\gamma$. The current causal regularity condition on
  $\scrip$ implies that $\Gamma$ has an end point $q$ on
  $\partial{I^+(C; \bar M)}\cap \scrip$. If $q$ were not a past point
  of $\partial{I^+(C; \bar M)}\cap \scrip$ one could extend $\gamma$
  as a causal curve in $\partial{I^+(C; \bar M)}\cap \scrip$, which
  would contradict the maximality of $\gamma$. Hence $q$ is a past
  point of $\partial{I^+(C; \bar M)}\cap \scrip$.
\end{remark}
\begin{remark}
  When $\Scri^+$ is \emph{spacelike} throughout the condition of
  $\cH$--\emph{regularity} is equivalent to the requirement of
  existence of a neighborhood $\cal O$ of $\cH$ such that for every
  compact set $C\subset \cal O$ satisfying $I^+(C;\bar M) \cap \Scri^+
  \ne \emptyset$, the set $\overline{I^+(C; \bar M)}$ does not contain
  all of $\scrip$.  We note that the standard conformal completions of
  de Sitter space--time, as well as those of the positive-$\Lambda$
  generalized Kottler space--times \cite{GibbonsHawkingCEH}, satisfy
  our $\cH$--regularity condition.
\end{remark}

Let us present our first set of conditions which guarantees that the
causal hypotheses of Proposition \ref{Ptheta1} hold:

\begin{prop}\label{Pglobal1}
  Let $(M,g)$ be a spacetime with a $\cH$--regular $\scrip$, and
  suppose that there exists in $M$ a partial Cauchy surface $\Sigma$
  such that
\begin{enumerate}
\item[(i)] $I^+(\Sigma; M) \cap I^-(\Scrip; \bar M) \subset
  D^+(\Sigma; M)$, and
\item[(ii)]  $\Scrip\subset  I^+(\Sigma; \bar M)$.
\end{enumerate}
Then there is a neighborhood $\cal O$ of $\cH$ such that
if $C$ is a compact subset of $J^+(\Sigma;M) \cap \cal O$  that
meets $I^-(\Scrip; \bar M)$, there exists a future inextendible
(in $M$) null geodesic $\eta \subset \partial I^+(C; M)$ starting at a
point on $C$ and having future end point on $\scrip$.
\end{prop}

\begin{remark}
  The conditions (i) and (ii) in Proposition \ref{Pglobal1} form a
  version of \emph{``asymptotic predictability''}.  They \emph{do not}
  imply that global hyperbolicity extends to the horizon or $\Scrip$.
  These conditions are satisfied in the set-up of~\cite{HE}.
\end{remark}

\begin{proof}
  Choose $\cal O$ as in the definition of $\cH$--regularity; then for
  compact $C\subset \cal O$ satisfying $I^+(C;\bar M) \cap \Scri^+ \ne
  \emptyset$ there is a past point $q$ on $\partial{I^+(C; \bar
    M)}\cap \Scri^+$.  By \cite[Theorem 8.1.6, p.~194]{Waldbook}
  (valid in the present context) there exists a causal curve $\eta
  \subset \partial I^+(C;\bar M)$ with future end point $q$ which
  either is past inextendible in $\bar M$ or has a past end point on
  $C$. As $q$ is a past point with respect to $\scrip$, $\eta$ must
  meet $M$. Note that, due to the potentially unusual shape of
  $\scrip$, in principle $\eta$ may meet $\scrip$ after $q$ when
  followed backwards in time even infinitely often in a finite
  interval. The following argument avoids this difficulty altogether:
  we consider any connected component $\eta_0$ of $\eta\cap M$;
  $\eta_0$ is a null geodesic with future end point on $\scrip$.  Now,
  conditions (i) and (ii) imply that $\eta_0$ enters the interior of
  $D^+(\Sigma;M)$.  If $\eta_0$ is past inextendible in $M$, then,
  following $\eta_0$ into the past, $\eta_0$ must meet and enter the
  timelike past of $\Sigma$.  This leads to an achronality violation
  of $\Sigma$: Choose $p \in \eta_0 \cap I^-(\Sigma;M)$.  Then, since
  $I^+(\partial I^+(C;M);M) \subset I^+(C;M)$, by moving slightly to
  the future of $p$, we can find a point $p'\in I^+(C;M)\cap
  I^-(\Sigma;M)\subset I^+(\Sigma;M)\cap I^-(\Sigma;M)$.  We conclude
  that $\eta_0$ has a past end point on $C$.  \qed\end{proof}

The conditions of the proposition that follows present an alternative
to those of Proposition \ref{Pglobal1}; they form a version of
\emph{``strong asymptotic predictability''}:
\begin{prop} \label{Pglobal2}

  Let $(M,g)$ be a spacetime with a 
  $\cH$--regular $\Scrip$, and suppose that $M$ contains a causally
  simple domain $V$ such that $$ \overline {I^-(\Scrip; \bar M)}\cap
  M\subset V\ .$$ Then there is a neighborhood $\cal O$ of $\cH$ such
  that if $C$ is a compact subset of $V \cap \cal O$ that meets
  $I^-(\Scrip; \bar M)$, there exists a future inextendible (in $M$)
  null geodesic $\eta \subset \partial I^+(C; V)$ starting at a point
  on $C$ and having future end point on $\Scrip$.
\end{prop}
\begin{remark}
  Recall, an open set $V$ in $M$ is causally simple provided for all
  compact subsets $K\subset V$, the sets $J^{\pm}(K;V)$ are closed in
  $V$.  Causal simplicity is implied by global hyperbolicity; note,
  however, that the latter is not a sensible assumption in the
  timelike $\Scrip$ case: For example, anti--de Sitter space is not
  globally hyperbolic; nevertheless it is causally simple.
\end{remark}

\begin{remark}
  Proposition~\ref{Pglobal2} assumes that causal simplicity extends to
  the horizon, but not necessarily to $\Scrip$ --- it is not assumed
  that the closure $\overline V$ of $V$ in $\bar M$ is causally
  simple.\mnote{rewording}  Its hypotheses are satisfied in the set-up of
  \cite{Waldbook} and in the (full) set-up of \cite{HE}. (In both
  those references global hyperbolicity of $V$ is assumed.)  We note
  that replacing $M$ by $V$ we might as well assume that $M$ is
  causally simple.
\end{remark}
\begin{remark}
  If $C$ is a smooth spacelike hypersurface-with-boundary, with
  $\partial C =S$, then (i) $\eta$ meets $C$ at a point on $S$, (ii)
  $\eta$ meets $S$ orthogonally, and (iii) $\eta$ is outward pointing
  relative to $C$.
\end{remark}
\smallskip

\begin{proof}
  Choose ${\cal O}\subset V$ as in the definition of
  $\cH$--regularity, and let $q$ be a past point of $\partial
  I^+(C;\bar M)\cap \scrip$ with respect to $\scrip$.  Then $q\in
  \partial I^+(C;\bar V)$, where $\bar V = V \cup \Scrip$.  Arguing as
  in Proposition~\ref{Pglobal1} we obtain a causal curve $\eta\subset
  \partial I^+(C;\bar V)$ with future end point $q\in\scrip$, which is
  past inextendible in $\partial I^+(C;\bar V)$.  As before, $\eta$
  meets $M$, and we let $\eta_0$ be a component of $\eta\cap M$.  Then
  $\eta_0$ is a null geodesic in $V$ with future end point on $\scrip$
  which is past inextendible in $\partial I^+(C; V)$.  Since $V$ is
  causally simple, $\partial I^+(C;V) = J^+(C;V) \setminus I^+(C;V)$,
  which implies that $\eta_0$ has past end point on $C$.
  \qed\end{proof}

If $M$ itself is globally hyperbolic, taking $V=M$ in Proposition
\ref{Pglobal2} one obtains:

\begin{Corollary}
  \label{Cglobal} Under the hypotheses of point a) of Theorem
  \ref{Tarea}, the conclusions of Proposition \ref{Ptheta1} hold.
\end{Corollary}

\subsection{Complete generators}
\label{sScomplete}

Let us turn our attention to horizons $\cH$ the generators of which
are future complete. We emphasize that $\cH$ is not necessarily an
event horizon, and the space--time does not have to satisfy any
causality conditions.  We start with some terminology: consider a
$C^2$ spacelike manifold $S\subset M$ of co--dimension two, let $p_0\in
S$. We can choose coordinates in a globally hyperbolic neighborhood
$\cO$ of $p_0$ such that the paths $s\to (x^1,\ldots,x^n,x^{n+1}=s)$
are timelike and such that
\begin{equation}
  \label{Sgr}
  S\cap \cO = \{x^{n}=x^{n+1}=0\}\ .
\end{equation}
 Suppose further that 
 \begin{equation}
   \label{Stg}
   p_0\in S\cap \cH\ , \qquad T_{p_0}S\subset T_{p_0}\cH \ ,
 \end{equation}
 and that $p_0$ is an Alexandrov point of $\cH$. Passing to a subset
 of $\cO$ if necessary we may assume that $\cO\cap \cH$ is a graph of
 a function $f$, with $p_0=(x_0,f(x_0))$. We shall say that \emph{$S$
   is second order tangent to $\cH$} if the above conditions hold and
 if in the coordinate system of Equation~\eq{Sgr} we have
\begin{equation}
  \label{S2tg}
  \forall \ X,Y \in T_{p_0}S \qquad D^2f(X,Y)=0 \ .
\end{equation}
Here $D^2f$ is the Alexandrov second derivative of $f$ at $x_0$, as in
Equation~\eq{alex2}.  The notion of $S\subset\calS$ being \emph{second
  order tangent to a section $\calS\cap\cH$} at an Alexandrov point of
this section is defined in an analogous way, with coordinates adapted so that, locally, $\calS =\{x^n=0\}$
or $\calS =\{x^{n+1}=0\}$.\mnote{comment added}

We note the following result:
\begin{Lemma} \label{Ltheta2} Let $S$ be a
  $C^2$ spacelike manifold of co--dimension two which is tangent to a
  horizon $\cH$ at $p_0\in \cH\cap S$. Suppose that
\begin{enumerate}
\item   $p_0$ is an Alexandrov point of $\cH$, and $S$  is
    second order tangent to  $\cH$ there, or
 \item   $p_0$ is an Alexandrov point of $\calS\cap\cH$, with
   $\calS=\{x^{n}=0\}$ in the coordinate system of \eq{Sgr}, and $S$  is
    second order tangent to  $\calS\cap\cH$ there, or
 \item   $p_0$ is an Alexandrov point of $\calS\cap\cH$, with
   $\calS=\{x^{n+1}=0\}$ in the coordinate system of \eq{Sgr}, and $S$  is
    second order tangent to  $\calS\cap\cH$ there.
\end{enumerate}
Let $\Gamma$ be a null geodesic containing a generator $\gamma$ of
$\cH$ through $p_0$, with $\Gamma(0)=p_0$. If $\Gamma(1)$ is a focal
point of $S$, then there exists $a\in[0,1]$ such that $\Gamma(a)$ is
an endpoint of $\gamma$ on $\cH$.
\end{Lemma}

\mnote{remark commented out}
\begin{remark}
We stress that we haven't assumed anything about the time orientation
of $\Gamma$.
\end{remark}

\begin{proof}
If $\Gamma(0)$ is an end point of $\cH$ there is nothing to
prove, so we can suppose that $p_0$ is an interior point of the
generator $\gamma$. Suppose that $\Gamma(1)$ is a focal point of $S$,
it is well known\footnote{\label{footgr}The only complete proof of
  this fact known to us is to be found in \cite[Section
  2]{Galloway:fitopology}.  For the case at hand one should use the
  variation defined there with $c_\ell=0$. We note the following
  misprints in \cite[Section 2]{Galloway:fitopology}: In Equation
  (2.7) $\eta^\prime_0$ should be replaced by $N_0$ and
  $\eta^\prime_\ell$ should be replaced by $N_\ell$; in Equation
  (2.10) $N_0$ should be replaced by $\eta^\prime_0$ and $N_\ell$
  should be replaced by $\eta^\prime_\ell$.} that for any $b>1$ there
exists a one parameter family of timelike paths $\Gamma_v\:[0,b]\to M$,
$|v|\le v_0$, such that
\begin{eqnarray}
&   \Gamma_v(0) \in S\ , \qquad \Gamma_v(b)=\Gamma(b)\ ,
& \nonumber \\
&\forall u \in [0,b] \qquad d_{\kaux}(\Gamma_v(u),\Gamma(u))\le C|v|\ , &
 \label{dist}
\\
& \displaystyle{g\bigg(\frac{\partial\Gamma_v }{\partial u} ,
  \frac{\partial\Gamma_v 
  }{\partial u} \bigg)} \le - Cv^2\ , \label{est} 
\end{eqnarray}
with some constant $C$. Here $d_{\kaux}$ denotes the distance on $M$
measured with respect to some auxiliary complete Riemannian metric
${\kaux}$. 

Let $X$ be any smooth vector field which vanishes at $\Gamma(b)$ and
which equals $\partial/\partial x^{n+1}$ in a neighborhood of $p_0$,
where $x^{n+1}$ is the $({n+1})$st coordinate in the coordinate system
of Equation~\eq{Sgr}. Let $\phi_s$ denote the (perhaps locally
defined) flow of $X$, for $|s|\le 1$ we have the straightforward
estimate
\begin{equation}
  \label{lineq}
  \Big|g\bigg(\frac{\partial(\phi_s(\Gamma_v(u)) )}{\partial u} ,
  \frac{\partial(\phi_s(\Gamma_v (u)) )}{\partial u}
  \bigg) - g\bigg(\frac{\partial\Gamma_v (u) }{\partial u} , 
  \frac{\partial\Gamma_v     (u)  }{\partial u}\bigg)\Big| \le C_1 |s|\ ,
\end{equation}
for some constant $C_1$, whenever $\phi_s(\Gamma_v)(u)$ is defined.
Set
$$
\eta(\delta) = \sup_{q\in B^{n-1}(0,\delta)}
\frac{|f(q)|}{(d_\kaux(q,0))^2}\ \ , 
$$ where $f$ is again the graphing function of $\cH\cap \cO$, and
$B^{n-1}(0,\delta)$ is as in the proof of Proposition~\ref{Ptheta1};
$d_\kaux(q,0)$ should be understood in the obvious way.  Under the
hypotheses 1.\  and 2., by Equations~\eq{alex2} and \eq{S2tg} we have
\begin{equation}
  \label{lim}
  \eta(\delta)\to_{\delta \to 0} 0 \ .
\end{equation}
On the other hand, suppose that 3.\  holds, let $g$ be the graphing
function of $\calS\cap\cH$ in the coordinate system of \eq{Sgr}, the
hypothesis that $S$ is second order tangent to $\calS\cap\cH$ gives
$$g(x)=o(|x|^2)\ ,$$ where $x$ is a shorthand for
$(x^1,\ldots,x^{n-1})$. Now $(x,g(x))\in \cH$ so that $f(x,g(x))=0$,
hence
$$|f(x,0)|=|f(x,0)-f(x,g(x))|\le L|g(x)|=o(|x|^2)\ ,$$ where $L$ is
the Lipschitz continuity constant of $f$ on $B^{n}(0,\delta)$. Thus
Equation~\eq{lim} holds in all cases.

Equation~\eq{dist} shows that for $|v| \le \delta/C_2$, for some
constant $C_2$, we will have $\Gamma_v(0) \in B^{n-1}(0,\delta)\times
\{0\}\times \{0\} \subset S\cap \cO$, so that $\Gamma_v(0)$ can be
written as
$$\Gamma_v(0)= (x(v),x^{n+1}=0)\ , \qquad x(v)=(q(v), x^{n}=0)\ ,
\qquad q(v)\in B^{n-1}(0,\delta) \ . $$
Consider the point $p(v)\equiv (x(v), f(x(v)))$; by definition of $X$
we have
$$p(v)=\phi_{f(x(v))}(\Gamma_v(0)) \in \cH\ .$$ From Equations
\eq{dist}, \eq{est} and \eq{lineq}, together with the definition of
$\eta$ we obtain
$$\forall u \in [0,b] \qquad
g\bigg(\frac{\partial(\phi_{f(x(v))}(\Gamma_v(u)) )}{\partial u} ,
\frac{\partial(\phi_{f(x(v))}(\Gamma_v (u)) )}{\partial u} \bigg) \leq
-C v^2 + C_1 C^2 \eta(\delta)v^2\ . $$ 
This and Equation~\eq{lim} show that for $\delta$ small enough
$\phi_s(\Gamma_v)$ will be a timelike path from $p(v)\in \cH$ to
$\Gamma_v(b)$. Achronality of $\cH$ implies that
$\Gamma_v(b)\not\in\cH$, thus $\Gamma$ leaves $\cH$ somewhere on
$[0,b)$.  As $b$ is arbitrarily close to $1$, the result follows.
\qed\end{proof}

As a Corollary of Lemma~\ref{Ltheta2} one immediately obtains:

\begin{Proposition}
  \label{Ptheta2} Under the hypothesis b) of Theorem~\ref{Tarea}, we
  have
  $$\tA\ge 0\ . $$ 
Further, if $\calS$ is any spacelike or timelike $C^2$ hypersurface
that meets $\cH$ \transversally, then
$$\tAcS\ge 0\ .$$
\end{Proposition}  
\begin{proof}
  Suppose that $\tA(p_0)<0$, for $\delta$ small enough let
  $S=S(\delta)$ be the manifold $S_{0,0,\delta}$ defined in
  Equation~\eq{Sgraphdef}.  Consider the future directed maximally
  extended null geodesic $\Gamma$ normal to $S$ which coincides, for
  some values of its parameters, with a generator $\gamma$ of $\cH$
  through $p_0$.  Now, $\gamma$ is complete to the future by
  hypothesis, thus so must be the case with $\Gamma$.  By well known
  results$^{\mbox{\scriptsize\ref{footgr}}}$ \cite[Theorem~43,
  p.~292]{ONeill}, $S$ has a focal point along $\Gamma$ at finite
  affine distance, say $1$. By Lemma~\ref{Ltheta2} the generator
  $\gamma$ has a future end point, which contradicts the definition of
  a future horizon.  Finally, if $\calS$ is a spacelike hypersurface,
  or a timelike hypersurface properly transverse to $\cH$, then the
  same argument with $S_{0,0,\delta}$ given by
  Equation~\eq{SgraphdefS} establishes $\tAcS\ge 0$.  \qed\end{proof}

\section{Propagation of Alexandrov points along generators, optical equation}
\label{Sfr}

The aim of this section is to show that, roughly speaking, Alexandrov
points \emph{``propagate to the past''} along the generators.  Recall,
now, that the Weingarten map $b$ of a smooth null hypersurface $S$
satisfies a Ricatti equation
\begin{eqnarray}\label{eq:2bn}
b'+b^2 +R = 0 .  
\end{eqnarray}   
Here $'$ and $R$ are defined in Appendix~\ref{apR}, see
Equation~\eq{eq:2bprim} and the paragraph that follows there.
Equation~\eq{eq:2bn} leads to the well known Raychaudhuri
equation in general relativity: by taking the trace of~\eq{eq:2bn} we
obtain the following formula for the derivative of the null mean
curvature $\theta=\theta(s)$ along $\eta$,
\begin{eqnarray}
\theta' = -{\rm Ric}(\eta',\eta') - \sigma^2 - \frac1{n-2}\theta^2,
\label{eq:2c} 
\end{eqnarray}
where $\sigma$, the shear scalar\footnote{\label{conflict}This is one
  of the very few occurrences of the shear scalar (traditionally
  denoted by $\sigma$ in the physics literature) in our paper, we hope
  that this conflict of notation with the auxiliary Riemannian metric
  also denoted by $\kaux$ will not confuse the reader.}, is the trace
of the square of the trace free part of $b$. Equation~\eq{eq:2c} is
the well-known Raychaudhuri equation (for an irrotational null
geodesic congruence) of relativity.  This equation shows how the Ricci
curvature of spacetime influences the null mean curvature of a null
hypersurface.  We will refer to Equation~\eq{eq:2bn} as the
\emph{optical equation}.

Let $\kaux$ be any auxiliary complete smooth Riemannian metric on $M$
and define $UM\subset TM$ to be the bundle of $\kaux$--unit vectors
tangent to $M$. Following~\cite{ChGalloway} for $p\in\cH$ we define
$\cN_p\subset U_pM$ as the collection of $\kaux$--unit, future
directed vectors tangent to generators of $\cH$ through $p$. Those
vectors are necessarily lightlike and will be called
\emph{semi--tangents} to $\cH$.  We set
$$\cN= \cup_{p\in\cH}\cN_p\ . $$
The main result of this section is the following:

\begin{Theorem}\label{Tfr} 
  Let $\Gint$ denote the set of interior points of a generator
  $\Gamma$ of $\cH$ (\emph{i.e.}, $\Gamma$ without its end--point, if
  any). If $p_0\in\Gint$ is an Alexandrov point of a section
  $\calS\cap\cH$, where $\calS$ is an embedded \emb spacelike or timelike $C^2$
  hypersurface that intersects $\cH$ \transversally, then
    \begin{equation}       \label{Gintincl}
     \Gint\cap J^-(p_0)\setminus \{p_0\}\subset\dH\ .
    \end{equation}
    Further the Alexandrov derivative $D^2f$ of any graphing function
    of $\cH$ varies smoothly over $\Gint\cap J^-(p_0)$, and the null
    Weingarten map $b_\Al$ constructed out of $D^2f$ in a way
    analogous to that presented in Appendix~\ref{sec:Greg-null}
    satisfies the optical equation~\eq{eq:2bn} and the 
    Raychaudhuri equation~\eq{eq:2c} there.
\end{Theorem}

\begin{remark} It would be of interest to find out whether or not the
  inclusion~\eq{Gintincl} can be strengthened to $\Gint\subset\dH$. It
  follows from Theorem \ref{TCfr} below that this last inclusion will
  hold for $\Hnmk$ almost all generators passing through any given
  section $\calS\cap\cH$ of $\cH$, when the generators are counted by
  counting the points at which they meet $\calS\cap\cH$.
\end{remark}
\begin{remark} 
  There exist horizons containing generators on which no points are
  Alexandrov. As an example, let $\cH$ be the boundary of the future
  of a square lying in the $t=0$ hypersurface of three dimensional
  Minkowski space--time. Then $\cH$ is a union of portions of null
  planes orthogonal to the straight segments lying on the boundary of
  the square together with portions of light cones emanating from the
  corners of the square. Consider those generators of $\cH$ at which
  the null planes meet the light cones: it is easily seen that no
  point on those generators is an Alexandrov point of $\cH$.
\end{remark}

As constructing support functions (or support hypersurfaces) to the
horizon $\cH$ is generally easier than doing analysis directly on
$\cH$ we start by giving a criteria in terms of upper and lower
support functions for a function to have a second
order expansion~\eq{alex2}.

\begin{lemma}\label{pinch}
  Let $U\subset \bbR^n$ be an open neighborhood of $x_0$ and $f\:
  U\to \bbR$ a function.  Assume for $\ell =1,2,\dots$ that there are
  open neighborhoods $U^+_\ell$ and $U^-_\ell $ of $x_0$ and $C^2$
  functions $f^\pm_\ell\:U^\pm_\ell  \to \bbR$ so that $f_\ell^-\le f$ on
  $U^-_\ell$ and $f\le f_\ell^+$ on $U^+_\ell$, with
  $f(x_0)=f_\ell^\pm(x_0)$.  Also assume
  that there is a symmetric bilinear form $Q$ so that
$$
\lim_{\ell\to \infty}D^2f^+_\ell(x_0)=\lim_{\ell\to
\infty}D^2f^-_\ell(x_0)=Q\ .
$$ Then $f$ has a second order  expansion at
$x_0$:
$$
f(x)=f(x_0)+df(x_0)(x-x_0) +\frac12Q(x-x_0,x-x_0) +o(|x-x_0|^2)\;,
$$
where $df(x_0)=df^{\pm}_\ell(x_0)$ (this is independent of $\ell$ and the
choice of $+$ or $-$).  Thus the Alexandrov second derivative of $f$
exists at $x=x_0$ and is given by $D^2f(x_0)=Q$.
\end{lemma}

\begin{remark}
The extra generality will not be needed here, but we remark that the
proof shows that the hypothesis the functions $f_\ell^+$ are $C^2$ can
be weakened to only requiring that they all have a second order 
expansion at $x_0$ with no regularity being needed away from $x=x_0$.
\end{remark}

\begin{proof}
Without loss of generality we may assume that $x_0=0$. By replacing
$U_\ell^+$ and $U_\ell^-$ by $U_\ell:=U_\ell^+\cap U_\ell^-$ we assume
for each $\ell$ that $f^+_\ell$ and $f_\ell^-$ have the same domain.
For any $k,\ell$ we have $f^+_\ell-f^-_k\ge 0$ has a local minimum at
$x_0=0$ and thus a critical point there.  Whence
$df^+_\ell(0)=df^-_k(0)$ for all $k$ and $l$. This shows the linear
functional $df(0)=df^+_\ell(0)$ is well defined.  We now replace $f$
by $x\mapsto f(x)-df(0)x-\frac12Q(x,x)$ and $f^\pm_\ell$ by $x\mapsto
f^\pm_\ell(x)-df(0)x-\frac12Q(x,x)$.  Then $df(0)=df_\ell^\pm(0)=0$,
$\lim_{\ell\to 0}D^2f_\ell^\pm(0)=0$ and to finish the proof it is
enough to show that $f(x)=o(|x|^2)$.  That is we need to find for
every $\e>0$ a $\delta_\e>0$ so that $|x|<\delta_\e$ implies
$|f(x)|\le \e|x|^2$.  To do this choose any $r_0>0$ so that
$B(0,r_0)\subset U$.  Then choose an $\ell$ large enough that
$$
-\e |x|^2 \le D^2f^-_\ell(0)(x,x)\;,\qquad D^2f^+_\ell(0)(x,x)\le \e|x|^2\;,
$$
for all $x\in B(0,r_0)$.  Now choose $r_1\le r_0$ so that
$B(0,r_1)\subset U_\ell$.
For this $\ell$ we can use the Taylor
expansions of $f^+_\ell$ and $f^-_\ell$ at $0$ to find a $\delta_\e>0$
with $0<\delta_\e\le r_1$ so that
$$
\Big|f^\pm_\ell(x)- \frac12Df_\ell^\pm(0)(x,x)\Big|\le
\frac12\e|x|^2
$$
for all $x$ with $|x|<\delta_1$.  Then $f\le f_\ell^+$ implies that if
$|x|<\delta_\e$ then
$$
f(x)\le \frac12D^2f^+_\ell(0)(x,x)+\Big(f^+_\ell(x)
        -\frac12D^2f^+_\ell(0)(x,x)\Big)
        \le \frac{\e}{2}|x|^2+\frac{\e}{2}|x|^2=\e|x|^2\ ,
$$
with a similar calculation, using $f_\ell^-\le f$, yielding $-\e|x|^2\le
f(x)$.  This completes the proof.\qed
\end{proof}

  \noindent {\sc Proof of Theorem~\ref{Tfr}: }  This proof uses, 
essentially, the    same geometric facts about horizons that are used 
in the proof of 
  Proposition~\ref{Ricatti}.  Let $p_0\in\Gint$ be an Alexandrov point
  of a section $\calS\cap\cH$, where $\calS$ is a spacelike or
  timelike $C^2$ hypersurface that intersects $\cH$ \transversally,
  and with $p_0\in \Gint$.  By restricting to a suitable neighborhood
  of $p_0$ we can assume without loss of generality that $M$ is
  globally hyperbolic, that $\Gamma$ maximizes the distance between
  any two of its points, and that $\cH$ is the boundary of $J^+(\cH)$
  and $J^-(\cH)$ (that is $\cH$ divides $M$ into two open sets, its
  future $I^+(\cH)$ and its past $I^-(\cH)$).

  We will simplify notation a bit and assume that $\calS$ is
  spacelike, only trivial changes are required in the case $\calS$ is
  timelike.  We can find $C^2$ coordinates $x^1,\dots, x^{n}$ on
  $\calS$ centered at $p_0$ so that in these coordinates $\calS\cap
  \cH$ is given by a graph $x^n=h(x^1,\dots, x^{n-1})$.  By
  restricting the size of $\calS$ we can assume that these coordinates
  are defined on all of $\calS$ and that their image is
  $B^{n-1}(r)\times (-\delta,\delta)$ for some $r,\delta>0$ and that
  $h\: B^{n-1}(r)\to (-\delta,\delta)$.  As $p_0$ is an Alexandrov
  point of $\calS\cap \cH$ and $h(0)=0$ the function $h$ has a second
  order expansion
$$
h(\vec{x})=dh({0})\vec{x}+\frac12D^2h({0})(\vec{x},\vec{x})
  +o(|\vec x|^2)\ ,
$$
where $\vec{x}=(x^1,\dots,x^{n-1})$.  By possibly changing the sign of
$x^n$ we may assume that
$$
\{(\vec{x},y)\in \calS \ |\ y\ge h(\vec{x})\}\subset J^+(\cH)\;,\qquad
\{(\vec{x},y)\in \calS \ |\ y\le  h(\vec{x})\}\subset J^-(\cH)\ .
$$
For $\e\ge0$  define
$$
h_\e^\pm(\vec{x})=dh(0)\vec{x}+\frac{1}{2}D^2h(0)(\vec{x},\vec{x})
\pm\frac{\e}{2}|\vec x|^2\ .
$$
Then $h^\pm_\e$ is a $C^2$ function on $B^{n-1}(r)$, and for $\e=0$
the function $h_0=h_0^+=h_0^-$ is just the second order 
expansion of $h$ at $\vec x=0$.  For each $\e>0$ the second order
Taylor expansion for $h$ at $0$ implies that there is an open
neighborhood $V_\e$ of $0$ in $B^{n-1}(r)$ so that
$$
h_\e^-\le h\le h^+_\e \quad \mbox{on}\quad V_\e\ .
$$
Set
$$
N:=\{(\vec x,h_0(\vec x))\ |\ \vec x \in B^{n-1}(r)\}\ ,\qquad
N_\e^\pm:=\{(\vec x,h_\e^\pm(\vec x))\ |\ \vec x \in V_\e\}\ .
$$
Let $\nor$ be the future pointing timelike unit normal to $\calS$ and
let $\eta\:(a,b)\to \Gint$ be the affine parameterization of $\Gint$
with $\eta(0)=p_0$ and $\la \eta'(0),\nor(p_0)\ra=-1$ (which implies
that $\eta$ is future directed).  Let $\nnor$ be the unique $C^1$
future directed null vector field along $N$ so that
$\nnor(p_0)=\eta'(0)$ and $\la \nnor,\nor\ra=-1$.  Likewise let
$\nnor^\pm_\e$ be the $C^1$ future directed null vector field along
$N^\pm_\e$ with $\nnor(p_0)=\eta'(0)$ and $\la
\nnor_\e^\pm,\nor\ra=-1$.

Let $p$ be any point of $\Gint\cap J^-(p_0)\setminus\{p_0\}$.  Then
$p=\eta(t_0)$ for some $t_0\in (a,b)$.  To simplify notation assume
that $t_0\le 0$.  By Lemma~\ref{Ltheta2} $N$ has no focal points in
$\Gint$ and in particular no focal points on $\eta\big|_{(a,0]}$.
Therefore if we fix a $t_1$ with $t_1<t_0<b$ then there will be an
open neighborhood $W$ of $0$ in $B^{n-1}(r)$ so that
$$
\widetilde{\cH}
:=\{\exp(t\nnor(\vec x,h_0(\vec{x})))\ |\ \vec{x}\in W,\ t\in (t_1,0)\}
$$ is an embedded null hypersurface of $M$. By
Proposition~\ref{C2-null} the hypersurface is of smoothness class
$C^2$.  The focal points depend continuously on the second fundamental
form so there is an $\e_0>0$ so that if $\e<\e_0$ then none of the
submanifolds $N^{\pm}_\e$ have focal points along
$\eta\big|_{[t_1,0]}$.  Therefore if $0<\e<\e_0$ there is an open
neighborhood $W_\e$ and such that
$$
\widetilde{\cH}^\pm_\e
:=\{\exp(t\nnor(\vec x,h(\vec{x})))\ |\ \vec{x}\in W_\e,\ t\in(t_1,0)
\}
$$
is a $C^2$ embedded hypersurface of $M$.

We now choose smooth local coordinates $y^1,\dots, y^{n+1}$ for $M$ centered at
$p=\eta(t_0)$ so that $\f/\f y^{n+1}$ is a future pointing timelike
vector field and the level sets $y^{n+1}=\mbox{Const.}$ are
spacelike.  Then there will be an open neighborhood $U$ of $0$ in
$\bbR^n$ so that near $p$ the horizon $\cH$ is given by a graph
$y^{n+1}=f(y^1,\dots, y^n)$.  Near $p$ the future and past of $\cH$
are given by $J^+(\cH)=\{ y^{n+1}\ge f(y^1,\dots,y^{n})\}$ and
$J^-(\cH)=\{y^{n+1} \le f(y^1,\dots,y^n)\}$.  There will also be open
neighborhoods $U_0$ and $U^\pm_\e$ of $0$ in $\bbR^{n}$ and functions
$f_0$ and $f^{\pm}_\e$ defined on these sets so that near
$p$
\begin{eqnarray*}
\widetilde{\cH}&=&\{ (y^1,\dots,y^n, f_0(y^1,\dots,y^n))\ |\
(y^1,\dots,y^n)\in U_0\}\ ,\\
\widetilde{\cH}^\pm_\e&=& \{ (y^1,\dots,y^n, f_\e^\pm(y^1,\dots,y^n))\ |\
(y^1,\dots,y^n)\in U_\e^\pm\}\ .
\end{eqnarray*}
The hypersurfaces $\widetilde{\cH}$ and $\widetilde{\cH}^\pm_\e$ are
$C^2$ which implies the functions $f_0$ and $f^{\pm}_\e$ are all
$C^2$.

Since $N_{\e}^-\subset
J^-(\cH)$, a simple achronality argument shows that $\widetilde{\cH}_\e^-
\subset J^{-}(\cH)$.
(This uses the properties of $\cH$ described in the first paragraph of the
proof.)
By choosing $N^+_{\e}$ small enough,  we can assume that
$\widetilde\cH^+_{\e}$ is achronal.
We now show that, relative to some neighborhood of $p$, $\cH \subset
J^-(\widetilde\cH^+_{\e})$.
Let $\cal O_\e$ be a neighborhood of $p$, disjoint from
$\mathscr{S}$, such that $\widetilde\cH^+_{\e}\cap \cal O_\e$
separates $\cal O_\e$ into the disjoint open sets
$I^+(\widetilde\cH^+_{\e} \cap \cal O_\e;\cal O_\e)$ and
$I^-(\widetilde\cH^+_{\e}\cap \cal O_\e;\cal O_\e)$.
Now, by taking $\cal O_\e$ small enough, we claim that $\cH$ does not
meet $I^+(\widetilde\cH^+_{\e}\cap \cal O_\e;\cal O_\e)$.  If that is
not the case, there is a sequence $\{p_{\ell}\}$ such that $p_{\ell}
\in \cH \cap I^+(\widetilde\cH^+_{\e}\cap \cal O_\e;\cal O_\e)$ and
$p_{\ell} \to p$.  For each ${\ell}$, let $\Gamma_{\ell}$ be a future
inextendible null generator of $\cH$ starting at $p_{\ell}$.  Since
$p$ is an interior point of $\Gamma$, the portions of the
$\Gamma_{\ell}$'s to the future of $p_\ell$ must approach the portion
of $\Gamma$ to the future of $p$.  Hence for ${\ell}$ sufficiently
large, $\Gamma_{\ell}$ will meet $\mathscr{S}$ at a point $q_{\ell}\in
\mathscr{S}\cap \cH$, say, such that $q_{\ell}\to p_0$.  For such
${\ell}$, the segment of $\Gamma_{\ell}$ from $p_{\ell}$ to $q_{\ell}$
approaches the segment of $\Gamma$ from $p$ to $p_0$ uniformly as
${\ell}\to \infty$. Since $\widetilde\cH^+_{\e}$ separates a small
neighborhood of the segment of $\Gamma$ from $p$ to $p_0$, it follows
that for ${\ell}$ large enough, the segment of $\Gamma_{\ell}$ from
$p_{\ell}$ to $q_{\ell}$ will meet $\widetilde\cH^+_{\e}$.  But this
implies for such ${\ell}$ that $p_{\ell} \in I^-(N^+_{\e})$, which
contradicts the achronality of $\widetilde\cH^+_{\e}$.  We conclude,
by choosing $\cal O_\e$ small enough, that $\cH \cap
I^+(\widetilde\cH^+_{\e}\cap \cal O_\e;\cal O_\e) = \emptyset$, and
hence that $\cH \cap {\cal O}_\e \subset J^-(\widetilde\cH^+_{\e}\cap
\cal O_\e;\cal O_\e)$.  Shrinking $U^+_{\e}$ if necessary, this
inclusion, and the inclusion $\widetilde{\cH}_\e^- \subset J^{-}(\cH)$
imply that on $U^-_{\e}$ we have $f^-_{\e} \le f$ and that on
$U^+_{\e}$ we have $f\le f^+_{\e}$.  Therefore if we can show that
$\lim_{\e\searrow 0}D^2f_\e^\pm(0)=D^2f_0(0)$ then Lemma~\ref{pinch}
implies that $0$ is an Alexandrov point of $f$ with Alexandrov second
derivative given by $D^2f(0)= D^2f_0(0)$.

To see that $\lim_{\e\searrow 0}D^2f_\e^\pm(0)=D^2f_0(0)$ let $b_0$
and $b_\e^\pm$ be the Weingarten maps of $\widetilde{\cH}$ and
$\widetilde{\cH}^\pm_\e$ along $\eta$.  By Proposition~\ref{Ricatti}
they all satisfy the same Ricatti equation~\eq{eq:2b}.  The initial
condition for $b_0$ is calculated algebraically from the second
fundamental form of $N$ at $p_0$ (\emph{cf.} Section~\ref{Ssections})
and likewise the initial condition for $b_\e^\pm$ is calculated
algebraically from the second fundamental form of $N^\pm_\e$ at $p_0$.
From the definitions we clearly have
$$ 
\mbox{Second Fundamental Form of $N^\pm_\e$ at $p_0$}\quad
\longrightarrow\quad \mbox{Second Fundamental Form  of $N$ at $p_0$}
$$ as $\e \searrow 0$.  Therefore continuous dependence of solutions
to ODE's on initial conditions implies that $\lim_{\e\searrow0}
b^\pm_\e=b_0$ at all points of $\eta\big|_{[0,t_0]}$.  As
$D^2f^\pm_\e(0)$ and $D^2f_0(0)$ are algebraic functions of $b_\e^\pm$
and $b_0$ at the point $p=\eta(t_0)$ this implies that
$\lim_{\e\searrow 0}D^2f_\e^\pm(0)=D^2f_0(0)$ and completes the proof
that $p$ is an Alexandrov point of $\cH$.

Finally it follows from the argument that at all points of $\Gint$ the
null Weingarten map $b$ for $\cH$ is the same as the Weingarten for
the $C^2$ null hypersurface $\widetilde{\cH}$. So $b$ will
satisfy~\eq{eq:2b} by Proposition~\ref{Ricatti}.  \qed

\medskip

We end this section with one more regularity result, Theorem
\ref{TCfr} below. Its proof requires some techniques which are
introduced in Section \ref{Smonotonicity} only --- more precisely, an
appropriate generalization of Lemma \ref{phi-lip} is needed; this, in
turn, relies on an appropriate generalization of Lemma
\ref{A-delta-C11new}. For this reason we defer the proof of Theorem
\ref{TCfr} to an appendix, Appendix \ref{ACfr}.  
\begin{Theorem}
  \label{TCfr} Let $\calS$ be any $C^2$ hypersurface intersecting $\cH$
  properly transversally, and define
  \begin{eqnarray} \label{trdef}
    S_0&=&\{q\in\calS\cap\cH: \ \mbox{$q$ is an interior point of a
      generator of $\cH$}\}\ ,  \\ S_1&=&\{q\in S_0: \ 
    \mbox{all interior points of the generator} \nonumber\\
      \label{trdef2}   & & \phantom{xxxxxxxxx} \mbox{ through $q$ are
       Alexandrov points of $\cH$}\}\ , \\ \label{trdef3} S_2&=&\{q\in
    S_0: \ \mbox{ $q$ is an Alexandrov point of $\cH$}\}\ 
    ,\end{eqnarray} Then $S_1$ and $S_2$ have full $n-1$ dimensional
  Hausdorff measure in $S_0$.
\end{Theorem}
\begin{remark}
  $S_0$ does not have to have full measure in $\calS\cap\cH$, it can
  even be empty. This last case occurs indeed when $\calS=\{t=0\}$ in
  the example described around Equation \eq{defs} in Section
  \ref{Ssections}. Note, however, that if $\calS$ is a level set of a
  properly transverse foliation $\calS(t)$, then (as already
  mentioned) for almost all $t$'s the sets $S(t)_0$ (as defined in
  \eq{trdef} with $\calS$ replaced by $\calS(t)$) will have full
  measure in $\calS(t)$.  We shall call a generator an {\em Alexandrov
    generator} if all its interior points are Alexandrov points. It
  follows that for generic sections (in the measure sense above) for
  almost all points through those sections the corresponding
  generators will be Alexandrov.  The discussion here thus gives a
  precise meaning to the statement that {\em almost all generators of
    a horizon are Alexandrov generators}.
\end{remark}

\section{Area monotonicity}

\label{Smonotonicity}
In this section we shall show the monotonicity of the area, assuming
that the Alexandrov divergence $\tA$ of the generators of $\cH$, or
that of a section of $\cH$, is non--negative. The result is local in
the sense that it only depends on the part of the event horizon $\hor$
that is between the two sections $S_1$ and $S_2$ whose area we are
trying to compare. We consider a spacetime $(M,g)$ of dimension~$n+1$.
We have the following:

\begin{thm}\label{thm:local-area} Suppose that $(M,g)$ is an $(n+1)$
  dimensional spacetime with a future horizon $\hor$.  Let $\Sigma_a$,
  $a=1,2$ be two embedded \emb achronal spacelike hypersurfaces of $C^2$
  differentiability class, set $S_a= \Sigma_a\cap\cH$.  Assume that
\begin{equation}\label{condi}
S_1\subset J^-(S_2)\ ,\qquad 
S_1\cap 
S_2=\emptyset \ ,
\end{equation} and that either
\begin{enumerate}
\item  the divergence $\tA$ of $\cH$ defined ($\Hnk$--almost
  everywhere) by Equation \eq{theta} is non--negative on $J^+(S_1)\cap
  J^-(S_2)$, or
\item the divergence $\tA^{S_2} $ 
  of $S_2$ defined ($\Hnmk$--almost everywhere) by Equation
  \eq{thetaS} is non--negative, with the null energy condition
  \eq{nec} holding on $J^+(S_1)\cap J^-(S_2)$.
\end{enumerate}
Then
\begin{enumerate}
\item \begin{equation}\label{area-ineq}
\Arm(S_1)\le \Ar(S_2)\ , 
\end{equation}
with $\Arm(S_1)$ defined in Equations \eq{nmult}--\eq{aream} and
$\Ar(S_2)$ defined in Equation \eq{areah}. This implies the inequality
$$\Ar(S_1)\le \Ar(S_2)\ .$$
\item If equality holds in (\ref{area-ineq}), then $(J^+(S_1)\setminus
  S_1)\cap (J^-(S_2)\setminus S_2)$ (which is the part of $\hor$
  between $S_1$ and $S_2$) is a \emph{smooth totally geodesic}
  (analytic if the metric is analytic) null hypersurface in $M$.
  Further, if $\gamma$ is a null generator of $\hor$ with
  $\gamma(0)\in S_1$ and $\gamma(1)\in S_2$, then the curvature tensor
  of $(M,g)$ satisfies $R(X,\gamma'(t))\gamma'(t)=0$ for all $t\in
  [0,1]$ and $X\in T_{\gamma(t)}\cH$.
\end{enumerate}
\end{thm}
\begin{remark}\label{R5.2}
Note that if $S_1\cap S_2\ne \nothing$ then the inequality
$\Arm(S_1)\le \Ar(S_2)$ need not hold. (The inequality
$\Ar(S_1)\le \Ar(S_2)$ will still hold.) For example, if
$\Ar_{S_1}{(S_1)}>\Ar(S_1)$, then letting $S_2=S_1$ gives a
counterexample.  If $S_1\cap S_2\ne \nothing$ then the correct
inequality is 
$\Arm(S_1\setminus S_2)\le\Ar(S_2\setminus S_1)$.
\end{remark}

\mnote{Krolak moved to appendix}
\begin{proof} 
Let us start with an outline of the proof,
without the technical details --- these will be supplied later in the
form of a series of lemmas.  Let $\mathcal{N}_\hor(S_1)$ be the
collection of generators of $\hor$ that meet $S_1$.  Let $A\subseteq
S_2$ be the set of points that are of the form $S_2\cap \gamma$ with
$\gamma\in \mathcal{N}_\hor(S_1)$; replacing $\Sigma_2$ by an
appropriate submanifold thereof if necessary, $A$ will be a closed
subset of $S_2$ (Proposition~\ref{gregnew}). The condition
\eq{condi} together with achronality of $\cH$ imply that every $q\in
A$ is on exactly one of the generators $\gamma\in
\mathcal{N}_\hor(S_1)$, we thus have a well defined function $\phi \:
A\to S_1$ given by
\begin{equation}\label{phi-del}
\phi(q) = S_1\cap \gamma\quad \text{where $\gamma\in \mathcal{N}_\hor(S_1)$
and $q=S_2\cap \gamma$}.
\end{equation} 
Let us for simplicity assume that the affine distance from $S_1$ to
$S_2$ on the generators passing through $S_1$ is bounded from below,
and that the affine existence time of those generators to the future
of $S_2$ is also bounded from below (in what follows we will see how
to reduce the general case to this one).  In this case $A$ can be embedded in a
$C^{1,1}$ hypersurface $N$ in $\Sigma_2$ (Lemma~\ref{A-delta-C11new})
and $\phi$ can be extended to a locally Lipschitz function
$\widehat\phi$, from $N$ to $\Sigma_1$ (Lemma~\ref{phi-lip}). $A$ is
$\hmtwo^{n-1}$ measurable (closed subsets of manifolds are Hausdorff
measurable$^{\mbox{\scriptsize \ref{fBorel}}})$, 
\mnote{repetition concerning $\phi$ complained about by the referee
  removed} so we can apply the generalization of the
change--of--variables theorem known as the area formula
\cite[Theorem~3.1]{Federer:measures} (with $m$ and $k$ there equal to
$n-1$) to the extension $\widehat\phi$ of $\phi$ to get
\begin{equation}
\int_{S_1}\#
\phi^{-1}[p]\,d\hmone^{n-1}(p)=\int_{A}J(\phi)(q)\,d\hmtwo^{n-1}(q)\ ,
\label{covf}
\end{equation}
where $J(\phi)$ is the restriction of the Jacobian\footnote{It should
  be pointed out that the Jacobian $J(\widehat \phi)$ is \emph{not}
  the usual Jacobian which occurs in the change--of--variables theorem
  for Lebesgue measure on $\R^n$, but contains the appropriate
  $\sqrt{\det g_{ij}}$ factors occurring in the definition of the
  measure associated with a Riemannian metric, see \cite[Def.~2.10,
  p.~423]{Federer:measures}. A clear exposition of the Jacobians that
  occur in the area and co--area formulas can be found in
  \cite[Section 3.2]{EvansGariepy} in the flat $\R^n$ case. See also
  ~\cite[Appendix p.~66]{Howard:kinematic} for the smooth Riemannian
  case.} of $\widehat\phi$ to $A$, and $h_a$, $a=1,2$, denotes the
metric induced on $\Sigma_a$ by $g$.
We observe that
$$
\# \phi^{-1}[p]=N(p,S_2)
$$ for all $q\in S_1$.  Indeed, if $\gamma$ is a null generator of
$\hor$ and $p\in \gamma\cap S_1 $ then the point $q \in S_2$ with
$q\in \gamma$ satisfies $\phi(q)=p$.  Thus there is a bijective
correspondence between $\phi^{-1}[p]$ and the null generators of
$\hor$ passing through $p$.  (This does use that $S_1\cap
S_2=\nothing$.)  This implies that
$$
\int_{S_1}\#
\phi^{-1}[p]\,d\hmone^{n-1}(p)=\int_{S_1}
N(p,S_2)\,d\hmone^{n-1}(p) \ ,
$$
which can be combined with the area formula \eq{covf} to give that
\begin{equation}\label{pre-area}
\Arm(S_1)=\int_{A}J(\phi)(q)\,d\hmtwo^{n-1}(q).
\end{equation}
We note that \cite[Theorem~3.2.3, p.~243]{FedererMeasureTheory} also
guarantees\footnote{The result in \cite[Theorem~3.2.3,
  p.~243]{FedererMeasureTheory} is formulated for subsets of $\R^q$,
  but the result generalizes immediately to manifolds by considering
  local charts, together with a partition of unity argument.} that
$A\ni p \to N(p,S_2)$ is measurable, so that $\Arm(S_1)$ is well
defined.  Now a calculation, that is straightforward when $\hor$ is
smooth (Proposition~\ref{Gregs}, Appendix~\ref{apR}), shows that
having the null mean curvatures $\tA$ -- or $\tA^{S_2}$ nonnegative
together with the null energy condition -- implies that $J(\phi)\le 1$
almost everywhere with respect to~$\hmtwo^{n-1}$ on $A$ --- this is
established in Proposition~\ref{Pjacob}. Using this
in~(\ref{pre-area}) completes the proof.

Our first technical step is the following: 
\begin{prop} \label{gregnew}
  Let the setting be as in Theorem~\ref{thm:local-area}.  Then there
  exists an open submanifold $\Sigma_2'$ of $\Sigma_2$ such that $A$
  is a closed subset of $S_2' = \Sigma_2' \cap \cH$.  Replacing
  $\Sigma_2$ with $\Sigma_2'$ we can thus assume that $A$ is closed in
  $S_2$.
\end{prop}

\begin{proof} 
  Let $W_2$ be the subset of $S_2$ consisting of all points
  $p\in S_2$ for which there exists a semi--tangent $X$ at $p$ of $\cH$
  such that the null geodesic $\eta$ starting at $p$ in the direction
  $X$ does not meet $\Sigma_1$ when extended to the past (whether or
  not it remains in $\cH$).  Let $\{p_k\}$ be a sequence of points in
  $W_2$ such that $p_k \to p\in S_2$.  We show that $p\in W_2$, and
  hence that $W_2$ is a closed subset of $S_2$.  For each $k$, let
  $X_k$ be a semi--tangent at $p_k$ such that the null geodesic
  $\eta_k$ starting at $p_k$ in the direction $X_k$ does not meet
  $\Sigma_1$ when extended to the past.  Since the collection of
  $\kaux$-normalized null vectors is locally compact, by passing to a
  subsequence if necessary, we may assume that $X_k\to X$, where $X$
  is a future pointing null vector at $p$.  Let $\eta$ be the null
  geodesic starting at $p$ in the direction $X$.  Since $(p_k,X_k)\to
  (p,X)$, $\eta_k \to \eta$ in the strong sense of geodesics. Since,
  further, the $\eta_k$'s remain in $\cH$ to the future and $\cH$ is
  closed, it follows that $\eta$ is a null generator of $\cH$ and $X$
  is a semi-tangent of $\cH$ at $p$. Since $\eta_k \to \eta$, if
  $\eta$ meets $\Sigma_1$ when extended to the past then so will
  $\eta_k$ for $k$ large enough.  Hence, $\eta$ does not meet
  $\Sigma_1$, so $p\in W_2$, and $W_2$ is a closed subset of $S_2$.
  Then $S_2' : = S_2 \setminus W_2$ is an open subset of $S_2$. Thus,
  there exists an open set $\Sigma_2'$ in $\Sigma_2$ such that $S_2' =
  S_2 \cap \Sigma_2' = \Sigma_2' \cap \cH$. Note that $p\in S_2'$ iff
  for each semi-tangent $X$ of $\cH$ at $p$, the null geodesic
  starting at $p$ in the direction $X$ meets $\Sigma_1$ when extended
  to the past. In particular, $A \subset S_2'$.  To finish, we show
  that $A$ is a closed subset of $S_2'$.  Let $\{p_k\}$ be a sequence
  in $A$ such that $p_k \to p \in S_2'$.  Let $\eta_k$ be the unique
  null generator of $\cH$ through $p_k$. Let $X_k$ be the
  $\kaux$-normalized tangent to $\eta_k$ at $p_k$. Let $q_k$ be the
  point where $\eta_k$ meets $S_1$.  Again, by passing to a
  subsequence if necessary, we may assume $X_k \to X$, where $X$ is a
  semi--tangent of $\cH$ at $p$.  Let $\eta$ be the null geodesic at
  $p$ in the direction $X$.  We know that when extended to the past,
  $\eta$ meets $\Sigma_1$ at a point $q$, say.  Since $\eta_k \to
  \eta$ we must in fact have $q_k \to q$ and hence $q\in S_1$.  It
  follows that $\eta$ starting from $q$ is a null generator of $\cH$.
  Hence $p\in A$, and $A$ is closed in $S_2'$.~\qed
\end{proof}

We note that in the proof above we have also shown the following:
\begin{Lemma}
  \label{Lclosed} The collection $\cN= \cup_{p\in\cH}\cN_p$ of
  semi--tangents is a closed subset of $TM$.\qed
\end{Lemma}
 Recall that we
have fixed a complete Riemannian metric ${\kaux}$ on $M$.  For each
$\delta>0$ let
\begin{eqnarray}A_\delta: & =& \left\{ p\in A\ | \ \sdist(p,S_1)\ge \delta\,
    ,\mbox{\rm \ and the generator $\gamma$ through} \right.
  \nonumber\\ & & \left.\mbox{\rm \phantom{x} \ $p$ can be extended
      at least a ${\kaux}$--distance $\delta$ to the future}\right\}.
  \label{adelta}
  \end{eqnarray}
  We note that if the $\kaux$ distance from $S_1$ to $S_2$ on the
  generators passing through $S_1$ is bounded from below, and if the
  length of the portions of those generators which lie to the future
  of $S_2$ is also bounded from below (which is trivially fulfilled
  when the generators are assumed to be future complete), then
  $A_\delta$ will coincide with $A$ for $\delta $ small enough.
 
 \begin{Lemma}
           \label{Lred} Without loss of generality we may assume
    \begin{equation}
      \label{eqAAd0}
  \overline S_1\cap 
S_2 =  \nothing 
    \end{equation}
  and 
\begin{equation}
      \label{eqAAd}
      A=\cup_{\delta>0} A_\delta\ .
    \end{equation}
  \end{Lemma}
  \begin{proof} We shall show how to reduce the general situation in which
    $S_1\cap S_2 = \nothing $ to one in which Equation~\eq{eqAAd0}
    holds.  Assume, first, that $\Sigma_1$ is connected, and introduce
    a complete Riemannian metric on $\Sigma_1$.  With respect to this
    metric $\Sigma_1$ is a complete metric space such that the closed
    distance balls $\overline B(p,r)$ (closure in $\Sigma_1$) are
    compact in $\Sigma_1$, and hence also in $M$.  Choose a $p$ in
    $\Sigma_1$ and let $$\Sigma_{1,i} = B(p,i)\ 
  $$ ($B(p,i)$ -- open balls). Then $\Sigma_1=\cup_i \Sigma_{1,i}
  =\cup_i \overline\Sigma_{1,i})$ (closure either in $\Sigma_1$ or in
  $M$) is an increasing union of compact sets. The $ \Sigma_{1,i}$'s
  are spacelike achronal hypersurfaces which have the desired property
  $$ \overline S_{1,i}\cap S_2 \subset \Sigma_{1} \cap \Sigma_2=
  \nothing \ $$ (closure in M), $S_{1,i}\equiv\Sigma_{1,i}\cap \cH$.
 Suppose that we have shown that
point 1 of Theorem~\ref{thm:local-area} is true for $S_{1,i}$, thus
$$ \Arm(S_{1,i})\le \Ar(S_2)\ . $$ As $S_{1,i}\subset S_{1,i+1}$,
$S_1=\cup S_{1,i}$, the monotone convergence theorem gives
$$\lim_{i\to\infty}\Arm(S_{1,i}) = \Arm(S_{1})\ , $$ whence the
result.  

If $\Sigma_1$ is not connected, one can carry the above procedure out
on each component (at most countably many), obtaining a sequence of
sets $\Sigma_{1,i}$ for each component of $\Sigma_1$. The resulting
collection of sets is countable, and an obvious modification of the
above argument establishes that~\eq{eqAAd0} holds. 

But if Equation~\eq{eqAAd0} holds, that is if $\overline S_1\cap S_2 =
\nothing$,  we have $\sdist(p,S_1)=\sdist(p,\overline S_1) > 0$ for
all $p\in S_2$.  Therefore~\eq{eqAAd} holds.  This completes the
proof.~\qed
\end{proof} 

%

In order to continue we need the following variation of the Whitney
extension theorem.  As the proof involves very different ideas than
the rest of this section we postpone the proof to
Appendix~\ref{app:C11extend}.

\begin{prop}\label{C11-extend}
Let $A\subset \R^n$ and $f\: A\to \R$.  Assume there is a constant
$C>0$ and for each $p\in A$ there is a vector $a_p\in \R^n$ so that the
inequalities
\begin{equation}\label{f-support}
f(p)+\la x-p,a_p\ra -\frac{C}{2}\|x-p\|^2\le f(x) \le f(p)+\la x-p,a_p\ra +\frac{C}{2}\|x-p\|^2
\end{equation}
hold for all $x\in A$.  Also assume that for all $p,q\in A$ and all
$x\in \R^n$ the inequality
\begin{equation}\label{dis-spt}
f(p)+\la x-p,a_p\ra -\frac{C}{2}\|x-p\|^2\le 
        f(q)+\la x-q,a_q\ra +\frac{C}{2}\|x-q\|^2
\end{equation}
holds.  Then there is a function $F\: \R^n\to \R$ of
class $C^{1,1}$ so that $f$ is the restriction of $F$ to $A$.
\end{prop}

\begin{remark}\label{para-remark}
  Following~\cite[Prop.~1.1~p.~7]{CaffarelliCabre} the
  hypothesis~(\ref{f-support}) of Proposition~\ref{C11-extend} will be
  stated by saying $f$ has \emph{global upper and lower support
    paraboloids of opening $C$} in $K$.  The condition~(\ref{dis-spt})
  can be expressed by saying \emph{the upper and lower support
    paraboloids of $f$ are disjoint}.
\end{remark} 

\begin{remark}\label{rmk:not-closed}
  Unlike most extension theorems of Whitney type this result does not
  require that the set being extend from be closed (here $A$ need not
  even be measurable) and there is no continuity assumption on $f$ or
  on the map that sends $p$ to the vector $a_p$ (cf.~\cite[Chap.~VI
  Sec.~2]{Stein:singular-integrals}, \cite[Vol.~1, Thm~2.3.6,
  p.~48]{BigHormander} where to get a $C^{1,1}$ extension the
  mapping $p\mapsto a_p$ is required to be Lipschitz.)  The usual
  continuity assumptions are replaced by the ``disjointness''
  condition~(\ref{dis-spt}) which is more natural in the geometric
  problems considered here.
\end{remark}

Proposition~\ref{C11-extend} is a key element for the proof of the
result that follows \cite{Howard}:
\begin{lemma}\label{A-delta-C11new}
  For any $\delta>0$ there is a $\Cloc^{1,1}$ hypersurface $N_\delta$
  of $\Sigma_2$ (thus $N_\delta$ has co--dimension two in~$M$) with
  $$A_\delta\subseteq N_\delta\ .$$
\end{lemma}
\begin{remark} We note that $N_\delta$ does not have to be connected.
\end{remark} 

\begin{proof} The strategy of the proof is as follows: We start by showing
that all points $q$ in $A_\delta$ possess space--time neighborhoods in
which all the generators of $\cH$ passing through $A_\delta$ are
contained in a $C^{1,1}$ hypersurface in $M$. This hypersurface is not
necessarily null, but  is transverse to $\Sigma_2$ on $A_\delta$.
$N_\delta$ will be obtained by a globalization argument that uses the
intersections of those locally defined hypersurfaces with
$\Sigma_2$.

By the local character of the arguments that follow, one can assume
without loss of generality that $M$ is globally hyperbolic; otherwise,
where relevant, one could restrict to a globally hyperbolic
neighborhood and define the pasts, futures, \emph{etc,\/} with respect
to this neighborhood.

For $q\in A_\delta$ define $q^\pm\in J^\pm(q)\cap \cH$ as those points
on the generator $\gamma_q$ of $\cH$ through $q$ which lie a
$\kaux$--distance $\delta$ away from $q$.  Achronality of $\cH$
implies that $\gamma_q$ has no conjugate points, hence the
hypersurfaces $\partial J^\mp(q^\pm)$ are smooth in a neighborhood of
$q$, tangent to $\cH$ there.

Let $q_0$ in $A_\delta$ and choose a basis $B=\{E_1,\ldots,E_n,E_{n+1}\}$
of $T_{q_0^+}M$ such that $g(E_i,E_j)=\delta_{ij}$ and
$g(E_i,E_{n+1})=0$ for $i,j=1,\ldots,n$, while $g(E_{n+1},E_{n+1})=-1$.
We shall denote by $(y_1,\ldots,y_{n+1})$ the normal coordinates
associated with this basis.

We note that $A_{\delta^\prime}\subset A_\delta$ for
$\delta<\delta^\prime$ so that it is sufficient to establish our
claims for $\delta$ small. Choosing $\delta$ small enough the cone
$$
\{y_{n+1}=-\sqrt{y_1^2+\cdots+y_n^2}\ |\ y_i\in\R, i=1,\ldots,n\}
$$ 
coincides with $\partial J^-(q_0^+)$ in a neighborhood of $q_0$.
Reducing $\delta$ again if necessary we can suppose that $q_0^-$
belongs to a normal coordinate neighborhood of $q_0^+$.  Let
$y^0_{n+1}$ be the $(n+1)^{st}$ coordinate of $q_0$ and let
$y^{0-}_{n+1}$ be the $(n+1)^{st}$ coordinate of $q_0^-$.  Denote by
$C^-$ the smooth hypersurface with compact closure obtained by
intersecting the previous cone by the space--time slab lying between
the hypersurfaces $y_{n+1}=\frac{1}{2}y^0_{n+1}$ and
$y_{n+1}=\frac{1}{2}(y^0_{n+1}+y^{0-}_{n+1})$.  Let us use the symbol
$\omega$ to denote any smooth local parameterization
$\omega\longrightarrow \bn(\omega)\in T_{q_0^+}M$ of this
hypersurface.
We then obtain a smooth parameterization by $\omega$ of $\partial
J^-(q_0^+)$ in a neighborhood of $q_0$ by using the map
$\omega\longrightarrow \exp_{q_0^+}(\bn(\omega))$. Reducing $\delta$
again if necessary, without loss of generality we may assume that
there are no conjugate points on $\exp_{q_0^+}(\bn(\omega))\subset
C^-$.

Using parallel transport of $B$ along radial geodesics from $q_0^+$,
we can then use $\omega$ to obtain a smooth parameterization of
$\partial J^-(m)$ in a neighborhood of $q_0$, for all $m$ in a
neighborhood of $q_0^+$. For all $m$ in this neighborhood let us denote
by $\bn_m(\omega)$ the corresponding vector in $T_mM$. The map
$(m,\omega)\longrightarrow \exp_m(\bn_m(\omega))$ is smooth for $m$
close to $q_0^+$ and for $\omega\in C^-$. This implies the continuity
of the map which to a couple $(m,\omega)$ assigns the second
fundamental form $I\!I(m,\omega)$, defined with respect to an auxiliary
Riemannian metric $\kaux$, of $\partial J^-(m)$ at
$\exp_m(\bn_m(\omega)) $.  If we choose the $m$'s in a sufficiently
small compact coordinate neighborhood of $q_0^+$, compactness of $C^-$
implies that the $\kaux$ norm of $I\!I$ can be bounded in this
neighborhood by a constant, independently of $(m,\omega)$.

Let, now $(x_1,\ldots,x_{n+1})$ be a coordinate system covering a
globally hyperbolic neighborhood of $q_0$, centered at $q_0$, and of
the form ${\cO}=B^n(2r) \times (-a,a)$.  We can further require that
$g(\partial_{n+1},\partial_{n+1})<C<0$ over ${\cO}$.  Transversality
shows that, reducing $r$ and $a$ if necessary, the hypersurfaces
$\exp_m(\bn_m(\omega))$ are smooth graphs above $B^n(2r)$, for $m$
close to $q_0^+$; we shall denote by $f^-_m$ the corresponding
graphing functions. Now, the first derivatives of the $f^-_m$'s can be
bounded by an $m$--independent constant: the vectors $V_i=\partial_i+
D_if^-_m\partial_{n+1}$ are tangent to the graph, hence non--timelike,
and the result follows immediately from the equation
$g(V_i,V_i)\geq0$.  Next, the explicit formula for the second
fundamental form of a graph (\emph{cf., e.g.}, \cite[Eqns.~(3.3) and
(3.4), p.~604]{AGHCPAM})
gives a uniform bound for the second derivatives of the $f^-_m$'s over
$B^n(2r) $. We make a similar argument, using future light cones
issued from points m near to $q_0^-$, and their graphing functions
over $B^n(2r)$ denoted by $f_m^+$.  Let $f$ be the graphing function
of $\cH$ over $B^n(2r) $; we can choose a constant $C$ such that, for
all $p$ in
\begin{equation}
  \label{Gdef}
  G_\delta:=\{x\in\Gamma\ |\  \Gamma \mbox{ is a  generator of $\cH$ passing
  through } A_\delta\}\cap {\cO}\ ,
\end{equation}
 the graph of the function
$$
f_p(x)=f(x_p)+Df(x_p)(x-x_p)-C||x-x_p||^2
$$ lies under the graph of $f^-_{p^+}$, hence in the past of $p^+$.
Here we write $x_p$ for the space coordinate of $p$, thus $p=(x_p,
f(x_p))$, \emph{etc.} Similarly, for all $q$ in $G_\delta$ the graph
of
$$
f_q(x)=f(x_q)+Df(x_q)(x-x_q)+C||x-x_q||^2
$$ lies above the graph of $f^+_{q^-}$, thus is to the future of
$q^-$.  Achronality of the horizon implies then that the inequality
\eq{dis-spt} holds for $x\in B^n(2r)$. Increasing $C$ if necessary,
Equation~\eq{dis-spt} will hold for all $x_p,x_q\in B^n(r)$ and $x\in
\R^n$: Indeed, let us make $C$ large enough so that
$$ (C-1)r^2\geq
\Lip_{B^n(r)}^2(f)-\inf_{B^n(r)}(f)
+\sup_{B^n(r)}(f)\ ,
$$ where $\Lip_{B^n(r)}(f)$ is the Lipschitz continuity constant of
$f$ on $B^n(r)$.  For $x\not\in B^n(2r)$ we then have
\begin{eqnarray*}
  f(x_q)-f(x_p) + \frac{C}{2} (||x-x_q||^2+||x-x_p||^2)
 \phantom{xxxxxxxx}& & \\ 
 +\la x-x_q,a_q\ra -\la x-x_p,a_p\ra & \ge & \\
 f(x_q)-f(x_p)+ \frac{C}{2}(||x-x_q||^2+||x-x_p||^2)
 \phantom{xxxxxxxx}& & \\ -\frac{1}{2}(||x-x_q||^2+||x-x_p||^2 
+||a_q||^2+||a_p||^2)
& \ge & \\ \inf_{B^n(r)}(f)-\sup_{B^n(r)}(f) + (C-1)r^2-
\Lip^2_{B^n(r)}(f)
& \geq & 0\ .
\end{eqnarray*}
Here $a_p=df(x_p)$, $a_q=df(x_q)$, and we have used $\Lip_{B^n(r)}(f)$
to control $||a_p||$ and $||a_q||$. 

Let the set $A$ of Proposition~\ref{C11-extend} be the projection
on $B^n(r)$ of $G_\delta$; by that proposition there exists a
$C^{1,1}$ function 
from $B^n(r)$ to $(-a,a)$,
 the graph of which contains $G_\delta$. (It
may be necessary to reduce $r$ and $\cO$ to obtain this).  Note that
this graph contains $A_\delta\cap \cO$ and is transverse to $\Sigma_2$
there.  From the fact that a transverse intersection of a $C^2$
hypersurface with a $C^{1,1}$ hypersurface is a $C^{1,1}$ manifold we
obtain that $A_\delta\cap\cO$ is included in a $C^{1,1}$ submanifold
of $\Sigma_2$, which has space--time co--dimension two.

Now $A_\delta$ is a closed subset of the manifold $\Sigma_2'$ defined
in Proposition~\ref{gregnew}, and can thus be covered by a countable
union of compact sets. Further, by definition of $A_\delta$ any point
thereof is an interior point of a generator.  Those facts and the
arguments of the proof of Proposition~\ref{Pinter} show that
$A_\delta$ can be covered by a countable locally finite collection of
relatively compact coordinate neighborhoods $\cU_i$ of the form
$\cU_i=B^{n-1}(\epsilon_i) \times (-\eta_i,\eta_i)$ such that
$\cU_i\cap\cH$ is a graph of a semi--convex function $g_i$:
\begin{equation}
  \label{graph}
\cU_i\cap\cH= \{x^n=g_i(x^A)\, ,(x^A)\equiv (x^1,\ldots,x^{n-1})\in
B^{n-1}(\epsilon_i)  \}\ .
\end{equation}
The $\epsilon_i$'s will further be restricted by the requirement that
there exists a $C^{1,1}$ function
\begin{equation}
  \label{hidef}
  h_i\:B^{n-1}(\epsilon_i)\to\R
\end{equation}
such that the graph of $h_i$ contains $\cU_i\cap\A_\delta$. The
$h_i$'s are the graphing functions of the $C^{1,1}$ manifolds just
constructed in a neighborhood of each point of $A_\delta$.

At those points at which $S_2$ is differentiable let $\mygnor$ denote
a $h_2$--unit vector field normal to $S_2$ pointing towards
$J^-({\cH})\cap\Sigma_2$. Choosing the orientation of $x^n$
appropriately 
we can assume that at those points at which $\mygnor $ is defined we have
\begin{equation}
  \label{etacond}
  \mygnor (x^n-g_i)>0 \ .
\end{equation}
In order to globalize this construction we use an idea of
\cite{Howard}. Let $\phi_i$ be a partition of unity subordinate to the
cover $\{\cU_i\}_{i\in\N}$, define
\begin{eqnarray}
  \label{chidef}
  \chi_i(q)&=& \cases{ (x^{n}-h_i(x^A))\phi_i(q)\ ,  & $q=(x^A,x^n)\in \cU_i$\
  , \cr 0\ , & otherwise, } \\
\chi_\delta &=& \sum_i \chi_i \ .
\end{eqnarray}
Define a hypersurface $N_\delta\subset\cup_i\cU_i$ via the equations
\begin{equation}
  \label{ndef}
  N_\delta\cap \cU_i= \{\chi_\delta=0\ , d\chi_\delta\ne 0\}\ . 
\end{equation}
(We note that $N_\delta$ does not have to be connected, but this is
irrelevant for our purposes.) If $q\in \Ad$ we have $\chi_i(q)=0$ for
all $i$'s hence $\Ad\subset \{\chi_\delta=0\}$. Further, for $q\in
\Ad\cap\cU_i$ we have $\mygnor (\chi_i) = \phi_i \mygnor (x^{n} - g_i) \ge
0$ from Equation~\eq{etacond}, and since for each $q\in \Ad$ there
exists an $i$ at which this is strictly positive we obtain
$$ \mygnor (\chi_\delta )=  \sum_i \mygnor (\chi_i) > 0\ .$$
It follows that $\Ad\subset N_\delta$, and the result is established.
\qed
\end{proof}

\begin{lemma}\label{phi-lip}
  Under the condition \eq{eqAAd}, the map $\phi\: A\to S_1$ is locally
  Lipschitz.
\end{lemma}

\begin{proof}
  Let $q_0\in A$, we need to find a neighborhood $\cU_A\equiv \cU\cap
  A$ in $A$ so that $\phi$ is Lipschitz in this neighborhood.  By
the condition \eq{eqAAd} there exists $\delta>0$ such that $q_0\in
A_\delta$. By lower semi--continuity of the existence time of
geodesics we can choose a neighborhood $\cU$ of $q_0$ in $\Sigma_2$
small enough so that
  \begin{equation}
    \label{nodelta}
\cU\cap A_\delta=\cU\cap A\ .
  \end{equation}
 Denote by $N$ the
  $\Cloc^{1,1}$ hypersurface $N_\delta$ (corresponding to the chosen
  small value of $\delta$) given by Lemma~\ref{A-delta-C11new},
  so that $$A\cap\cU\subseteq N\cap\cU\ .$$ Let $\nor$ be the future
  pointing unit normal to $\Sigma_2$. (So $\nor$ is a unit timelike
  vector.)  Let $\gnor$ be the unit normal to $N$ in $\Sigma_2$ that
  points to $J^-(\cH)\cap\Sigma_2$. Let $\nnor:=-(\nor+\gnor)$.  Then
  $\nnor$ is a past pointing null vector field along $N$ and if $q\in
  A_\delta$ then $\nnor(q)$ is tangent to the unique generator of
  $\hor$ through $q$.  As $\Sigma_2$ is $C^2$ then the vector field
  $\nor$ is $C^1$, while the $\Cloc^{1,1}$ character of $N$ implies
  that $\gnor$ is locally Lipschitz.  It follows that $\nnor$ is a
  locally Lipschitz vector field along $N$.
  Now for each $q\in A\cap\cU$ there is a unique positive real number
  $r(q)$ so that $\phi(q)=\exp(r(q)\nnor(q))\in S_1$.  Lower
  semi--continuity of existence time of geodesics implies that,
  passing to a subset of $\cU$ if necessary, for each $q\in N\cap\cU$
  there is a unique positive real number $\hat r(q)$ so that
  $\widehat\phi(q)=\exp(\hat r(q)\nnor(q))\in \Sigma_1$.  As
  $\Sigma_1$ is a Lipschitz hypersurface, Clarke's implicit function
  theorem \cite[Corollary, p.~256]{Clarke:optimization} implies that
  $q\mapsto \hat r(q)$ is a locally Lipschitz function near $q_0$. Thus
  $\widehat{\phi}$ is Lipschitz near $q_0$ and the restriction of
  $\widehat{\phi}$ to $A_\delta$ is $\phi$.~\qed
\end{proof}

\begin{cor}\label{P:S1-rect}
The section $S_1$ is countably $(n-1)$~rectifiable.
\end{cor} 

\begin{remark}\label{R:S1-rect}
  By starting with an arbitrary section $S=S_1$ of the form $S=\cH\cap
  \Sigma$, where $\Sigma$ is a $C^2$ spacelike hypersurface or
  timelike hypersurface that meets $\cH$ properly transversely, and
  then constructing a section $S_2=\cH\cap \Sigma_2$ for a $C^2$
  spacelike hypersurface $\Sigma_2$ so that the hypotheses of
  Theorem~\ref{thm:local-area} hold, one obtains that every such
  section $S$ is countably $(n-1)$~rectifiable.  In the case that
  $\Sigma_1\equiv \Sigma$ is spacelike a more precise version of this
  is given in~\cite{Howard} (where it is shown that $S$ is countably
  $(n-1)$~rectifiable of class $C^2$).
\end{remark}

\begin{proof}
From Lemma~\ref{A-delta-C11new} it is clear for each $\delta>0$ that
$A_\delta$ is countably $(n-1)$~rectifiable.  By  Lemma~\ref{phi-lip}
the map $\phi\big|_{A_\delta}\: A_\delta\to S_1$ is locally Lipschitz
and therefore $\phi[A_\delta]$ is locally countable
$(n-1)$~rectifiable. But $S_1=\bigcup_{k=1}^\infty \phi[A_{1/k}]$ so
$S_1$ is a countable union of countably $(n-1)$~rectifiable sets and
therefore is itself countably $(n-1)$~rectifiable.~\qed 
\end{proof}

It follows from the outline given at the beginning of the proof of
Theorem~\ref{thm:local-area} that we have obtained:

\begin{Proposition}\label{P:pre-area}
The formula~(\ref{pre-area}) holds.
\end{Proposition}

\begin{cor}\label{C:Nfinite}
  For any section $S=\cH\cap \Sigma$ where $\Sigma$ is a $C^2$
  spacelike or timelike hypersurface that intersects $\cH$ properly
  transversely the set of points of $S$ that are on infinitely many
  generators has vanishing $(n-1)$~dimensional Hausdorff measure.
\end{cor}

\begin{proof}
  For any point $p$ of $S$ choose a globally hyperbolic neighborhood
  $\cO$ of the point and then choose a neighborhood $S_1$ of $p$ in
  $S$ small enough that the closure $\overline{S}_1$ is compact,
  satisfies $\overline{S}_1\subset S\cap \cO$, and so that there is a
  $C^\infty$ Cauchy hypersurface $\Sigma_2$ of $\cO$ such that
  $\overline{S}_1\subset I^-(\Sigma_2;\cO)$.  Let $S_2=\cH\cap
  \Sigma_2$ and let $A$ be the set of points of $S_2$ that are of the
  form $S_2\cap \gamma$ where $\gamma$ is a generator of $\cH$ that
  meets $\overline{S}_1$.  Compactness of $\overline{S}_1$ together
  with the argument of the proof of Proposition~\ref{gregnew} show
  that the set of generators of $\cH$ that meet $\overline{S}_1$ is a
  compact subset of the bundle of null geodesic rays of $M$, and that
  $A$ is a compact subset of $S_2$.  Then Lemma~\ref{A-delta-C11new}
  implies that $A$ is a compact set in a $C^{1,1}$ hypersurface of
  $\Sigma_2$ and so $\Hau^{n-1}(A)<\infty$.  Lemma~\ref{phi-lip} and
  the compactness of $A$ yields that $\phi\: A\to \overline{S}_1$
  given by~\eq{phi-del} (with $\overline{S}_1$ replacing $S_1$) is
  Lipschitz.  Therefore the Jacobian $J(\phi)$ is bounded on $A$.  By
  Proposition~\ref{P:pre-area}
  $\int_{\overline{S}_1}N(p,S_2)\,d\Hau^{n-1}(p)
  =\int_{A}J(\phi)(q)\,d\hmtwo^{n-1}(q)<\infty$ and so
  $N(p,S_2)<\infty$ except on a set of $\Hau^{n-1}$ measure zero.  But
  $N(p,S_2)$ is the number of generators of $\cH$ through $p$ so this
  implies the set of points of $S_1$ that are on infinitely many
  generators has vanishing $(n-1)$~dimensional Hausdorff measure.  Now
  $S$ can be covered by a countable collection of such neighborhoods
  $S_1$ and so the set of points on $S$ that are on infinitely many
  generators has vanishing $(n-1)$~dimensional Hausdorff measure.
  This completes the proof.\qed
\end{proof}


To establish~\eq{area-ineq} it remains to show that $J(\phi)\le 1$
$\hmtwo^{n-1}$-almost everywhere.  To do this one would like to use
the classical formula that relates the Jacobian of $\phi$ to the
divergence $\theta$ of the horizon, \emph{cf.\/}\ Proposition
\ref{Gregs}, Appendix~\ref{apR} below.  However, for horizons which
are not $C^2$ we only have the Alexandrov divergence $\theta_\Al$ at
our disposal, and it is not clear whether or not this formula holds
with $\theta$ replaced by $\theta_\Al$ for general horizons. The proof
below consists in showing that this formula remains true after such a
replacement for generators passing through almost all points of $A$.

\begin{Proposition}
$J(\phi)\le 1$ $\hmtwo^{n-1}$--almost everywhere on $A$.
\label{Pjacob}
\end{Proposition}

\begin{proof}
 The argument of the paragraph preceding Equation~\eq{nodelta}
shows that it is sufficient to show $J(\phi)\le 1$
$\hmtwo^{n-1}$--almost everywhere on $A_\delta$ for each $\delta>0$.
Let $N_\delta$ be the $C^{1,1} $ manifold constructed in Lemma
\ref{A-delta-C11new}, and let $\cU\subset \Sigma_2$ be a coordinate
neighborhood of the form $\cV\times (a,b)$, with $\cV\subset \R^{n-1}$
and $a,b,\in \R$, in which $\cU\cap N_\delta$ is the graph of a
$C^{1,1}$ function $g\:\cV\to\R$, and in which $\cH\cap \cU$ is the
graph of a semi--convex function $f\:\cV\to\R$.  By
\cite[Theorem~3.1.5, p.~227]{FedererMeasureTheory} for every $\epsilon
>0$ there exists a twice differentiable function
$g_{1/\epsilon}\:\cV\to\R$ such that
\begin{equation}
  \label{ggeps}
  \Lnm(\{g\ne g_{1/\epsilon}\})< \epsilon 
\ .
\end{equation}
Here $\Lnm$ denotes the $(n-1)$~dimensional Riemannian measure on $\cV$
associated with the pull--back ${h_{\cV}}$ of the space--time metric
$g$ to $\cV$.  Let $\pr A_\delta$ denote the projection on $\cV$ of
$A_\delta\cap\cU$, 
thus $A_\delta\cap\cU$ is the graph of $g$ over $\pr A_\delta$.  For
$q\in \cV$ let $\theta^{n-1}_*(\Lnm,\pr A_\delta,q)$ be the density
function of $\pr A_\delta$ in $\cV$ with respect to the measure
$\Lnm$, defined as in \cite[page 10]{SimonGMT} using geodesic
coordinates centered at $q$ with respect to the metric ${h_{\cV}}$.
Define
\begin{equation}
  \label{dens1}
  B = \{q\in\pr A_\delta\ |\ \theta^{n-1} _*(\Lnm,\pr A_\delta,q)=1 \}
\subset \cV\ . 
\end{equation}
By \cite[Corollary 2.9]{Morgan} or \cite[2.9.12]{FedererMeasureTheory}
the function $\theta^{n-1} _{*}(\Lnm,\pr A_\delta,\cdot)$ differs from
the characteristic function $\chi_{\mbox{\scriptsize \pr} A_\delta}$ of
$\pr A_\delta$ by a function supported on a set of vanishing measure,
which implies that $B$ has full measure in $\pr A_\delta$.  Let
$$B_{1/\epsilon}=B\cap\{g= g_{1/\epsilon}\}\ ;$$ Equation~\eq{ggeps} shows that
\begin{equation}
  \label{bbeps}
  \Lnm(B\setminus B_{1/\epsilon}) < \epsilon\ . 
\end{equation}
We define $\tilde B_{1/\epsilon}\subset B_{1/\epsilon} \subset\pr A_\delta
\subset \cV$ as follows:
\begin{equation}
  \label{dens2}
\tilde B_{1/\epsilon} =\{q\in B_{1/\epsilon}\ |\ 
\theta^{n-1} _*(\Lnm,B_{1/\epsilon},q)=1 \}\ . 
\end{equation}
 Similarly the set
$\tilde B_{1/\epsilon}$ has full measure in $B_{1/\epsilon}$, hence
\begin{equation}
  \label{btbeps}
  \Lnm(B\setminus \tilde B_{1/\epsilon}) < \epsilon\ . 
\end{equation}
Let $(\pr A_\delta)_\Al$ denote the projection on $\cV$ of the set of
those Alexandrov points of $\cH\cap\Sigma_2$ which are in
$A_\delta\cap\cU$; $(\pr A_\delta)_\Al$ has full measure in $\pr
A_\delta$. Let, further, $\cV_{\mbox{\scriptsize Rad}}$ be the set of
points at which $g$ is twice differentiable; by Rademacher's theorem
(\emph{cf., e.g.,}\ \cite[p.~81]{EvansGariepy})
$\cV_{\mbox{\scriptsize Rad}}$ has full measure in $\cV$. We set
\begin{eqnarray}
  \label{hbeps}
  \hat B_{1/\epsilon} & = &\tilde B_{1/\epsilon} \cap (\pr
  A_\delta)_\Al\cap\cV_{\mbox{\scriptsize Rad}} \ , \\ \hat B & = &
  \bigcup _{i\in\N}\hat B_{\myi}
\end{eqnarray}
It follows from \eq{btbeps} that $ \hat B$ has full measure in $\pr
A_\delta$. Since $g$ is Lipschitz we obtain that the graph of $g$ over
$\hat B$ has full $\hmtwo^{n-1}$--measure in $A_\delta\cap \cU$.

Consider any $x_0\in   \hat B$, then $x_0$ is an Alexandrov point of
$f$ so that we have the expansion
\begin{eqnarray}
  \label{stineq1}
 &  f(x) = f(x_0) + df(x_0)(x-x_0)+ \frac{1}{2}D^2f(x_0)(x-x_0,x-x_0)+
  o(|x-x_0|^2)  \ , &
\end{eqnarray}
Next, $g$ is twice differentiable at $x_0$ so that we also have
\begin{eqnarray}
  \label{stineq1.1}
  & g(x) = g(x_0) + dg(x_0)(x-x_0)+ \frac{1}{2}D^2g(x_0)(x-x_0,x-x_0)+
  o(|x-x_0|^2) \ , & \\ & dg(x) - dg(x_0) = D^2g(x_0)(x-x_0,\cdot)+
  o(|x-x_0|) \ . & \label{stineq2.1}
\end{eqnarray}
Further there exists $j\in\N$ such that $x_0\in \hat B_{\myj}\subset
B_{\myj}$; by definition of $B_{\myj}$ we then have
\begin{equation}
  \label{gieq}
  g_{\myj}(x_0) = g(x_0)=f(x_0)\ .
\end{equation}
We claim that the set of directions $\vn\in B^{n-1}(1)\subset T_{x_0}\cV$
for which there exists a sequence of points $q_i={x_0}+r_i \vn$ with
$r_i\to 0$ and with the property that $f(q_i)=g_{\myj}(q_i)$ is dense
in $B^{n-1}(1)$. Indeed, suppose that this is not the case, then there
exists $\epsilon >0$ and an open set $\Omega\subset B^{n-1}(1)\subset
T_{x_0}\cV$ of directions $\vn \in B^{n-1}(1)$ such that the solid cone
$K_{\epsilon}=\{{x_0}+r\vn\ | \ \vn\in \Omega\ , r\in [0,\epsilon]\}$
contains no points from $B_{\myj}$. It follows that the density
function $\theta^{n-1} _*(\Lnm,B_{\myj},x_0)$ is strictly smaller than
one, which contradicts the fact that $x_0$ is a density point of $\pr
A_\delta$ ($x_0\in \hat B_{\myj}\subset B$, \emph{cf.\/}\ 
Equation~\eq{dens1}), and that $x_0$ is a density point of $B_{\myj}$
($x_0\in \hat B_{\myj}\subset \tilde B_{\myj}$, \emph{cf.\/}\
Equation~\eq{dens2}).

Equation~\eq{gieq} leads to the following Taylor expansions at $x_0$:
\begin{eqnarray}
  \label{stq1}
 & g_{\myj}(x) = f(x_0) + dg_{\myj}(x_0)(x-x_0)+ \frac{1}{2}D^2g_{\myj}(x_0)(x-x_0,x-x_0)+
  o(|x-x_0|^2)  \ , &
 \\ & dg_{\myj}(x) - d g_{\myj}(x_0) = D^2 g_{\myj}(x_0)(x-x_0,\cdot)+
   o(|x-x_0|)  \ . & \label{stq2}
\end{eqnarray}
Subtracting Equation~\eq{stineq1} from Equation~\eq{stq1} at points
$q_i={x_0}+r_i \vn$ at which $f(q_i)=g_{\myj}(q_i)$ we obtain
\begin{equation}
  \label{stq3}
  dg_{\myj}(x_0)( \vn)-df(x_0)( \vn)= O(r_i)\ .
\end{equation}
Density of the set of $\vn$'s for which \eq{stq3} holds implies that
\begin{equation}
  \label{stq4}
  dg_{\myj}(x_0)=df(x_0)\ .
\end{equation} Again comparing Equation~\eq{stineq1} with Equation~\eq{stq1}
at points at which $f(q_i)=g_{\myj}(q_i)$ it now follows that
\begin{equation}
  \label{stq4.1}
  D^2g_{\myj}(x_0)( \vn, \vn)-D^2f(x_0)( \vn, \vn)=
  2\frac{g_{\myj}({x_0}+r_i \vn)-f({x_0}+r_i \vn)}{r_i^2}+o(1)= o(1)\ ,
\end{equation}
and density of the set of $\vn$'s together with the polarization
identity gives
\begin{equation}
  \label{stq5}
  D^2g_{\myj}(x_0)=D^2f(x_0)\ .
\end{equation}
Similarly one obtains
\begin{eqnarray}
  \label{stq6}
   dg(x_0)& = &df(x_0)\ ,
\\   \label{stq7}
  D^2g(x_0)&= & D^2f(x_0)\ .
\end{eqnarray}
Define $S_j$ to be the graph (over $\cV$) of $g_{\myj}$; 
Equations~\eq{stq4} and \eq{stq6} show that both $S_j$ and $N_\delta$ are
tangent to $\cH\cap\Sigma_2$ at $p_0\equiv(x_0,f(x_0))$:
\begin{equation}
  \label{graphtan}
  T_{p_0}(\cH\cap\Sigma_2)=  T_{p_0} N_\delta=T_{p_0} S_j 
\ .
\end{equation}
Let $ \bn_j$ denote the ($C^1$) field of $\kaux$--unit future directed
null normals to $ S_j$ such that $$ \bn_j({p_0})=\bn({p_0})\ ,$$ where
$\bn({p_0})$ is the semi--tangent to $\cH$ at $p_0$. Let
$\phi_j\:S_j\to\Sigma_1$ be the map obtained by intersecting the null
geodesics passing through points $q\in S_j$ with tangent parallel to
$\bn_j(q)$ there. Equation~\eq{graphtan} shows that
\begin{equation}
  \label{phiid}
  \phi_j(p_0)=\phi(p_0)=\widehat\phi(p_0)\ .
\end{equation}
The lower semi--continuity of the existence time of geodesics shows
that, passing to a subset of $\cU$ if necessary, $\phi_j$ is well
defined on $S_j$. By an argument similar to the one leading to
\eq{stq4}, it follows from Equations~\eq{stineq2.1}, \eq{stq2},
\eq{stq4} and \eq{stq5}--\eq{stq7} that the derivatives of $\phi_j$
and of $\widehat \phi$ coincide at $p_0$, in particular
\begin{equation}
  \label{jaceq}
  J(\phi)(p_0)\equiv  J(\widehat\phi)(p_0) =  J(\phi_j)(p_0)\ .
\end{equation}
Equation~\eq{stq5} further shows that $S_j$ is second order tangent to
$\cH$ at $p_0$ in the sense defined before Lemma~\ref{Ltheta2}, we
thus infer from that lemma that there are no focal points of $S_j$
along the segment of the generator $\Gamma$ of $\cH$ passing through
$p_0$ which lies to the future of $\Sigma_1$. Consider the set $\cH_j$
obtained as the union of null geodesics passing through $S_j$ and
tangent to $\bn_j$ there; standard considerations show that there
exists a neighborhood of $\Gamma\cap I^+(\Sigma_1)\cap I^-(\Sigma_2)$
in which $\cH_j$ is a $C^1$ hypersurface. It is shown in Appendix
\ref{apR} that 1) $\cH_j$ is actually a $C^2$
hypersurface,
\emph{cf.\/}\ Proposition~\ref{C2-null}, and 2) the null Weingarten
map $b=b_{\cH_j}$ of $\cH_j$ satisfies the Ricatti equation
\begin{equation}\label{eq:2cagain}
b'+b^2+R=0\ .
\end{equation}
Here a prime denotes a derivative with respect to an affine
parameterization $s\mapsto \eta(s)$ of $\Gamma$ that makes $\Gamma$
future directed.  Theorem~\ref{Tfr} implies that $\cH$ has a null
Weingarten map $b_{\Al}$ defined in terms of the Alexandrov second
derivatives of $\cH$ on all of the segment $\Gamma\cap
I^+(\Sigma_1)\cap I^-(\Sigma_2)$ and that this Weingarten map also
satisfies the Ricatti equation~\eq{eq:2cagain}. As the null Weingarten
map can be expressed in terms of the first and second derivatives of
the graphing function of a section Equations~\eq{stq4} and~\eq{stq5}
imply that $b_{\cH_j}(p_0)=b_{\Al}(p_0)$.  Therefore uniqueness of
solutions to initial value problems implies that $b_{\cH_j}=b_{\Al}$
on all of $\Gamma\cap I^+(\Sigma_1)\cap I^-(\Sigma_2)$.  But the
divergence (or null mean curvature) of a null hypersurface is the
trace of its null Weingarten map and thus on the segment $\Gamma\cap
I^+(\Sigma_1)\cap I^-(\Sigma_2)$ we have
$\theta_{\cH_j}=\mbox{trace}\, b_{\cH_j}=\mbox{trace}\,
b_{\Al}=\theta_{\Al}$.

We now finish the proof under the first of hypothesis of
Theorem~\ref{thm:local-area}, that is that the divergence the
$\theta_{\Al}\ge 0$ on $J^+(S_1)\cap J^-(S_2)$.  Then
$\theta_{\cH_j}=\theta_{\Al}\ge 0$ and Proposition~\ref{Gregs} implies
that $J(\phi)(p_0)=J(\phi_j)(p_0)\le 1$ as required.

The other hypothesis of Theorem~\ref{thm:local-area} is that
$\tA^{S_2} \ge 0$ and the null energy condition holding on
$J^+(S_1)\cap J^-(S_2)$.  Recall that $p_0=\Gamma\cap S_2$ is an
Alexandrov point of $\cH$; thus Theorem~\ref{Tfr} applies and shows
that $\theta=\theta_{\Al}$ exists along  the segment $\Gamma\cap
I^+(\Sigma_1)\cap I^-(\Sigma_2)$, and satisfies the Raychaudhuri equation
$$
\theta' = -{\rm Ric}(\eta',\eta') - \sigma^2 - \frac1{n-2}\theta^2\ .
$$ Here $\sigma^2$ is the norm squared of the shear (and should not be
confused with the auxiliary Riemannian metric which we have also
denoted by $\kaux$).  But the null energy condition implies ${\rm
  Ric}(\eta',\eta')\ge0$ so this equation and $\tA^{S_2} \ge 0$
implies $\theta=\theta_{\Al}\ge0$ on $\Gamma\cap I^+(\Sigma_1)\cap
I^-(\Sigma_2)$.  Then the equality $\theta_{\cH_j}=\theta_{\Al}$
yields $\theta_{\cH_j}\ge 0$ and again we can use
Proposition~\ref{Gregs} to conclude $J(\phi)(p_0)=J(\phi_j)(p_0)\le
1$.  This completes the proof.\qed
\end{proof}

To finish the proof of Theorem \ref{thm:local-area}, we need to
analyze what happens when
\begin{equation}
  \label{Areaeq}
\Arm(S_1)= \Ar(S_2)\ . 
\end{equation}
In this case Equation~\eq{covf} together with Proposition~\ref{Pjacob}
show that $A$ has full measure in $S_2$, that $N(p,S_2)=1$
$\hmone^{n-1}$--almost everywhere on $S_1$, and that $J(\phi)=1$
$\hmtwo^{n-1}$--almost everywhere on $S_2$. Next, the arguments of
the proof of that Proposition show that
\begin{equation}
  \label{vanish}
  \tA=0
\end{equation}
$\Hnk$--almost everywhere on $J^-(A)\cap J^+(S_1)=J^-(S_2)\cap
J^+(S_1) $.  The proof of Proposition \ref{Pjacob} further shows that
${\rm Ric}(\eta',\eta')=0$ $\Hnk$--almost everywhere on $\cH\cap
J^+(S_1)\cap J^-(S_2)$ (\emph{cf.\/}\ Equation~\eq{eq:2cagain}), hence
everywhere there as the metric is assumed to be smooth and the
distribution of semi--tangents is a closed set (Lemma \ref{Lclosed}).
Here $\eta'$ is any semi--tangent to $\cH$. We first note the
following observation, the proof of which borrows arguments from
\cite[Section IV]{BK2}:

\begin{Lemma}
  \label{Lnoend}
  Under the hypotheses of Theorem~\ref{thm:local-area}, suppose
  further that the equality~\eq{Areaeq} holds. Then there are no end
  points of generators of $\cH$ on
  \begin{equation}
    \label{defOmega}
    \Omega\equiv (J^+(S_1)\setminus S_1)\cap (J^-(S_2)\setminus S_2)\ 
    .
  \end{equation}
\end{Lemma}

\begin{proof} Suppose that there exists $q\in \Omega$ which is an end point
  of a generator $\Gamma$ of $\cH$, set $\{p\}=\Gamma\cap S_2$,
  extending $\Gamma$ beyond its end point and parameterizing it
  appropriately we will have
$$\Gamma(0)=q\ , \quad \Gamma(1)=p\ , \quad \Gamma(a)\in I^-(\cH) \ ,
$$ for any $a<0$ for which $\Gamma(a)$ is defined. Now $p$ is an
interior point of a generator, and semi--tangents at points in a
sufficiently small neighborhood of $p$ are arbitrarily close to the
semi--tangent $X_p$ at $p$.  Since $I^-(\cH)$ is open it follows from
continuous dependence of solutions of ODE's upon initial values that
there exists a neighborhood $\cV\subset S_2$ of $p$ such that every
generator of $\cH$ passing through $\cV$ leaves $\cH$ before
intersecting $\Sigma_1$ when followed backwards in time from $S_2$,
hence $A\cap \cV=\nothing$, and $A$ does not have full measure in
$S_2$. \qed
\end{proof}

To finish the proof we shall need the following result, which seems to
be of independent interest:

\begin{Theorem}
  \label{Tsmoothness} Let $\Omega$ be an open subset of a horizon
  $\hor$ which contains no end points of generators of $\hor$, and
  suppose that the divergence $\tA $ of $\cH$ defined ($\Hnk$--almost
  everywhere) by Equation~\eq{theta} vanishes $\Hnk$--almost
  everywhere. Then $\Omega$ is a smooth submanifold of $M$ (analytic
  if the metric is analytic).
\end{Theorem}

\begin{remark} \label{BKrem} 
The condition on $\Omega$ is equivalent to $\Omega$ being a $C^1$
hypersurface, \emph{cf.\/}\ \cite{BK2}.
\end{remark}

\begin{proof}
  Let $p_0\in \Omega$ and choose a smooth local foliation
  $\{\Sigma_\lambda\ |\ -\e < \lambda < \e\}$ of an open neighborhood
  $\mathcal{U}$ of $p_0$ in $M$ by spacelike hypersurfaces so that
  $\overline{\mathcal{U}\cap \Omega} \subset \Omega$ and so that
  $p_0\in \Sigma_0$.  Letting $\kaux$ be a the auxiliary Riemannian
  metric, by possibly making $\mathcal{U}$ smaller we can assume that
  the $\kaux$-distance of $\overline{\mathcal{U}\cap \Omega}$ to
  $\overline{\Omega}\setminus \Omega$ is $< \delta$ for some
  $\delta>0$.  Let $\Omega_{\Al}$ be the set of Alexandrov points of
  $\Omega$ and let $\mathcal{B}=\Omega\setminus\Omega_{\Al}$ be the
  set of points of $\Omega$ where the Alexandrov second derivatives do
  not exist.  We view $\lambda$ as a function $\lambda\: \mathcal{U}\to
  \bbR$ in the natural way.  Now $\lambda$ is smooth on $\mathcal{U}$
  and by Remark~\ref{BKrem} $\Omega$ is a $C^1$ manifold so the
  restriction $\lambda\big|_{\Omega}$ is a $C^1$ function.  Letting
  $h_{\kaux}$ be the pull back of our auxiliary Riemannian metric
  $\kaux$ to $\Omega$ we apply the co-area formula to
  $\lambda\big|_{\Omega}$ and use that by Alexandrov's theorem
  $\Hau_{h_{\kaux}}^{n}(\mathcal{B})=0$ to get
$$
\int_{-\e}^\e \Hau_{h_{\kaux}}^{n-1}(\mathcal{B}\cap
\Sigma_\lambda)\,d\lambda 
=\int_{\mathcal{B}}J(\lambda\big|_{\Omega})\,d\Hau_{h_{\kaux}}^{n}=0\ .
$$
This implies that for almost all $\lambda \in (-\e,\e)$ that
$\Hau_{h_{\kaux}}^{n-1}(\mathcal{B}\cap \Sigma_\lambda)=0$.  Therefore
we can choose a $\lambda$ just a little bigger than $0$ with
$\Hau_{h_{\kaux}}^{n-1}(\mathcal{B}\cap \Sigma_\lambda)=0$ and so that
$p_0\in J^-(\Omega\cap\Sigma_\lambda)$.  To simplify notation we denote
$\Sigma_\lambda$ by $\Sigma$.  Then from the choice of $\Sigma$ we
have that $\Hau_{h_{\kaux}}^{n-1}$ almost every point
of $\Sigma$ is an Alexandrov point of $\Omega$.  

By transversality and that $\Omega$ is $C^1$ the set
$\Sigma\cap\Omega$ is a $C^1$ submanifold of $\Sigma$.  Recalling that
$\delta$ is less than the $\kaux$-distance of
$\overline{\mathcal{U}\cap \Omega}$ to $\overline{\Omega}\setminus
\Omega$ we see that for any $p\in \Sigma$ that the unique (because
$\Omega$ is $C^1$) generator $\Gamma$ of $\Omega$ through $p$ extends
in $\Omega$ a $\kaux$-distance of at least $\delta$ both to the future
and to the past of $\Sigma$. Letting $A=\Sigma\cap \Omega$ and using
the notation of Equation~\eq{adelta}, this implies that $A=A_\delta$.
As $A=\Sigma\cap \Omega$ is already a $C^1$ submanifold of $\Sigma$
Lemma~\ref{A-delta-C11new} implies that $\Sigma\cap \Omega$ is a
$C^{1,1}$ hypersurface in $\Sigma$.  Let $g$ by any Lipschitz local
graphing function of $A$ in $\Sigma$. From
Rademacher's theorem (\emph{cf., e.g.,}\ \cite[p.~81]{EvansGariepy})
it follows that $C^{1,1}=W^{2,\infty}_{\loc}$, further the Alexandrov
second derivatives of $g$ coincide with the classical ones almost
everywhere.
By \cite[p.~235]{EvansGariepy} the second distributional
derivatives of $g$ equal the second classical derivatives of $g$
almost everywhere.  It follows that the equation
\begin{equation}
  \label{semilinear}
  \theta_\Al = 0
\end{equation}
can be rewritten, by freezing the coefficients of the second
derivatives at the solution $g$, as a linear elliptic weak
(distributional) equation with Lipschitz continuous coefficients for
the graphing function $g\in W^{2,\infty}_{\loc}$. Elliptic regularity
shows that $g$ is, locally, of $C^{2,\alpha}$ differentiability class
for any $\alpha \in (0,1)$. Further, Equation~\eq{semilinear} is a
quasi--linear elliptic equation for $g$ (\emph{cf., e.g.,}\ 
\cite{Greg:nullsplit}), a standard bootstrap argument shows that $g$
is smooth (analytic if the metric is analytic) and it easily follows
that $\Omega$ in a neighborhood of $\Sigma\cap \Omega$ containing
$p_0$ is smooth (or analytic). As $p_0$ was
an arbitrary point of $\Omega$ this completes the proof.  \qed
\end{proof}

Returning to the proof of Theorem~\ref{thm:local-area}, we note that
Lemma~\ref{Lnoend} shows that all points of $\Sigma\cap \Omega$, where
$\Omega$ is given by Equation~\eq{defOmega}, are interior points of
generators of $\cH$. Simple arguments
together with the invariance of the domain theorem (\emph{cf., e.g.,}\ 
\cite[Prop. 7.4, p. 79]{Dold}) show that $\Omega$ is an open
submanifold of $\cH$, and Equation \eq{vanish} shows that we can use
Theorem \ref{Tsmoothness} to conclude.
\qed\end{proof}

\section{Conclusions}
\label{Sconclusions}

Let us present here some applications of Theorems~\ref{Trigidity} and
\ref{Tsmoothness}, proved above. The first one is to the theory of
stationary black holes (\emph{cf., e.g.}, \cite{Heusler:book,ChAscona}
and references therein): in that theory the question of
differentiability of event horizons arises at several key places.
Recall that smoothness of event horizons has been established in 1)
static \cite{Vishveshwara,CarterJMP} and 2) \cite{CarterJMP}
stationary--axisymmetric space--times. However, staticity or
stationarity--axisymmetry are often not known \emph{a priori} --- that
is indeed the case in \emph{Hawking's rigidity theorem}
\cite{HE}\footnote{This theorem is actually wrong as stated in
  \cite{HE}; a corrected version, together with a proof, can be found
  in \cite{Ch:rigidity}.}. Now, the rigidity theorem asserts that a
certain class of stationary black holes have axi--symmetric domains of
outer communication; its hypotheses include that of analyticity of the
metric \emph{and of the event horizon}. The examples of black holes
(in analytic vacuum space--times) the horizons of which are nowhere
$C^2$ constructed in \cite{ChGalloway} show that the hypothesis of
analyticity of the event horizon and that of analyticity of the metric
are logically independent. It is thus of interest to note the
following result, which is a straightforward corollary of Theorem
\ref{Trigidity} and of the fact that isometries preserve area:

\begin{Theorem}
  \label{Tbh} Let $\phi$ be an  isometry of a
  black--hole space--time $(M,g)$ satisfying the hypotheses of
  Theorem~\ref{Trigidity}. If $\phi$ maps $\cH$ into $\cH$, then for
  every spacelike hypersurface $\Sigma$ such that
\begin{equation}
  \label{bhhcond}\phi(\Sigma\cap\cH)\subset
  J^{+}(\Sigma\cap\cH)\end{equation} 
the set
\begin{equation}
  \label{bhh}
    \left(J^{-}(\phi(\Sigma\cap\cH))\setminus\phi(\Sigma\cap\cH)
\right)\cap\left(J^{+}(\Sigma\cap\cH)\setminus(\Sigma\cap\cH)\right)
\subset \cH
\end{equation}
is a smooth (analytic if the metric is analytic) null submanifold of
$M$ with vanishing null second fundamental form.
\end{Theorem}

As already pointed out, the application we have in mind is that to
stationary black holes, where $\phi$ actually arises from a one
parameter group of isometries $\phi_t$.  We note that in such a
setting the fact that isometries preserve the event horizon, as well
as the existence of hypersurfaces $\Sigma$ for which \eq{bhhcond}
holds with $\phi$ replaced by $\phi_t$ (for some, or for all $t$'s),
can be established under various standard conditions on the geometry
of stationary black holes, which are of no concern to us here.

Recall, next, that the question of differentiability of \emph{Cauchy}
horizons often arises in considerations concerning \emph{cosmic
  censorship} issues (\emph{cf., e.g.,}\ 
\cite{AndMonrev,RendallLiving}). An interesting result in this
context, indicating non--genericity of occurrence of \emph{compact}
Cauchy horizons, is the Isenberg--Moncrief theorem, which asserts that
\emph{analytic compact Cauchy horizons with periodic generators} in
\emph{analytic, electro--vacuum} space-times are \emph{Killing}
horizons, for a Killing vector field defined on a neighborhood of the
Cauchy horizon \cite{VinceJimcompactCauchy}.  We note that if all the
generators of the horizon are periodic, then the horizon has no
end--points, and 
analyticity follows\footnote{The proof proceeds as follows: Theorem
  \ref{Tfr} shows that the optical equations hold on almost all
  generators of the Cauchy horizon; periodicity of the generators
  together with the Raychaudhuri equation shows then that $\tA=0$
  almost everywhere, hence Theorem \ref{Tsmoothness} applies.} from
Theorem~\ref{Tsmoothness}.  Hence the hypothesis of analyticity of the
event horizon is not needed in \cite{VinceJimcompactCauchy}. We also
note that there exists a (partial) version of the Isenberg--Moncrief
theorem, due to Friedrich, R\'acz and Wald \cite{FRW}, in which the
hypotheses of analyticity of \cite{VinceJimcompactCauchy} are replaced
by those of smoothness both of the metric and of the Cauchy horizon.
Theorem~\ref{Tsmoothness} again shows that the hypothesis of
smoothness of the Cauchy horizon is not necessary in \cite{FRW}.

To close this section let us note an interesting theorem of Beem and
Kr\'olak \cite[Section IV]{BK2}, which asserts that if a compact
Cauchy horizon in a space--time satisfying the null energy condition
contains, roughly speaking, an open dense subset $\cO$ which is a
$C^2$ manifold, then there are no end points of the generators of the
event horizon, and the divergence of the event horizon vanishes.
Theorem~\ref{Tsmoothness} again applies to show that the horizon must
be as smooth as the metric allows. Our methods here could perhaps
provide a proof of a version of the Beem--Kr\'olak theorem in which
the hypothesis of existence of the set $\cO$ will not be needed; this
remains to be seen.

\appendix

\section{The Geometry of $C^2$ Null Hypersurfaces}\label{sec:Greg-null}
\label{apR}
In this appendix we prove a result concerning the regularity of null
hypersurfaces normal to a $C^k$ submanifold in space--time. We also
review some aspects of the geometry of null hypersurfaces, with the
presentation adapted to our needs. We follow the exposition of
\cite{Greg:nullsplit}.

Let $(M,g)$ be a spacetime, \emph{i.e.}, a smooth, paracompact
time-oriented Lorentzian manifold, of dimension $n+1\ge 3$.  We denote
the Lorentzian metric on $M$ by $g$ or $\la\,,\ra$.  A \emph{($C^2$)
  null hypersurface} in $M$ is a $C^2$ co-dimension one embedded
submanifold $\nullhyp $ of $M$ such that the pullback of the metric
$g$ to $\nullhyp $ is degenerate.  Each such hypersurface $\nullhyp $
admits a $C^1$ non-vanishing future directed null vector field $K\in
\Gamma T\nullhyp $ such that the normal space of $K$ at a point $p\in
\nullhyp $ coincides with the tangent space of $\nullhyp $ at $p$,
\emph{i.e.}, $K_p^{\perp} = T_p\nullhyp $ for all $p\in \nullhyp $.
(If $\nullhyp $ is $C^2$ the best regularity we can require for $K$ is
$C^1$.)  In particular, tangent vectors to $\nullhyp $ not parallel to
$K$ are spacelike.  It is well-known that the integral curves of $K$,
when suitably parameterized, are null geodesics.  These integral
curves are called the {\it null geodesic generators\/} of $\nullhyp $.
We note that the vector field $K$ is unique up to a positive scale
factor.

Since $K$ is orthogonal to $\nullhyp $ we can introduce the null
Weingarten map and null second fundamental form of $\nullhyp $ with
respect $K$ in a manner roughly analogous to what is done for
spacelike hypersurfaces or hypersurfaces in a Riemannian manifold, as
follows: We start by introducing an equivalence
relation on tangent vectors: for $X, X'\in T_p\nullhyp $, $X'=X \mbox{
  mod } K$ if and only if $X' - X = \lambda K$ for some $\lambda \in
\mathbb R$.  Let $\ol X$ denote the equivalence class of $X$.  Simple
computations show that if $X'=X \mbox{ mod } K$ and $Y'=Y \mbox{ mod }
K$ then $\la X',Y'\ra = \la X,Y\ra$ and $\la \Dgreg_{X'} K,Y'\ra = \la
\Dgreg_X K,Y\ra$, where $\Dgreg$ is the Levi-Civita connection of $M$.
Hence, for various quantities of interest, components along $K$ are
not of interest.  For this reason one works with the tangent space of
$\nullhyp $ modded out by $K$, \emph{i.e.}, $T_p\nullhyp /K = \{\ol X
\ |\ X\in T_p\nullhyp \}$ and $T\nullhyp /K = \cup_{p\in \nullhyp
  }T_p\nullhyp /K$.  $T\nullhyp /K$ is a rank $n-1$ vector bundle over
$\nullhyp $.  This vector bundle does not depend on the particular
choice of null vector field $K$.  There is a natural positive definite
metric $h$ in $T\nullhyp /K$ induced from $\la,\ra$: For each $p\in
\nullhyp $, define $h\:T_p\nullhyp /K\times T_p\nullhyp /K\to \mathbb
R$ by $h(\ol X,\ol Y) = \la X,Y\ra$.  From remarks above, $h$ is
well-defined.

The {\it null Weingarten map\/} $b=b_K$ of $\nullhyp $ with respect to
$K$ is, for each point $p\in \nullhyp $, a linear map $b\: T_p\nullhyp
/K\to T_p\nullhyp /K$ defined by $b(\ol X) = \ol{\Dgreg_X K}$.  It is
easily verified that $b$ is well-defined and, as it involves taking a
derivative of $K$, which is $C^1$ the tensor $b$ will be $C^0$ but no
more regularity can be expected.  Note if $\widetilde K = fK$, $f\in
C^{1}(\nullhyp )$, is any other future directed null vector field
tangent to $\nullhyp $, then $\Dgreg_X \widetilde K= f\Dgreg_X K
\mbox{ mod } K$.  Thus $b_{fK}=fb_K$.  It follows that the Weingarten
map $b$ of $\nullhyp $ is unique up to positive scale factor and that
$b$ at a given point $p\in \nullhyp $ depends only on the value of $K$
at $p$ when we keep $\nullhyp$ fixed but allow $K$ to vary while
remaining tangent to the generators of $\nullhyp$.

A standard computation shows, $h(b(\ol X), \ol Y) = \la \Dgreg_X K,
Y\ra = \la X, \Dgreg_Y K\ra = h(\ol X,b(\ol Y))$.  Hence $b$ is
self-adjoint with respect to $h$.  The {\it null second fundamental
  form\/} $B=B_K$ of $\nullhyp $ with respect to $K$ is the bilinear form
associated to $b$ via $h$: For each $p\in \nullhyp $, $B\:T_p\nullhyp /K\times
T_p\nullhyp /K\to \mathbb R$ is defined by $B(\ol X,\ol Y) = h(b(\ol X),\ol
Y)= \la\Dgreg_X K,Y\ra$.  Since $b$ is self-adjoint, $B$ is symmetric.
In a manner analogous to the second fundamental form for spacelike
hypersurfaces, a null hypersurface is totally geodesic if and only if
$B$ vanishes identically \cite[Theorem~30]{Kupeli}.

The {\it null mean curvature\/} of $\nullhyp $ with respect to $K$ is
the continuous scalar field $\theta\in C^{0}(\nullhyp )$ defined by
$\theta ={\rm tr}\,b$; in the general relativity literature $\theta$
is often referred to as the \emph{convergence} or \emph{divergence} of
the horizon.  Let $e_1,e_2, \ldots, e_{n-1}$ be $n-1$ orthonormal
spacelike vectors (with respect to $\la,\ra$) tangent to $\nullhyp $
at $p$. Then $\{\ol e_1,\ol e_2, \ldots, \ol e_{n-1}\}$ is an orthonormal
basis (with respect to $h$) of $T_p\nullhyp /K$.  Hence at $p$,
\begin{eqnarray}
\theta
& =  {\rm tr}\,b = \sum_{i=1}^{n-1} h(b(\ol e_i),\ol e_i) \nonumber\\
& =   \sum_{i=1}^{n-1} \la \Dgreg_{e_i}K,e_i\ra.  \label{eq:2a}
\end{eqnarray}

Let $\Sigma$ be the intersection, transverse to $K$, of a hypersurface
in $M$ with $\nullhyp $. Then $\Sigma$ is a $C^2$ $(n-1)$~dimensional
spacelike submanifold of $M$ contained in $\nullhyp $ which meets $K$
orthogonally. From Equation~\eq{eq:2a}, $\theta|_{\Sigma} = {\rm
  div}_{\Sigma}K$, and hence the null mean curvature gives a measure
of the divergence of the null generators of $\nullhyp $. Note that if
$\widetilde K = fK$ then $\widetilde\theta =f\theta$.  Thus the null
mean curvature inequalities $\theta\ge 0$, $\theta\le 0$, are
invariant under positive scaling of $K$.  In Minkowski space, a future
null cone $\nullhyp =\f I^+(p) -\{p\}$ (respectively, past null cone
$\nullhyp =\f I^-(p) -\{p\}$) has positive null mean curvature,
$\theta >0$ (respectively, negative null mean curvature, $\theta <0$).


The null second fundamental form of a null hypersurface obeys a
well-defined comparison theory roughly similar to the comparison
theory satisfied by the second fundamental forms of a family of
parallel spacelike hypersurfaces (\emph{cf.\/}\
Eschenburg~\cite{Eschenburg:hypersurfaces}, which we follow in
spirit).

Let $\eta \: (a,b)\to M$, $s\to \eta(s)$, be a future directed affinely
parameterized null geodesic generator of $\nullhyp $.  For each $s\in (a,b)$,
let
$$
b(s) = b_{\eta'(s)}: T_{\eta(s)}\nullhyp /\eta'(s) \to
T_{\eta(s)}\nullhyp /\eta'(s) 
$$ be the Weingarten map based at $\eta(s)$ with respect to the null
vector $K=\eta'(s)$.  Recall that the null Weingarten map $b$ of a
smooth null hypersurface $\nullhyp $ satisfies a Ricatti equation
(\emph{cf.}~\cite[p.~431]{Beem-Ehrlich:Lorentz2}; for completeness we
indicate the proof below).
\begin{eqnarray}\label{eq:2b}
b'+b^2 +R = 0 .  
\end{eqnarray}   Here $'$ denotes covariant
differentiation in the direction $\eta'(s)$, with $\eta$ -- an
affinely parameterized null geodesic generator of $\nullhyp $; more
precisely, if $X=X(s)$ is a vector field along $\eta$ tangent to
$\nullhyp $, then\footnote{Here $b(\ol X)$ is an equivalence class of
  vectors, so it might be useful to give a practical prescription how
  its derivative $b(\ol X)'$ can be calculated. Let $s \to c(s)$ be a
  null generator of $\cH$.  Let $s \to V(s)$ be a $T\cH/K$--vector
  field along $c$, {\em i.e.}, for each $s$, $V(s)$ is an element of $
  T_{c(s)}\cH/K$.  Say $s\to V(s)$ is smooth if (at least locally)
  there is a smooth --- in the usual sense --- vector field $s \to
  Y(s)$ along $c$ such that $V(s) = {\bar Y}(s)$ for each $s$.  Then
  define the covariant derivative of $s \to V(s)$ along $c$ by: $V'(s)
  = \overline{ Y'(s)}$, where $Y'$ is the usual covariant derivative.
  It is easily shown, using the fact that $\nabla_KK$ is proportional
  to $K$, that $V'$ so defined is independent of the choice of $Y$.
  This definition applies in particular to $b(\bar X)$.}
\begin{eqnarray}\label{eq:2bprim}
b'(\ol X) = b(\ol X)'- b(\ol{X'}).
\end{eqnarray} Finally $R\:T_{\eta(s)}\nullhyp /\eta'(s)\to
T_{\eta(s)}\nullhyp /\eta'(s)$ is the curvature endomorphism defined
by $R(\ol X) = \ol{R(X,\eta'(s))\eta'(s)}$, where $(X,Y,Z)\to R(X,Y)Z$
is the Riemann curvature tensor of $M$ (in our conventions, $R(X,Y)Z=
\Dgreg_X\Dgreg_YZ-\Dgreg_Y\Dgreg_XZ - \Dgreg_{[X,Y]}Z$).  

We indicate the proof of Equation
\eq{eq:2b}.  Fix a point $p = \eta(s_0)$, $s_0\in (a,b)$, on $\eta$.
On a neighborhood $U$ of $p$ in $\hor$ we can scale the null vector
field $K$ so that $K$ is a geodesic vector field, $\nabla_KK=0$, and
so that $K$, restricted to $\eta$, is the velocity vector field to
$\eta$, {\em i.e.}, for each $s$ near $s_0$, $K_{\eta(s)}= \eta'(s)$. Let
$X\in T_pM$.  Shrinking $U$ if necessary, we can extend $X$ to a
smooth vector field on $U$ so that $[X,K] =\nabla_XK - \nabla_KX = 0$.
Then, $R(X,K)K = \nabla_X\nabla_KK-\nabla_K\nabla_XK -\nabla_{[X,K]}K
= -\nabla_K\nabla_KX$.  Hence along $\eta$ we have, $X'' =
-R(X,\eta')\eta'$ (which implies that $X$, restricted to $\eta$, is a
Jacobi field along $\eta$).  Thus, from Equation \eq{eq:2bprim}, at
the point $p$ we have,
\begin{eqnarray}\nonumber
b'(\overline X) & = & \overline{\nabla_XK}\,' - b(\overline{\nabla_KX})   =
\overline{\nabla_KX}\,'-b(\overline{\nabla_XK})\\ \nonumber
   & = &  \overline{X''} - b(b(\overline X))  =  -
\overline{R(X,\eta')\eta'} - b^2(\overline X) \\
& = & - R(\overline X) - b^2(\overline X), \label{calculation}
\end{eqnarray}
which establishes Equation~\eq{eq:2b}.

Equation~\eq{eq:2b} leads to the well known Raychaudhuri equation for
an irrotational null geodesic congruence in general relativity: by
taking the trace of \eq{eq:2b} we obtain the following formula for the
derivative of the null mean curvature $\theta=\theta(s)$ along $\eta$,
\begin{eqnarray}
\theta' = -{\rm Ric}(\eta',\eta') - \sigma^2 - \frac1{n-2}\theta^2,
\label{eq:2capp} 
\end{eqnarray}
where $\sigma$, the shear scalar, is the trace of the square of the
trace free part of $b$. This equation shows how the Ricci curvature of
spacetime influences the null mean curvature of a null hypersurface.
We note the following:

\begin{prop}\label{Ricatti}
Let $\nullhyp $ be a $C^2$ null hypersurface in the $(n+1)$~dimensional
spacetime $(M,g)$ and let $b$ be the one parameter family of
Weingarten maps along an affine parameterized null generator $\eta$.
Then the covariant derivative 
$b'$ defined by Equation~\eq{eq:2bprim} exists and
satisfies Equation~\eq{eq:2b}.
\end{prop}

\begin{remark} When $\nullhyp $ is smooth this is a standard result,
  proved by the calculation \eq{calculation}. However
  when $\nullhyp $ is only $C^2$ all we know is that $b$ is a $C^0$
  tensor field so that there is no reason {\it a priori\/} that the
  derivative $b'$ should exist.  A main point of the proposition is
  that it does exist and satisfies the expected differential equation.
  As the function $s\mapsto R_{\eta(s)}$ is $C^\infty$ then the
  Riccati equation implies that actually the dependence of
  $b_{\eta(s)}$ on $s$ is $C^\infty$.  This will be clear from the
  proof below for other reasons.
\end{remark}

\begin{proof}
  Let $\eta\: (a,b)\to \nullhyp $ be an affinely parameterized null
  generator of $\nullhyp $.  To simplify notation we assume that $0\in
  (a,b)$ and choose a $C^\infty$ spacelike hypersurface $\Sigma$ of
  $M$ that passes through $p=\eta(0)$ and let $N=\nullhyp \cap
  \Sigma$.  Then $N$ is a $C^2$ hypersurface in $\Sigma$.  Now let
  $\tilde{N}$ be a $C^\infty$ hypersurface in $\Sigma$ so that
  $\tilde{N}$ has second order contact with $N$ at $p$.  Let
  $\tilde{K}$ be a smooth null normal vector field along $\tilde{N}$
  such that at $p$, $\tilde{K} = \eta'(0)$. Consider the hypersurface
  $\tilde{\mathscr H}$ obtained by exponentiating normally along
  $\tilde{N}$ in the direction $\tilde{K}$; by Lemma \ref{Ltheta2}
  there are no focal points along $\eta$ as long as $\eta$ stays on
  $\cH$. Passing to a
  subset of $\tilde{N}$ if necessary to avoid cut points,
  $\tilde{\mathscr H}$ will then be a $C^{\infty}$ null hypersurface
  in a neighborhood of $\eta$.  Let $B(s)$ and $\tilde{B}(s)$ be the
  null second fundamental forms of $\mathscr{H}$ and $\tilde{\mathscr
    H}$, respectively, at $\eta(s)$ in the direction $\eta'(s)$. We
  claim that $\tilde{B}(s) = B(s)$ for all $s\in (a,b)$.  Since the
  null Weingarten maps $\tilde{b} =\tilde{b}(s)$ associated to
  $\tilde{B} = \tilde{B}(s)$ satisfy Equation~\eq{eq:2b}, this is
  sufficient to establish the lemma.

  We first show that $\tilde{B}(s) = B(s)$ for all $s\in [0,c]$ for
  some $c\in (0,b)$.  By restricting to a suitable neighborhood of $p$
  we can assume without loss of generality that $M$ is globally
  hyperbolic.  Let $X\in T_p\Sigma$ be the projection of $\eta'(0)\in
  T_pM$ onto $T_p\Sigma$.  By an arbitrarily small second order
  deformation of $\tilde{N}\subset \Sigma$ (depending on a parameter
  $\epsilon$ in a fashion similar to Equation~\eq{fde}) we obtain a
  $C^{\infty}$ hypersurface $\tilde{N}^+_{\epsilon}$ in $\Sigma$ which
  meets $N$ only in the point $p$ and lies to the side of $N$ into
  which $X$ points.  Similarly, we obtain a $C^{\infty}$ hypersurface
  $\tilde{N}^-_{\epsilon}$ in $\Sigma$ which meets $N$ only in the
  point $p$ and lies to the side of $N$ into which $-X$ points.  Let
  $\tilde{K}^{\pm}_{\epsilon}$ be a smooth null normal vector field
  along $\tilde{N}^{\pm}_{\epsilon}$ which agrees with $\eta'(0)$ at
  $p$.  By exponentiating normally along $\tilde{N}^{\pm}_{\epsilon}$
  in the direction $\tilde{K}^{\pm}_{\epsilon}$ we obtain, as before,
  in a neighborhood of $\eta\big|_{[0,c]}$ a $C^{\infty}$ null hypersurface
  $\tilde{\mathscr H}^{\pm}_{\epsilon}$, for some $c\in (0,b)$.  Let
  $\tilde{B}^{\pm}_{\epsilon}(s)$ be the null second fundamental form
  of $\tilde{\mathscr H}^{\pm}_{\epsilon}$ at $\eta(s)$ in the
  direction $\eta'(s)$.

  By restricting the size of $\Sigma$ if necessary we find open sets
  $W$, $W^{\pm}_{\epsilon}$ in $\Sigma$, with $W^-_{\epsilon} \subset
  W \subset W^+_{\epsilon}$, such that $N \subset \partial_{\Sigma}W$
  and $\tilde{N}^{\pm}_{\epsilon} \subset
  \partial_{\Sigma}W^{\pm}_{\epsilon}$.  Restricting to a sufficiently
  small neighborhood of $\eta\big|_{[0,c]}$, we have $\mathscr{H}\cap
  J^+(\Sigma)\subset \partial J^+(W)$ and $\tilde{\mathscr
    H}^{\pm}_{\epsilon}\cap J^+(\Sigma) \subset \partial
  J^+(W^{\pm}_{\epsilon})$. Since $J^+(\overline{W^-_{\epsilon}})
  \subset J^+(\overline{W}) \subset J^+(\overline{W^+_{\epsilon}})$,
  it follows that $\tilde{\mathscr H}^-_{\epsilon}$ is to the future
  of $\mathscr{H}$ near $\eta(s)$ and $\mathscr{H}$ is to the future
  of $\tilde{\mathscr H}^+_{\epsilon}$ near $\eta(s)$, $s\in [0,c]$.
  Now if two null hypersurfaces $\cH_1$ and $\cH_2$ are tangent at a
  point $p$, and $\cH_2$ is to the future of $\cH_1$, then the
  difference of the null second fundamental forms $B_2-B_1$ is
  positive semidefinite at $p$. We thus obtain
  $\tilde{B}^-_{\epsilon}(s)\ge B(s)\ge \tilde{B}^+_{\epsilon}(s)$.
  Letting $\epsilon \to 0$, (i.e., letting the deformations go to
  zero), we obtain $\tilde B(s) = B(s)$ for all $s\in [0,c]$.  A
  straightforward continuation argument implies, in fact, that $\tilde
  B(s) = B(s)$ for all $s\in [0,b)$.  A similar argument establishes
  equality for $s \in (a,0]$.  \qed\end{proof}

In the last result above the hypersurface $\nullhyp $ had to be of at
least $C^2$ differentiability class. Now, in our applications we have
to consider hypersurfaces $\nullhyp $ obtained as a collection of null
geodesics normal to a $C^2$ surface. A naive inspection of the problem
at hand shows that such $\nullhyp $'s could in principle be of $C^1$
differentiability only. Let us show that one does indeed have $C^2$
differentiability of the resulting hypersurface:

\begin{prop}\label{C2-null}
  Consider a $C^{k+1}$ spacelike submanifold $N\subset M$ of
  co--dimension two in an $(n+1)$~dimensional spacetime $(M,g)$, with
  $k\ge 1$.  Let $\nnor$ be a non-vanishing $C^k$ null vector field
  along $N$, and let $\mathcal{U}\subseteq \R\times N\to M$ be the set
  of points where the function
$$
f(t,p):=\exp_p(t\nnor(p))
$$ is defined.  If $f_{(t_0,p_0)*}$ is injective then there is an open
neighborhood $\mathcal{O}$ of $(t_0,p_0)$ so that the image
$f[\mathcal{O}]$ is a $C^{k+1}$ embedded hypersurface in $M$.
\end{prop}

\begin{remark}\label{rk:C1para}
In our application we only need the case $k=1$.  This result
is somewhat surprising as the function $p\mapsto \nnor(p)$ used in the
definition of $f$ is only $C^k$. We emphasize that we are \emph{not}
assuming that $f$ is injective. We note that $f$ will not be of
$C^{k+1}$ differentiability class in general, which can be seen as
follows: Let $t\to r(t)$ be a $C^{k+1} $ curve in the $x$-$y$ plane of
Minkowski 3-space which is \emph{not} of $C^{k+2}$ differentiability
class.  Let $t \to \bn(t)$ be the spacelike unit normal field along the
curve in the $x$-$y$ plane, then $t\to \bn(t)$ is $C^k$ and is \emph{not}
$C^{k+1}$.  Let $T = (0,0,1)$ be the unit normal to the $x$-$y$ plane.
Then $K(t) = \bn(t) + T$ is a $C^k$ normal null field along $t \to
r(t)$.  The normal exponential map $f\:R^2 \to \R^3$ in the direction
$K$ is given by $f(s,t) = r(t) + s [\bn(t) + T]$, and hence $df/dt =
r'(t) + sn'(t)$, showing explicitly that the regularity of $f$ can be
no greater than the regularity of $\bn(t)$, and hence no greater than
the regularity of $r'(t)$.
\end{remark}

\begin{proof}
  This result is local in $N$ about $p_0$ so there is no loss of
  generality, by possibly replacing $N$ by a neighborhood of $p_0$ in
  $N$, in assuming that $N$ is a embedded submanifold of $M$.  The map
  $f$ is of class $C^k$ and the derivative $f_{(t_0,p_0)*}$ is
  injective so the implicit function theorem implies $f[\mathcal{U}]$
  is a $C^k$ hypersurface near $f(t_0,p_0)$.  Let $\eta$ be any
  nonzero timelike $C^\infty$ vector field on $M$ defined near $p_0$
  (some restrictions to be put on $\eta$ shortly) and let $\Phi_s$ be
  the flow of $\eta$.  Then for sufficiently small $\e$ the map
  $\tilde{f}\: (-\e,\e)\times N\to M$ given by
$$
\tilde{f}(s,p):=\Phi_s(p)
$$ is injective and of class $C^{k+1}$.  Extend $\nnor$ to any $C^k$
vector field $\tilde{\nnor}$ along $\tilde{f}$.  (It is not assumed
that the extension $\tilde{k}$ is null.)  That is $\tilde{\nnor}\:
(-\e,\e)\times N\to TM$ is a $C^k$ map and $\tilde{\nnor}(s,p)\in
T_{\tilde{f}(s,p)}M$.  Note that we can choose $\tilde{\nnor}(s,p)$ so
that the covariant derivative $\frac{\nabla \tilde{\nnor}}{\f
  s}(0,p_0)$ has any value we wish at the one point $(0,p_0)$.  
Define a map $F\:
(t_0-\e,t_0+\e)\times(-\e,\e)\times N\to M$ by
$$
F(t,s,p)=\exp(t\tilde{\nnor}(s,p)).
$$
We now show that $F$ can be chosen to be a local diffeomorphism near
$(t_0,0,p_0)$.  Note that $F(t,0,p)=f(t,p)$ and by assumption
$f_{*(t_0,p_0)}$ is injective.  Therefore the restriction of
$F_{*(t_0,0,p_0)}$  to 
$T_{(t_0,p_0)}(\bbR\times N) \subset T_{(t_0,0,p_0)}(\bbR\times
\bbR\times N)$ is injective.  Thus by the inverse function theorem it
is enough to show that $F_{*(t_0,0,p_0)}(\f/\f s)$ is linearly
independent of the subspace $F_{*(t_0,0,p_0)}[T_{(t_0,p_0)}(\bbR\times
N)]$.  Let
$$
V(t)=\frac{\f F}{\f s}(t,s,p_0)\bigg|_{s=0}\ .
$$ Then $V(t_0)=F_{*(t_0,0,p_0)}(\f/\f s)$ and our claim that $F$ is a
local diffeomorphism follows if $V(t_0)\notin
F_{*(t_0,0,p_0)}[T_{(t_0,p_0)}(\bbR\times N)]$. For each $s,p$ the map
$t\mapsto F(s,t,p)$ is a geodesic and therefore $V$ is a Jacobi field
along $t\mapsto F(0,t,p_0)$. (Those geodesics might change type as $s$
is varied at fixed $p_0$, but this is irrelevant for our purposes.)
The initial conditions of this geodesic are
$$
V(0)=\frac{\f}{\f s}F(0,s,p_0)\bigg|_{s=0}=\frac{\f}{\f
s}\Phi_s(p_0)\bigg|_{s=0}=\eta(p_0)
$$
and 
$$
\frac{\nabla V}{\f t}(0)=\frac{\nabla}{\f t}\frac{\nabla}{\f
s}F(t,s,p_0)\bigg|_{s=0,t=0}
=\frac{\nabla}{\f s}\frac{\nabla}{\f t}F(t,s,p_0)\bigg|_{s=0,t=0}
= \frac{\nabla \tilde{\nnor}}{\f s}(0,p_0)\ .
$$
From our set up we can choose $\eta(p_0)$ to be any timelike vector
and $\frac{\nabla\tilde{\nnor}}{\f s}(0,p_0)$ to be any vector.  As
the linear map from $T_{p_0}M\times T_{p_0}M\to T_{f(t_0,p_0)}M$ which
maps the initial conditions $V(0)$, $\frac{\nabla V}{\f t}(0)$ of a
Jacobi field $V$ to its value $V(t_0)$ is surjective\footnote{If $v\in
T_{f(t_0,p_0)}N$ there is a Jacobi field with $V(t_0)=v$ and
$\frac{\nabla V}{\f t}(t_0)=0$, which implies subjectivity.}  it is an
open map.  Therefore we can choose $\eta(p_0)$ and
$\frac{\nabla\tilde{\nnor}}{\f s}(0,p_0)$ so that $V(t_0)$ is not in
the nowhere dense set $F_{*(t_0,0,p_0)}[T_{(t_0,p_0)}(\bbR\times N)]$.
Thus we can assume $F$ is a local $C^k$ diffeomorphism on some small
neighborhood $\mathcal{A}$ of $(t_0,0,p_0)$ onto a small neighborhood
$\mathcal{B}:=F[\mathcal{A}]$ of $F(t_0,0,p_0)$ as claimed.

Consider the vector field $F_*(\f/\f t)=\f F/\f t$ along $F$.  Then
the integral curves of this vector field are the geodesics $t\mapsto
F(t,s,p)=\exp(t\tilde{\nnor}(s,p))$. (This is true even when $F$ is
not injective on its entire domain.)  These geodesics and their
velocity vectors depend smoothly on the initial data.  In the case at
hand the initial data is $C^k$ so $\f F/\f t$ is a $C^k$ vector field
along $F$.  Therefore the one form $\alpha$ defined by $\alpha(X):=\la
X,\f F/\f t\ra$ on the neighborhood $\mathcal{B}$ of $q_0$ is $C^k$.
The definition of $F$ implies that $f(t,p)=F(t,0,p)$ and therefore the
vector field $\f F/\f t$ is tangent to $f[\mathcal{O}]$ and the null
geodesics $t\mapsto f(t,p)=F(t,0,p)$ rule $f[\mathcal{O}]$ so that
$f[\mathcal{O}]$ is a null hypersurface.  Therefore for any vector $X$
tangent to $f[\mathcal{O}]$ we have $\alpha(X)=\la X,\f F/\f t\ra=0$.
Thus $f[\mathcal{O}]$ is an integral submanifold for the distribution
$\{X\ | \ \alpha(X)=0\}$ defined by $\alpha$.  But, as is easily seen
by writing out the definitions in local coordinates, an integral
submanifold of a $C^k$ distribution is a $C^{k+1}$ submanifold.  (Note
that in general there is no reason to believe that the distribution
defined by $\alpha$ is integrable. However, we have shown directly
that $f[\mathcal{O}]$ is an integral submanifold of that
distribution.)  \qed\end{proof}

We shall close this appendix with a calculation, needed in the main
body of the paper, concerning Jacobians. Let us start by recalling the
definition of the Jacobian needed in our context. Let $\phi\:M \to N$
be a $C^1$ map between Riemannian manifolds, with $\dim M \le \dim
N$. Let $n=\dim M$ and let $e_1,\ldots,e_n$ be an orthonormal basic of
$T_pM$ then the Jacobian of $\phi$ at $p$ is $ J(\phi)(p) =
\|\phi_{*p}e_1\wedge \phi_{*p}e_2\wedge\cdots \wedge \phi_{*p}e_n\|$. 
When $\dim M =\dim N$ and both $M$ and $N$ are oriented with
$\omega_M$ being the volume form on $M$, and $\omega_N$ being the volume
form on $N$, then $J(\phi)$ can also be described as the positive
scalar satisfying: $\phi^*(\omega_N)= \pm J(\phi)\,\omega_M$.

Let $S$ be a $C^2$ co-dimension two acausal spacelike submanifold of a
smooth spacetime $M$, and let $K$ be a past directed $C^1$ null vector
field along $S$.  Consider the normal exponential map in the direction
$K$, $\Phi\:\bbR \times S\to M$, defined by $\Phi(s,x)= \exp_xsK$.
($\Phi$ need not be defined on all of $\bbR \times S$.)  Suppose the
null geodesic $\eta\: s\to \Phi(s,p)$ meets a given acausal spacelike
hypersurface $\GregS $ at $\eta(1)$.  Then there is a neighborhood $W$
of $p$ in $S$ such that each geodesic $s\to \Phi(s,x)$, $x\in W$ meets
$\GregS $, and so determines a $C^1$ map $\phi\: W \to \GregS $, which
is the projection into $\Sigma$ along these geodesics.  Let $J(\phi)$
denote the Jacobian determinant of $\phi$ at $p$.  $J(\phi)$ may be
computed as follows.  Let $\{X_1, X_2, \ldots ,X_k\}$ be an
orthonormal basis for the tangent space $T_pS$. Then ,
$$
J(\phi) = \|\phi_{*p}X_1\wedge \phi_{*p}X_2\wedge\cdots \wedge \phi_{*p}X_k\|.
$$ Suppose there are no focal points to $S$ along $\eta|_{[0,1]}$.
Then by shrinking $W$ and rescaling $K$ if necessary, $\Phi\:
[0,1]\times W\to M$ is a $C^1$ embedded null hypersurface $N$ such
that $\Phi(\{1\}\times W) \subset \GregS $.  Extend $K$ to be the
$C^1$ past directed null vector field, $K = \Phi_*(\frac{\Gregd
  }{\Gregd s})$ on $N$.  Let $\theta=\theta(s)$ be the null mean
curvature of $N$ with respect to $-K$ along $\eta$. For completeness
let us give a proof of the following, well known result:

\begin{Proposition} \label{Gregs}
With $\theta = \theta(s)$ as described above,\nopagebreak
\begin{enumerate}
\nopagebreak\item If there are no focal points to $S$ along $\eta|_{[0,1]}$, then
\begin{equation}
  \label{jacob}
  J(\phi)=\exp\bigg(-\int_0^1 \theta(s) ds\bigg)\ .
\end{equation}
\item If $\eta(1)$ is the first focal point to $S$ along
  $\eta|_{[0,1]}$, then
  $$J(\phi)=0\ .$$
\end{enumerate}
\end{Proposition}
\begin{remark} In particular, if $N$ has nonnegative null mean curvature
with respect to the future pointing null normal, \emph{i.e.}, if
$\theta \ge 0$, we obtain that $J(\phi)\le 1$.
\end{remark}

\begin{remark}  Recall that $\theta$ was only defined
  when a normalization of $K$ has been chosen. We stress that in
  \eq{jacob} that normalization is so that $K$ is tangent to an
  affinely parameterized geodesic, with $s$ being an affine distance
  along~$\eta$, and with  $p$ corresponding to $s=0$ and  $\phi(p)$
  corresponding to $s=1$.
\end{remark}
\begin{proof}
 1. To relate $J(\phi)$ to the null mean curvature of $N$, extend the
  orthonormal basis $\{X_1, X_2, \ldots  ,X_k\}$ to Lie parallel vector
  fields $s\to X_i(s)$, $i= 1,\ldots,k$, along $\eta$, $\mathcal{L}_KX_i
  = 0$ along $\eta$.  Then by a standard computation, 
\Gregbeq J(\phi) & =
  &\|\phi_{*p}X_1\wedge \phi_{*p}X_2\wedge\cdots \wedge \phi_{*p}X_k\|
  \nonumber\\ & = & \|X_1(1)\wedge X_2(1)\wedge\cdots \wedge X_k(1)\|
  \nonumber\\ & = & \sqrt{g}\,\bigg|_{s=1} \, , \nonumber \Gregeeq 
where $g =
  \det[g_{ij}]$, and $g_{ij} = g_{ij}(s)= \la X_i(s),X_j(s)\ra$.  We
  claim that along $\eta$,
$$
\theta = -\frac1{\sqrt{g}}\frac{d}{ds}\sqrt{g} \, .
$$ The computation is standard.  Set $b_{ij} = B(\overline
X_i,\overline X_j)$, where $B$ is the null second fundamental form of
$N$ with respect to $-K$, $h_{ij} = h(\overline X_i,\overline X_j) =
g_{ij}$, and let $g^{ij}$ be the $i,j$th entry of the inverse matrix
$[g_{ij}]^{-1}$. Then $\theta = g^{ij}b_{ij}$.  Differentiating
$g_{ij}$ along $\eta$ we obtain, \Gregbeq \frac{d}{ds}g_{ij} = K\la
X_i,X_j\ra & = & \la \Gregn _KX_i, X_j\ra + \la X_i,\Gregn _K X_j\ra
\nonumber \\ & = & \la \Gregn _{X_i}K, X_j\ra + \la X_i,\Gregn
_{X_j}K\ra \nonumber \\ & = &-( b_{ij} + b_{ji}) = -2b_{ij} \,
.\nonumber \Gregeeq Thus,
$$
\theta = g^{ij}b_{ij} =- \frac12 g^{ij}\frac{d}{ds}g_{ij} =-
\frac12\,\frac1{g}\frac{dg}{ds}
=- \frac1{\sqrt{g}}\frac{d}{ds}\sqrt{g} \, ,
$$
as claimed. Integrating along $\eta$ from $s=0$ to $s=1$ we obtain,
$$
J(\phi) = \sqrt{g}\,\bigg|_{s=1} 
= \sqrt{g}\,\bigg|_{s=0}\cdot\exp\bigg(-\int_0^1 \theta\,ds\bigg)
= \exp\bigg(-\int_0^1 \theta\,ds\bigg)\, .
$$

2. Suppose now that $\eta(1)$ is a focal point to $S$ along $\eta$, but
that there are no focal points to $S$ along $\eta$ prior to that.
Then we can still construct the $C^1$ map $\Phi\: [0,1]\times W\to M$,
with $\Phi(\{1\}\times W) \subset \GregS $, such that $\Phi$ is an
embedding when restricted to a sufficiently small open set in
$[0,1]\times W$ containing $[0,1)\times\{p\}$.  The vector fields $s
\to X_i(s)$, $s \in[0,1)$, $i= 1, .., k$, may be constructed as above,
and are Jacobi fields along $\eta|_{[0,1)}$, which extend smoothly to
$\eta(1)$.  Since $\eta(1)$ is a focal point, the vectors $\phi_*X_1=
X_1(1)$, \ldots, $\phi_*X_k = X_k(1)$ must be linearly dependent, which
implies that $J(\phi) = 0$.
\qed\end{proof}

\newcommand{\mycals}{{\mycal S}}
\newcommand{\mycalm}{{\mycal M}}
\newcommand{\mycalu}{{\mycal U}}
\newcommand{\Mclosed}{\hskip10pt {\overline{\phantom{I}}\hskip-15pt{\mycal M}}}
\section{Some comments on the area theorem of Hawking and Ellis}
\label{HEarea}
In this appendix we wish to discuss the status of our
$\cH$--regularity condition with respect to the conformal completions
considered by Hawking and Ellis \cite{HE} in their treatment of the
area theorem.  For the convenience of the reader let us recall here
the setting of~\cite{HE}. One of the conditions of the Hawking--Ellis
area theorem \cite[Proposition~9.2.7, p.~318]{HE}  is  that
spacetime $({\mycal M},g)$ is \emph{weakly asymptotically simple and
  empty} (``WASE'', \cite[p.~225]{HE}).  This means that there exists
an open set ${\mycal \mycalu}\subset \mycalm$ which is isometric to
$\mycalu'\cap \mycalm'$, where $\mycalu'$ is a neighborhood of null
infinity in an asymptotically simple and empty (ASE) spacetime
$(\mycalm',g')$ \cite[p.~222]{HE}.  It is further assumed that
$\mycalm$ admits a partial Cauchy surface $\mycals$ with respect to
which $\mycalm$ is \emph{future asymptotically predictable}
(\cite{HE}, p.~310). This is defined by the requirement that $\scrip$
is contained in closure of the future domain of dependence ${\cal
  D}^+(\mycals; \mycalm)$ of $\mycals$, where the closure is taken in
the conformally completed manifold $\Mclosed =\mycalm \cup
\scrip\cup\scri^{-}$, with both $\scrip$ and $\scri^{-}$ being null
hypersurfaces. Next, one says that $(\mycalm,g)$ is \emph{strongly
  future asymptotically predictable} (\cite{HE}, p.~313) if it is
future asymptotically predictable and if $J^{+}(\mycals)\cap \bar
J^{-}(\scrip;\Mclosed )$ is contained in ${\cal D}^+(\mycals;
\mycalm)$.  Finally (\cite{HE}, p.~318), $(\mycalm ,g)$ is said to be
a \emph{regular predictable space} if $(\mycalm ,g)$ is strongly
future asymptotically predictable and if the following three
conditions hold:
\begin{enumerate}
\item[($\alpha$)] $\mycals \cap  \bar
J^{-}(\scrip;\Mclosed  )$ is homeomorphic to $\R^3\setminus$(an open
set with compact closure).
\item [($\beta$)] $\mycals$ is simply connected.
\item [($\gamma$)] the family of hypersurfaces $\mycals(\tau)$
  constructed in \cite[Proposition~9.2.3, p.~313]{HE} has the property
  that for sufficiently large $\tau$ the sets $\mycals(\tau)\cap
  \bar J^{-}(\scrip;\Mclosed  )$ are contained in $\bar
  J^{+}(\scri^{-};\Mclosed  )$.
\end{enumerate}
It is then asserted in \cite[Proposition~9.2.7, p.~318]{HE} that the
area theorem holds for regular predictable spaces satisfying the null
energy condition.

Now in the proof of \cite[Proposition~9.2.1, p.~311]{HE} 
(which is one of the results used in the proof of
\cite[Proposition~9.2.7, p.~318]{HE}) Hawking and Ellis write: ``This
shows that if $\mycal{W}$ is any compact set of $\mycals $, every
generator of $\Scri^+$ leaves $J^+({\mycal W};\Mclosed)$.''  The
justification of this given in \cite{HE} is wrong\mnote{ to be
  complete one should attempt to decide whether or not the claim of
  Proposition 9.2.1 can be justified if the hypothesis that
  $(\mycalm,g)$ is a regular predictable space is made \\ - \\ both
  PTC and Greg have tried some, without any conclusions}\footnote{In
  the proof of \cite[Proposition~9.2.1, p.~311]{HE} it is claimed that
  ``...  Then $\mycals'\setminus \mycalu'$ is compact...''. This
  statement is incorrect in general, as shown by the example
  $(\mycalm,g)= (\mycalm',g')= (\R^{4},\, $diag$ (-1,+1,+1,+1))$,
  $\mycals=\mycals'=\{t=0\}$, $\mycalu=\mycalu'=\{t\ne 0\}$. This
  example does not show that the claim is wrong, but that the proof
  is; we do not know whether the claim in Proposition~9.2.1 is correct
  as stated under the hypothesis of future asymptotic predictability
  of $(\mycalm,g)$ made there.  Let us note that the conditions
  $(\alpha)$--$(\gamma)$ do not seem to be used anywhere in the proof
  of Proposition~9.2.7 as presented in \cite{HE}, and it is
  conceivable that the authors of \cite{HE} had in mind some use of
  those conditions in the proof of Propostion~9.2.1. We have not
  investigated in detail whether or not the assertion made there can
  be justified if the supplementary hypothesis that $(\mycalm,g)$ is a
  regular predictable space is made, as the approach we advocate in
  Section \ref{Sarea2} allows one to avoid the ``WASE'' framework
  altogether. }. If one is willing to impose this as a supplementary
hypothesis, then this condition can be thought of as the
Hawking--Ellis equivalent of our condition of $\cH$--regularity of
$\Scri^+$. When such a condition is imposed in addition to the
hypothesis of strong asymptotic predictability and weak asymptotic
simplicity (``WASE'') of $(\mycalm,g)$, then the hypotheses of
Proposition \ref{Ptheta1} hold, and the conclusions of our version of
the area theorem, Theorem~\ref{Tarea}, apply.

\begin{figure}[t]
 \input{HEcond.pstex_t}
 \caption[]{The set $\Omega\equiv
   J^+(\mycals;\mycalm ) \cap \partial\mycalu$ and its image under
   $\psi$.}
 \label{condHE.fig}
\end{figure}
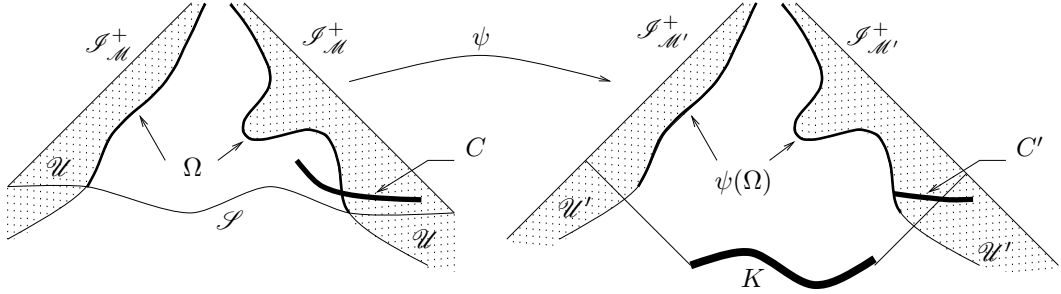

 An alternative way to guarantee that the hypotheses of Proposition
 \ref{Ptheta1} will hold in the ``future asymptotically predictable
 WASE'' set--up of \cite{HE} (for those sets $C$ which lie to the
 future of $\mycals$) is to impose some mild additional conditions on
 $\mycalu$ and $\mycals$. There are quite a few possibilities, one
 such set of conditions is as follows: Let $\psi : \mycalu \to
 \mycalu' \cap \mycalm '$ denote the isometry arising in the
 definition of the WASE spacetime $\mycalm $.  First, we require that
 $\psi$ can be extended by continuity to a continuous map, still
 denoted by $\psi$, defined on $\overline \mycalu$.  Next, suppose
 there exists a compact set $K \subset \mycalm '$ such that,
\begin{equation}
  \label{condHE2}
  \psi(J^+(\mycals;\mycalm ) \cap \partial\mycalu) \subset J^+(K;\mycalm ')\;,
\end{equation}
see Figure \ref{condHE.fig}. 
Let us show that, under the future asymptotically predictable WASE
conditions together with \eq{condHE2}, for every compact set $C\subset
J^+(\mycals;\mycalm )$ that meets $I^-(\scrip; \Mclosed )$ there
exists a future inextendible (in $\mycalm $) null geodesic $\eta
\subset \partial J^+(C; \mycalm )$ starting on $C$ and having future
end point on $\scrip$.  First, we claim that
\begin{equation}
  \label{condHEp}
  \psi(J^+(C;\mycalm ) \cap
\mycalu) \subset J^+(K \cup C'; \mycalm ')\;,
\end{equation}
where $C'= \psi(C\cap \bar \mycalu)$. Indeed, let $p\in
\psi(J^+(C;\mycalm ) \cap \mycalu)$, therefore there exists a future
directed causal curve $\gamma$ from $C$ to $\psi^{-1}(p)\in\mycalu$.
If $\gamma\subset\mycalu$, then $p\in J^+( C'; \mycalm ')$. If not,
then $\gamma$ exits $\mycalu$ when followed from $\psi^{-1}(p)$ to the
past at some point in $J^+(\mycals;\mycalm ) \cap \partial\mycalu$,
and thus $p\in\psi(J^+(\mycals;\mycalm ) \cap \partial\mycalu) \subset
J^+(K;\mycalm ')$, which establishes \eq{condHEp}. Since $K\cup C'$ is
compact, by Lemma 4.5 and Proposition~4.13 in \cite{Newman:Scri}, each
generator of $\Scri_{\mycalm '}^+$ meets $\partial J^+(K\cup
C';\Mclosed ')$ exactly once.  It follows that, under the natural
identification of $\Scri_{\mycalm }^+$ with $\Scri_{\mycalm '}^+$, the
criteria for $\cH$-regularity discussed in Remark~\ref{r4.5} are
satisfied.  Hence, we may apply Proposition \ref{Pglobal1} to obtain
the desired null geodesic $\eta$.

There exist several other proposals how to modify the WASE conditions
of \cite{HE} to obtain better control of the space--times at hand
\cite{Newman:coscen,Krolak:erpfsps,ClarkedeFelice2}, but we have not
investigated in detail their suitability to the problems considered
here.

\section{Some comments on the area theorems of Kr\'olak}
\label{krolak}
Kr\'olak has previously extended the definition of a black hole to
settings more general, in various ways, than the standard setting
considered in Hawking and Ellis \cite{HE}.  In each of the
papers~\cite{Krolak:bh,Krolak:scc,Krolak:coscen} Kr\'olak obtains an
area theorem, under the implicit assumption of piecewise smoothness.
It follows from the results presented here that the area theorems of
Kr\'olak still hold without the supplementary hypothesis of piecewise
smoothness, which can be seen as follows. First, in each of the
papers~\cite{Krolak:bh,Krolak:scc,Krolak:coscen} the event horizon
$\cH$ is defined as the boundary of a certain past set, which implies
by \cite[Prop.~6.3.1 p.~187]{HE} that $\cH$ is an achronal closed
embedded $C^0$ hypersurface.  Moreover, by arguments in
\cite{Krolak:bh,Krolak:scc,Krolak:coscen} $\cH$ is ruled by future
inextendible null geodesics and hence, in all the
papers~\cite{Krolak:bh,Krolak:scc,Krolak:coscen} $\cH$ is a future
horizon as defined here. Now, because in~\cite{Krolak:scc,Krolak:bh}
the null generators of $\cH$ are assumed to be future complete, one
can apply Theorem~\ref{Tarea} to conclude that the area theorem holds,
under the explicit assumptions of~\cite{Krolak:scc,Krolak:bh}, for the
horizons considered there, with no additional regularity conditions.

On the other hand, in~\cite{Krolak:coscen} completeness of generators
is not assumed, instead a regularity condition on the horizon is
imposed.  Using the notation of~\cite{Krolak:coscen}, we shall say
that a horizon $\mathcal{H}_{\mathscr{T}}$ (as defined
in~\cite{Krolak:coscen}) is \emph{weakly regular} iff for any point
$p$ of $\mathcal{H}_{\mathscr{T}}$ there is an open neighborhood
$\mathcal{U}$ of $p$ such that for any compact set $K$ contained in
$\mathcal{U}\cap \overline{\mathcal{W}_{\mathscr{T}}}$ the set
$J^+(K)$ contains a $N^\infty$-TIP from
$\widehat{\mathcal{W}}_{\mathscr{T}}$. (The set
${\mathcal{W}}_{\mathscr{T}}$ may be thought of as the region outside
of the black hole, while $\widehat{\mathcal{W}}_{\mathscr{T}}$
represents null infinity.) This differs from the definition
in~\cite[p.~370]{Krolak:coscen} in that Kr\'olak requires the compact
set $K$ to be in $\mathcal{U}\cap {\mathcal{W}_{\mathscr{T}}}$ rather
than in the somewhat larger set $\mathcal{U}\cap
\overline{\mathcal{W}_{\mathscr{T}}}$. \footnote{In Kr\'olak's
  definition $K$ is not allowed to touch the event horizon
  $\mathcal{H}_{\mathscr{T}}$.  But then in the proof
  of~\cite[Theorem~5.2]{Krolak:coscen} when $\mathscr{T}$ is deformed,
  it must be moved completely off of the horizon and into
  $\mathcal{W}_{\mathscr{T}}$.  So the deformed $\mathscr{T}$ will, in
  general, have a boundary in $\mathcal{W}_{\mathscr{T}}$.  The
  generator $\gamma$ in the proof of~\cite[Theorem~5.2]{Krolak:coscen}
  may then meet $\mathscr{T}$ at a boundary point, which introduces
  difficulties in the focusing argument used in the proof. The
  definition used here avoids this problem.}  Under this slightly
modified regularity condition, the arguments of the proof of
\cite[Theorem~5.2]{Krolak:coscen} yield positivity of $\tA$: the
deformation of the set $\mathscr T$ needed in that proof in
\cite{Krolak:coscen} is obtained using our sets
$S_{\epsilon,\eta,\delta}$ from Proposition \ref{Ptheta1}. Our
Theorem~\ref{thm:local-area} then implies that area monotonicity holds
for the horizons considered in \cite{Krolak:coscen}, subject to the
minor change of the notion of \emph{weak regularity} discussed above,
with no additional regularity conditions.

\section{Proof of Theorem \protect\ref{TCfr}}
\label{ACfr}

For $q\in S_0$ let $\Gamma_q\subset \cH$ denote the generator of $\cH$
passing through $q$. Throughout this proof all curves will be
parameterized by signed $\kaux$--distance from $S_0$, with the
distance being negative to the past of $S_0$ and positive to the
future. We will need the following Lemma:

\begin{Lemma}
  \label{LBorel}
  $S_0$ is a Borel subset of $\calS$, in particular $S_0$ is $\Hnmk$
  measurable.
\end{Lemma}
 \begin{proof}
 For each
$\delta>0$ let
\begin{eqnarray}A_\delta: & =& \left\{ p\in S_0\ | \ \mbox{the domain
      of definition of }\ \Gamma_p \right.
\nonumber \\ & & \phantom{xxxxxxxxxxxxxxxx}
    \left.\mbox{contains the interval }[-\delta, \delta] \right\}\ .
  \label{adelta1}
  \\ B_\delta: & =& \left\{ p\in S_0\ | \ \mbox{the domain of
      definition of the inextendible geodesic }\right.  \nonumber \\ 
  & & \phantom{xxxxxxxx} \left.\ \mbox{ $\gamma_p$ containing
      $\Gamma_p$ contains the interval }\ [-\delta, \delta] \right\}\ 
  .
  \label{bdelta}
  \end{eqnarray}
  Lower semi--continuity of existence time of geodesics shows that the
  $B_\delta$'s are open subsets of $\calS$. Clearly
  $A_{\delta'}\subset B_\delta$ for $\delta'\ge \delta$. We claim that
  for $\delta'\ge\delta$ the sets $A_{\delta'}$ are closed subsets of
  the $B_\delta$'s.  Indeed, let $q_i\in A_{\delta'}\cap B_\delta$ be
  a sequence such that $q_i\to q_\infty\in B_\delta$. Since the
  generators of $\cH$ never leave $\cH$ to the future, and since
  $q_\infty\in B_\delta$, it immediately follows that the domain of
  definition of $\Gamma_{q_\infty}$ contains the interval $[0,
  \delta]$. Suppose, for contradiction, that $\Gamma_{q_\infty}(s_-)$
  is an endpoint on $\cH$ with $s_-\in(-\delta,0]$, hence there exists
  $s'\in(-\delta,0]$ such that $\gamma_{q_\infty}(s')\in I^-(\cH)$.
  As $q_\infty$ is an interior point of $\Gamma_{q_\infty}$, the
  $\kaux$--unit tangents to the $\Gamma_{q_i}$'s at $q_i$ converge to
  the $\kaux$--unit tangent to $\Gamma_{q_\infty}$ at $q_\infty$. Now
  $ I^-(\cH)$ is open, and continuous dependence of ODE's upon initial
  data shows that $\gamma_{q_i}(s')\in I^-(\cH)$ for $i$ large enough,
  contradicting the fact that $q_i\in A_{\delta'}$ with
  $\delta'\ge\delta$. It follows that $q_\infty\in A_{\delta'}$, and
  that $A_{\delta'}$ is closed in $B_\delta$. But a closed subset of
  an open set is a Borel set, hence $A_{\delta'}$ is Borel in $\calS$.
  Clearly
  $$S_0=\cup_{i}A_{1/i}\ , $$ which implies that $S_0$ is a Borel
  subset of $\calS$. The $\Hnmk$--measurability of $S_0$ follows now
  from \cite[p.~293]{Federer:lectures} or
  \cite[p.~147]{Edgar:fractals}$^{\mbox{\scriptsize
      \ref{fBorel}}}$.
  \qed\end{proof}

\smallskip

Returning to the proof of Theorem \ref{TCfr}, set
\begin{eqnarray*}
  \label{Cfr.1}
  \Gamma_q^+ & = & \Gamma_q \cap J^+(q)\ , \\ \label{Cfr.2}
  \cH_{\mbox{\scriptsize sing}} & = & \cH\setminus \cH_\Al\ , \\ 
  \label{Cfr.3} 
\Omega & = & \cup_{q\in S_0} \Gamma_q^+\ , 
\\ \label{Cfr.4} \Omega_{\mbox{\scriptsize sing}} & = & \Omega\cap
\cH_{\mbox{\scriptsize sing}}\ .\end{eqnarray*} By definition we have
$ \Omega_{\mbox{\scriptsize sing}}\subset \cH_{\mbox{\scriptsize
    sing}}$ and completeness of the Hausdorff measure\footnote{A
  measure is \emph{complete} iff all sets of outer measure zero are
  measurable.  Hausdorff measure is constructed from an outer measure
  using Carath\'eodory's definition of measurable sets \cite[p.
  54]{FedererMeasureTheory}.  All such measures are complete
  \cite[Theorem~2.1.3 pp. 54--55]{FedererMeasureTheory}.} together with
$\Hnk( \cH_{\mbox{\scriptsize sing}})=0$ implies that $
\Omega_{\mbox{\scriptsize sing}}$ is $n$--Hausdorff measurable, with
\begin{equation}
  \label{Cfr.5}
\Hnk(\Omega_{\mbox{\scriptsize sing}})=0\ . 
  \end{equation}
  Let $\phi\:\Omega\to S_0$ be the map which to a point $p\in
  \Gamma_q^+$ assigns $q\in S_0$. The arguments of the proofs of
  Lemmata \ref{A-delta-C11new} and \ref{phi-lip} show that $\phi$ is
  locally Lipschitz. This, together with Lemma \ref{LBorel}, allows us
  to use the co--area formula \cite[Theorem~3.1]{Federer:measures} to
  infer from \eq{Cfr.5} that
  $$0 = \int_{\Omega_{\mbox{\scriptsize sing}}} J(\phi) d\Hnk =
  \int_{S_0} \Hsone(\Omega_{\mbox{\scriptsize sing}}\cap \phi^{-1}(q))
  d \Hnmk(q)\ ,$$ where $J(\phi)$ is the Jacobian of $\phi$,
  \emph{cf.} \cite[p.~423]{Federer:measures}. Hence
\begin{equation}
  \label{Cfr.6}
\Hsone(\Omega_{\mbox{\scriptsize sing}}\cap \phi^{-1}(q))=0
  \end{equation}
for almost all $q$'s in $S_0$. A chase through the definitions shows
that \eq{Cfr.6} is equivalent to
  \begin{equation}
    \label{nullm}
    \Hsone{(\Gamma_q^+\setminus \cH_{\Al})}= 0\ , 
  \end{equation}
  for almost all $q$'s in $S_0$. Clearly the set of Alexandrov points
  of $\cH$ is dense in $\Gamma_q^+$ when \eq{nullm} holds.  Theorem
  \ref{Tfr} shows, for such $q$'s, that all interior points of
  $\Gamma_q$ are Alexandrov points of $\cH$, hence points $q$
  satisfying \eq{nullm} are in $S_1$. It follows that $S_1$ is $\Hnmk$
  measurable, and has full $(n-1)$--Hausdorff measure in $S_0$.  This
  establishes our claim about $S_1$. The claim about $S_2$ follows now
  from the inclusion $S_1\subset S_2$. \qed

\section{Proof of Proposition \protect\ref{C11-extend}}
\label{app:C11extend}

Because of the identities 
\begin{eqnarray*}
&&\Big(f(p)+\la x-p,a_p\ra -\frac{C}{2}\|x-p\|^2\Big)+\frac{C}{2}\|x\|^2\\
   &&=\Big(f(p)+\frac{C}{2}\|p\|^2\Big)+\la x-p,a_p+Cp\ra\;,
\end{eqnarray*}
and
\begin{eqnarray*}
&&\Big(f(q)+\la x-q,a_q\ra+\frac{C}{2}\|x-q\|^2\Big)+\frac{C}{2}\|x\|^2 \\
        &&=\Big(f(q)+\frac{C}{2}\|q\|^2\Big)+\la x-q,a_q+Cq\ra +C\|x-q\|^2\;,
\end{eqnarray*}
we can replace $f$ by $x\mapsto f(x)+C\|x\|^2/2$ and $a_p$ by $a_p+Cp$
and assume that for all $p,x\in A$ we have
\begin{equation}\label{nf-support}
f(p)+\la x-p,a_p\ra \le f(x) \le f(p)+\la x-p,a_p\ra + C\|x-p\|^2\;,
\end{equation}
and for all $p,q\in A$ and $x\in \R^n$
\begin{equation}\label{ndis-spt}
f(p)+\la x-p,a_p\ra \le 
        f(q)+\la x-q,a_q\ra + C\|x-q\|^2\;.
\end{equation}
These inequalities can be given a geometric form that is easier to
work with.  Let $P:=\{(x,y)\in \R^n\times\R : y> C\|x\|^2\}$.  Then
$P$ is an open convex solid paraboloid of $\R^{n+1}$. We will denote
the closure of $P$ by $\ol{P}$. From the identity
$$ f(q)+\la x-q,a_q\ra +C\|x-q\|^2=
\big(f(q)-(4C)^{-1}\|a_q\|^2\big)+C\|x-\big(q-(2C)^{-1}a_q\big)\|^2
$$ it follows that the solid open paraboloids $\{(x,y)\in \R^n\times
\R: y> f(q)+\la x-q,a_q\ra +C\|x-q\|^2\}$ are translates in $\R^{n+1}$
of $P$.  Let $G[f]:=\{(x,y)\in \R^n\times \R: x\in A,\ y=f(x)\}$ be
the graph of $f$.  The inequalities~(\ref{nf-support})
and~(\ref{ndis-spt}) imply that for each $p\in A$ there is an affine
hyperplane $H_p=\{(x,y)\in \R^n\times \R:y=f(p)+\la x-p,a_p\ra\}$ of
$\R^{n+1}$ and a vector $b_p\in \R^{n+1}$ so that
\begin{equation}\label{hyper-p}
(p,f(p))\in H_p,
\end{equation}
\begin{equation}\label{para-p}
(P+b_p)\cap G[f]=\emptyset\quad \text{but} \quad (p,f(p))\in b_p+\ol{P},
\end{equation}
and for all $p,q\in A$
\begin{equation}\label{hyper-para}
H_p\cap (b_q+ P)=\emptyset.
\end{equation}
As the paraboloids open up, this last condition implies that each
$b_q+P$ lies above all the hyperplanes $H_p$.

Let
$$
Q:=\mathop{\rm Convex\, Hull} \bigg( \bigcup_{p\in A}(b_p+P)\bigg).
$$
Because $P$ is convex if $\alpha_1,\ldots, \alpha_m\ge 0$ satisfy
$\sum_{i=1}^m\alpha_i=1$ then for any $v_1,\ldots, v_m\in \R^{n+1}$ 
$$
\alpha_1(v_1+P)+\alpha_2(v_2+P)+\cdots+\alpha_m(v_m+P)=
        (\alpha_1v_1+\cdots+\alpha_mv_m)+P.
$$
Therefore if
$$
B:=\mathop{\rm Convex\, Hull}\, \{b_p: p\in A\}
$$
then
$$
Q=\bigcup_{v\in B}(v+P)
$$ so that $Q$ is a union of translates of $P$.  Thus $Q$ is open.
Because $P$ is open we have that if $\lim_{\ell \to \infty}v_\ell=v$
then $v+P\subseteq \bigcup_{\ell}(v_\ell +P)$.  So if
$\ol{B}$ is the closure of $B$ in $\R^{n+1}$ then we also have
\begin{equation}\label{QP}
Q=\bigcup_{v\in \ol{B}}(v+P).
\end{equation}
For each $p$ the open half space above $H_p$ is an open convex set and
$Q$ is the convex hull of a subset of this half space.  Therefore $Q$
is contained in this half space.  Therefore $Q\cap H_p=\emptyset$ for
all $p$.

We now claim that for each point $z\in \f Q$ there is a supporting
paraboloid for $z$ in the sense that there is a vector $v\in \ol{B}$
with $v+P\subset Q$ and $z\in v+\ol{P}$.  To see this note that as
$z\in \f Q$ there is a sequence $\{b_\ell\}_{\ell=1}^\infty\subset B$ and
$\{w_\ell\}_{\ell=1}^\infty \subset P$ so that $\lim_{\ell\to
\infty}(b_\ell+w_\ell)=z$.  Fixing a $p_0\in A$ and using that all the
sets $b_\ell +P$ are above the hyperplane $H_{p_0}$ we see that both
the sequences $b_\ell$ and $w_\ell$ are bounded subsets of $\R^{n+1}$
and  by going to a subsequence we can assume that $v:=\lim_{\ell\to
\infty}b_\ell$ and $w:=\lim_{\ell\to \infty}w_\ell$ exist.  Then
$z=v+w$, $v\in \ol{B}$ and $w\in \ol{P}$.  Then~(\ref{QP}) implies
$v+P\subset Q$ and $w\in \ol{P}$ implies $z\in v+\ol{P}$.  Thus we
have the desired supporting paraboloid.

We also claim that the graph $G[f]$ satisfies $G[f]\subset \f Q$.
This is because for $p\in A$ the point then $(p,f(p))\in H_p$ and
$H_p$ is disjoint from $Q$.  Thus $(p,f(p))\notin Q$.  But
from~(\ref{para-p}) $(p,f(p))\in b_p+\ol{P}$ and as $b_p+\ol{P}\subset
\ol{Q}$ this implies $(p,f(p))\in \ol{Q}$.  Therefore $(p,f(p))\in \f
Q$ as claimed.

Let $F\: \R^n\to \R$ be the function that defines $\f Q$.  Explicitly 
$$
F(x)=\inf\{y : (x,y)\in Q\}.
$$
This is a function defined on all of $\R^n$ and $G[f]\subset \f
Q$ implies that this function extends $f$.  For any $x_0\in \R^n$
the convexity of $Q$ implies there is a supporting hyperplane $H$ for
$Q$ at its boundary point $(x_0,F(x_0))$ and we have seen there also a
supporting paraboloid $v+P$ for $\f Q=G[F]$ at $(x_0,F(x_0))$.
Expressing $H$ and $\f(v+P)$ as graphs over $\R^n$ these geometric
facts yield that there is a vector $a_{x_0}$ so that the 
inequalities
$$
F(x_0)+\la x-x_0,a_{x_0}\ra \le F(x) \le F(x_0)+\la
x-x_0,a_{x_0}\ra+C\|x-x_0\|^2
$$
hold for all $x\in \R^n$.  Because the function $F$ is defined on all
of $\R^n$ (rather than just the subset $A$), it follows
from~\cite[Prop.~1.1~p.~7]{CaffarelliCabre} that $F$ is of
class~$C^{1,1}$.\qed

\textbf{Acknowledgments} P.T.C.\ acknowledges useful discussions with
or comments from Guy Barles, Piotr Haj\l asz, Tom Ilmanen, Bernd
Kirchheim, Andrzej Kr\'olak, Olivier Ley and Laurent V\'eron.  R.H.\ 
benefited from discussions and/or correspondence with Joe Fu, Sergei
Konyagin, and Ron DeVore. We are grateful to Lars Andersson for
bibliographical advice.

\providecommand{\bysame}{\leavevmode\hbox to3em{\hrulefill}\thinspace}

\end{document}

%% file: HEcond.pstex_t
\begin{picture}(0,0)%
\epsfig{file=HEcond.pstex}%
\end{picture}%
\setlength{\unitlength}{0.00045800in}%
\begingroup\makeatletter\ifx\SetFigFont\undefined
\def\x#1#2#3#4#5#6#7\relax{\def\x{#1#2#3#4#5#6}}%
\expandafter\x\fmtname xxxxxx\relax \def\y{splain}%
\ifx\x\y   
\gdef\SetFigFont#1#2#3{%
  \ifnum #1<17\tiny\else \ifnum #1<20\small\else
  \ifnum #1<24\normalsize\else \ifnum #1<29\large\else
  \ifnum #1<34\Large\else \ifnum #1<41\LARGE\else
     \huge\fi\fi\fi\fi\fi\fi
  \csname #3\endcsname}%
\else
\gdef\SetFigFont#1#2#3{\begingroup
  \count@#1\relax \ifnum 25<\count@\count@25\fi
  \def\x{\endgroup\@setsize\SetFigFont{#2pt}}%
  \expandafter\x
    \csname \romannumeral\the\count@ pt\expandafter\endcsname
    \csname @\romannumeral\the\count@ pt\endcsname
  \csname #3\endcsname}%
\fi
\fi\endgroup
\begin{picture}(12024,3303)(289,-2741)
\put(8701,-2686){\makebox(0,0)[lb]{\smash{\SetFigFont{10}{12.0}{rm}$K$}}}
\put(2701,-2011){\makebox(0,0)[lb]{\smash{\SetFigFont{10}{12.0}{rm}$\mycal S$}}}
\put(5551,-1186){\makebox(0,0)[lb]{\smash{\SetFigFont{10}{12.0}{rm}$C$}}}
\put(11851,-1186){\makebox(0,0)[lb]{\smash{\SetFigFont{10}{12.0}{rm}$C'$}}}
\put(5626, 89){\makebox(0,0)[lb]{\smash{\SetFigFont{10}{12.0}{rm}$\psi$}}}
\put(1201, 14){\makebox(0,0)[lb]{\smash{\SetFigFont{10}{12.0}{rm}$\scrip_{\mycalm}$}}}
\put(7501,164){\makebox(0,0)[lb]{\smash{\SetFigFont{10}{12.0}{rm}$\scrip_{\mycalm'}$}}}
\put(9901, 89){\makebox(0,0)[lb]{\smash{\SetFigFont{10}{12.0}{rm}$\scrip_{\mycalm'}$}}}
\put(3676, 89){\makebox(0,0)[lb]{\smash{\SetFigFont{10}{12.0}{rm}$\scrip_{\mycalm}$}}}
\put(751,-1411){\makebox(0,0)[lb]{\smash{\SetFigFont{10}{12.0}{rm}$\mycalu$}}}
\put(6601,-1876){\makebox(0,0)[lb]{\smash{\SetFigFont{10}{12.0}{rm}$\mycalu'$}}}
\put(4936,-2236){\makebox(0,0)[lb]{\smash{\SetFigFont{10}{12.0}{rm}$\mycalu$}}}
\put(8401,-1561){\makebox(0,0)[lb]{\smash{\SetFigFont{10}{12.0}{rm}$\psi(\Omega)$}}}
\put(2326,-1411){\makebox(0,0)[lb]{\smash{\SetFigFont{10}{12.0}{rm}$\Omega$}}}
\put(11401,-2386){\makebox(0,0)[lb]{\smash{\SetFigFont{10}{12.0}{rm}$\mycalu'$}}}
\end{picture}